\documentclass{aa}
\usepackage[varg]{txfonts}
\usepackage[english]{babel}
\usepackage{booktabs}
\usepackage{graphicx}
\usepackage{caption}
\usepackage{fixltx2e}
\usepackage{amsmath,amssymb}
\usepackage{url}
\usepackage{rotating}
\usepackage{natbib}
\usepackage{bm}
\usepackage{xcolor}
\bibpunct{(}{)}{;}{a}{}{,}

\begin{document}

\title{Dynamics of DiskMass Survey galaxies in refracted gravity}

 \author{V.~Cesare\inst{1,2} \and A.~Diaferio\inst{1,2} \and T.~Matsakos\inst{1} \and G.~Angus\inst{1}}
 \institute{Dipartimento di Fisica, Università di Torino, Via P. Giuria 1, 10125 Torino, Italy, \email{valentina.cesare@unito.it}
        \and Istituto Nazionale di Fisica Nucleare (INFN), Sezione di Torino, Torino, Italy 
    }
\date{Received <date> / Accepted <date>}

 \abstract
{We aim to verify whether refracted gravity (RG) is capable of describing the dynamics of disk galaxies without resorting to the presence of dark matter. RG is a classical theory of gravity in which the standard Poisson equation is modified with the introduction of the gravitational permittivity, which is a universal monotonic function of the local mass density.  
We used the rotation curves and the radial profiles of the stellar velocity dispersion perpendicular to the galactic disks of 30 disk galaxies from the DiskMass Survey (DMS) to determine the gravitational permittivity. 
RG describes the rotation curves and the vertical velocity dispersions by requiring galaxy mass-to-light ratios that are in agreement with stellar population synthesis models, and disk thicknesses that are in agreement with observations, once observational biases are taken into account.
 Our results rely on setting the three free parameters of the gravitational permittivity for each individual galaxy. However, we show that the differences of these parameters from galaxy to galaxy can, in principle, be ascribed to statistical fluctuations. We adopted an approximate procedure to estimate a single set of parameters that may properly describe the kinematics of the entire sample and suggest that the gravitational permittivity is indeed a universal function.
Finally, we showed that the RG models of the individual rotation curves can only partly describe the radial acceleration relation (RAR) between the observed centripetal acceleration derived from the rotation curve and the Newtonian gravitational acceleration originating from the baryonic mass distribution. Evidently, the RG models underestimate the observed accelerations by 0.1 to 0.3 dex at low Newtonian accelerations. An additional problem that ought to be considered is the strong correlation, at much more than $5\sigma$, between the residuals of the RAR models and three radially-dependent properties of the galaxies, whereas the DMS data show a considerably less significant correlation, at more than $4\sigma$, for only two of these quantities. 
These correlations might be the source of the non-null intrinsic scatter of the RG models: this non-null scatter is at odds with the observed intrinsic scatter of other galaxy samples different from DMS, which is consistent with zero.
Further investigations are required to assess whether these discrepancies in the RAR originate from the DMS sample, which might not be ideal for deriving the RAR, or whether they are genuine failures of RG.}

\maketitle      

\section{Introduction}
\label{sec:intro}
        One of the most outstanding open questions in astrophysics is the mass discrepancy problem: the apparent amount of mass in the Universe is roughly ten times larger than the mass that is visible through its electromagnetic emission~\citep{Ost&Peeb73,OPY74,sanders2010dark}. This discrepancy occurs all the way from the largest scale of cosmic microwave background radiation to the galactic scale whenever we model the observed dynamics of astrophysical systems or their gravitational lensing features, using general relativity or its Newtonian weak field limit~\citep{Rubin&Ford70,Sanders1990,BullClust16,PlanckCMB}.
        Such observations are usually reconciled with expectations from the theory of gravity by assuming the existence of non-baryonic dark matter, the specific properties of which are still under debate~\citep{vAlbetal85,Vikhlininetal06,Kunzetal16,WDM,SICDM,WIMPandbeyond}.
        
        The most widely investigated interpretation is the cold dark matter (CDM) paradigm~\citep[e.g.][]{CDM}, where the cosmic structure forms as a result of the aggregation of smaller structures. In this context of substantially stochastic merging processes, some regularities in the observed properties of disk galaxies do not appear to occur naturally; they might, rather, require a substantial fine-tuning between the properties of the baryonic matter and the expected properties of the dark matter halo embedding the galaxy~\citep{McG05,Famaeyetal18}. For example, we might not naively expect a tight relation between the flat rotation velocity $v_\mathrm{f}$ and
        the baryonic  mass, the so-called baryonic Tully-Fisher relation~\citep{T&F77,McGetal00,Lellietal16BTFR}, because $v_\mathrm{f}$ is set by 
        the depth of the gravitational potential of the dark matter halo that contains $\sim$$90$\% of the total mass of the galaxy and should hardly
        be affected by the $10$\% baryonic mass: in the CDM framework, a very careful balance between star formation efficiency and stellar feedback might
        be necessary~\citep{McG12,Lellietal16BTFR}.
        Similarly, the observed centripetal acceleration implied by the rotation curve, $g_\text{obs}=v^2_\text{obs}(R)/R$, is tightly correlated with the Newtonian acceleration due to the baryonic matter distribution, $g_\text{bar}$, and the two accelerations perfectly
        coincide only above a single acceleration scale which is common to all galaxies \citep{McGetal16}, whereas, 
        at decreasing accelerations, the discrepancy between $g_\text{obs}$ and $g_\text{bar}$ monotonically increases. This radial acceleration 
        relation (RAR), although recently disputed~\citep{Rodriguesetal18}, could be particularly relevant because both $g_\text{obs}$ and $g_\text{bar}$ and their uncertainties are completely
        independent of each other \citep{Lietal18}.
        
        These observed regularities, although some of them appear reproducible in the CDM model~\citep[e.g.][]{Ludlowetal17}, might suggest that an alternative solution to the dark matter paradigm is a modification of the theory of gravity. MOdified Newtonian Dynamics (MOND,~\citealt{Milgrom83}) is capable of  modeling, even sometime predicting~\citep{Sand&McG02}, these observations by assuming a breakdown of the Newtonian gravity in low-acceleration
        environments, where the acceleration threshold is set to  $a_0=1.2\times10^{-10}$~m~s$^{-2}$ by observations~\citep[e.g.][]{McG04}.
        
        More recently, refracted gravity (RG,~\citealt{M&D16}) has proposed a different modification of the theory of gravity that is based on a completely different idea from MOND but it is expected to share most of its successes.  RG is a classic theory of gravity, whose modified Poisson equation includes the gravitational permittivity, $\epsilon(\rho)$, a monotonic function of the local mass density, $\rho$, that boosts the gravitational field in low-density environments. 
        RG can be reformulated as a scalar-tensor theory (Sanna et al., in preparation) and would thus share most of their general properties~\citep[e.g.][]{ST1,ST2}. 
        
         Specifically, the scalar field, which is non-minimally coupled
to the gravitational field, is responsible both for the gravitational
permittivity and could, thus, remove the need for dark matter and
for the accelerated expansion of the universe.
This feature is particularly attractive because both dark matter and dark energy can be mimicked by a single scalar field, similarly to other models that attempt to unify dark matter and dark energy, for example, $f(R)$ theories~\citep[e.g.][]{fofR}, quartessence theories~\citep[e.g.][]{quartessence}, mimetic gravity~\citep{MimeticGravity}, or generalised Chaplygin gas and beyond~\citep{Chaplyginandbeyond}. Since the scalar field in RG is a dynamical quantity, RG predicts a time evolution of the equation of state of the effective dark energy that can, in principle, be measured by upcoming space missions like Euclid.\footnote{\url{https://sci.esa.int/web/euclid/}}

    Here, we test the viability of RG by modelling the observed dynamics of disk galaxies. To provide the most stringent tests of the full dynamics of a disk galaxy, rather than modelling rotation curves alone, we consider a sample of galaxies where both the rotation curves and the velocity dispersion profiles, in the direction perpendicular to the disk, are available. 

        The DiskMass Survey (DMS,~\citealt{DMSi}) provides a sample that is fitting for our purpose. It contains 46 galaxies from the Uppsala General Catalogue (UGC) whose disks appear close to face-on; for 30 galaxies the measures of both the rotation curves and the vertical velocity dispersion profiles are publicly available. The DMS collaboration modelled the galaxy dynamics with Newtonian gravity by adopting the disk-scale heights derived from the observations of edge-on galaxies; they obtained sub-maximal disks, with mass-to-light ratios that are systematically smaller than what is expected from stellar population synthesis (SPS) models~\citep{B&deJ01}. 
        
        The DMS sample was also used by~\citet{Angus15} to test MOND; they found mass-to-light ratios consistent with the SPS values but with disk-scale heights systematically smaller than those inferred from the observations of edge-on galaxies. \citet{Milgrom15}, however, pointed out that this inconsistent disk thickness might originate from an observational bias: the measured velocity dispersion is inferred from the absorption lines near the $V$-band of the integrated spectra, which are dominated by the younger stellar population~\citep{Aniyan16}; this measured velocity dispersion is thus smaller than the velocity dispersion of the older stellar population which sets the estimate of the disk-scale height from near infrared photometry of edge-on galaxies~\citep{Kregel02,Pohlen00,S&D00,Xilouris97,Xilouris99,DMSii}. 
        
        This bias would also explain the low mass-to-light ratios estimated by the DMS collaboration because the disk surface mass density is proportional to the
        ratio between the velocity dispersion and the disk-scale height and it is thus underestimated~\citep{Aniyan16}.
The analysis of the dynamics of the DMS galaxies that we present here could also be affected by this bias. In Sect.~\ref{sec:VVD_RC_1pt55}  below, we estimate the amount of this bias and find that it is indeed consistent with the estimates of~\citet{Milgrom15} and~\citet{Aniyan16}.

        The outline of this paper is as follows. Section~\ref{sec:RG} summarises the main features of the RG theory. In Sect.~\ref{sec:Only_RC}, we test RG
        by modelling the rotation curves of the DMS galaxies alone, whereas in Sect.~\ref{sec:VVD_RC}, we model both their rotation curves and their vertical velocity dispersion profiles. We describe our model of the galaxy mass distribution and our Poisson solver in Appendixes~\ref{sec:SBgas} and~\ref{sec:SOR}, respectively.  
         
    Modelling the dynamics of each galaxy in RG requires two parameters for the galaxy, namely, its mass-to-light ratio and its disk-scale height, along with three RG free parameters. In Sect.~\ref{sec:Fit_all}, we show that the DMS sample could, in principle, be modelled by a single set of these three free RG parameters.
        In Sect.~\ref{sec:RAR}, we show that RG can also model the RAR of the DMS sample, although some tensions indeed exist. We present our conclusions in Sect.~\ref{sec:conclusions}. 
        We adopt the Hubble constant $H_0=73$~km~s$^{-1}$~Mpc$^{-1}$, as per~\citet{DMSvi}, throughout this paper.

        \section{Refracted Gravity}
        \label{sec:RG}
        
        RG is a classical theory of gravity inspired by the behaviour of electric fields in matter \citep{M&D16}: when an electric field line crosses a dielectric medium with a non-uniform permittivity, it suffers a change both in direction, namely, a refraction, and in magnitude. To mimic this behaviour in a gravitational field, RG adopts
        the modified Poisson equation:
        \begin{equation}
        \label{eq:PoissonRG}
                \nabla\cdot[\epsilon(\rho)\nabla\phi]=4\pi G\rho \; ,
        \end{equation}
        where $\phi$ is the gravitational potential, and $\epsilon(\rho)$ is the gravitational permittivity, an arbitrary monotonically increasing function of the mass density $\rho$.
        By adopting for $\epsilon(\rho)$ the asymptotic limits,       
        
         \begin{equation}
        \label{eq:eps2reg}
        \epsilon(\rho)=
        \begin{cases}
        1, & \rho\gg\rho_\mathrm{c}\\
        \epsilon_0, & \rho\ll\rho_\mathrm{c} \; ,
        \end{cases}
        \end{equation}
        the RG Poisson equation is reduced to the Newtonian form,
        \begin{equation}
        \label{eq:PoissonN}
        \nabla^2\phi=4\pi{G}\rho \; ,
        \\ \end{equation}
        in environments where the local density is much larger than the critical density $\rho_\mathrm{c}$. On the contrary, for a constant $\epsilon_0<1$, the gravitational field is boosted in low-density environments.

        For a spherically symmetric system, with mass $M(<$$r)$ within the radius $r$, the integration of Eq. \eqref{eq:PoissonRG} 
        yields $\partial\phi/\partial r = [G/\epsilon(\rho)]M(<$$r)/r^2$; therefore, the gravitational field has the same direction and the same dependence on $r$ as the Newtonian field, but it is enhanced by the factor $1/\epsilon(\rho)$. For systems that are not spherically symmetric, 
   expanding the divergence in Eq.~\eqref{eq:PoissonRG}, 
   \begin{equation}
   \label{eq:PoissonRG_inn_prod_1st_member}
           \frac{\partial\epsilon}{\partial\rho}\nabla\rho\cdot\nabla\phi+\epsilon(\rho)\nabla^2\phi = 4\pi G \rho \; ,
   \end{equation}
   shows that $\phi$ depends both on the density field $\rho$, 
   according to the second term in the left-hand side of the equation, as in the
   spherically symmetric case, and on its variation, according to the first term in the left-hand side of the equation. 
   
   We thus see that the analogy between the gravitational field and the electric field in a dielectric medium occurs for non-spherical systems. For example, in disk galaxies, the boost of the gravitational field can be visualised as a focussing of the gravitational field lines towards the disk plane, as they are refracted by the low-density regions above and below the disk. 
   According to this feature, RG predicts that increasingly flatter systems should require increasing dark matter content when interpreted in Newtonian gravity, as suggested by preliminary studies of elliptical galaxies~\citep{Deur2014}.
   
   As~\citet{M&D16} show, this focussing yields, for the gravitational acceleration $g$ in low-density regions at large distances $R$ from the disk centre, the asymptotic behaviour $g \sim (|g_\mathrm{N}|a_0)^{1/2}\propto R^{-1}$, where $g_\mathrm{N}$ is the Newtonian acceleration and $a_0$ coincides with the
   MOND critical acceleration that is set by the observed normalisation of the Tully-Fisher relation. This asymptotic behaviour is identical to the MOND limit in low-acceleration environments and suggests that the successes of MOND on the scale of galaxies should be shared by RG.

   Adopting RG, rather than MOND, as the modified theory of gravity is advantageous mostly from a theoretical point of view. 
   In MOND, the transition from a Newtonian regime to a regime of modified gravity is driven by the gravitational acceleration generated by the ordinary matter. The acceleration scale for this transition is indeed supported by extended observational evidence~\citep[e.g.][]{Milgrom83ii,Sand&McG02,McG04,McGetal16}. From a theoretical perspective, however, this feature has made the construction of a covariant formulation of MOND considerably challenging~\citep[e.g.][]{TeVeSSkordis,TeVeSproblems}. On the contrary, the adoption of a scalar quantity, like the density $\rho$, appears to simplify this task for RG. In addition, as suggested in~\citet{M&D16}, RG might in principle reproduce the phenomenology properly described by MOND without necessarily explicitly inserting an acceleration scale in the theory.

   For simplicity, RG assumes that the permittivity $\epsilon$ only depends on the density $\rho$ of ordinary matter. In principle, the gravitational sources are characterised by other scalar quantities, including their total mechanical and thermodynamical energy or their entropy. However, these quantities partly depend on the mass density and we might expect that adopting a more complex dependence of the permittivity $\epsilon$ might return a phenomenology that is comparable to the one we investigate here by assuming a simple dependence on $\rho$. 

   Clearly, all these issues remain unsettled at this stage of the investigation of RG: suggestions on how they could be properly tackled might originate from a better understanding of the connection between the permittivity $\epsilon$ and the scalar field $\varphi$ appearing in the covariant formulation of RG. Before exploring this connection, however, we need to investigate whether RG is indeed comparable to MOND in the description of the kinematic properties of disk galaxies. This is the task we intend to accomplish here.
        
\begin{figure}
                \resizebox{\hsize}{!}{\includegraphics{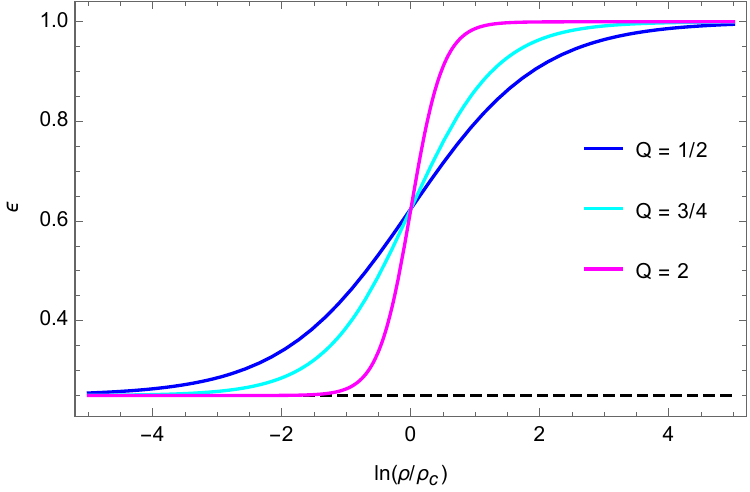}}
                \caption{Gravitational permittivity for different values of $Q$. The black dashed line shows $\epsilon_0=0.25$.}
                \label{fig:epsilonQ}
        \end{figure}

        As a test case for the quantitative analysis we describe in this work, following \citet{M&D16}, we adopt a smooth step function for the gravitational permittivity, 
        \begin{equation}
   \label{eq:eps}
   \epsilon(\rho)=\epsilon_0+(1-\epsilon_0)\frac{1}{2}\left\{\tanh\left[\text{ln}\left(\frac{\rho}{\rho_\mathrm{c}}\right)^Q\right]+1\right\} \; ,
   \end{equation}
   which depends on three parameters that we expect to be universal: the permittivity of the vacuum $\epsilon_0$, the power index $Q$, and the critical density $\rho_\mathrm{c}$. The parameter $\epsilon_0$ is limited in the range of $[0,1]$ by the definition of $\epsilon(\rho)$ and
   its asymptotic limits in Eq. \eqref{eq:eps2reg}. The parameter $Q$ sets the steepness of the transition between the Newtonian and the RG regimes. The critical density, $\rho_\mathrm{c}$ , sets the local density where this transition occurs. Figure~\ref{fig:epsilonQ} shows an example of the gravitational permittivity for different values of $Q$ and for $\epsilon_0=0.25$. We emphasise that Eq.~\eqref{eq:eps} is just an arbitrary expression for $\epsilon(\rho)$ that we choose here to test the viability of RG. Other
   expressions for $\epsilon(\rho)$, ones that can still increase monotonically with $\rho$ and have the asymptotic limits of Eq. \eqref{eq:eps2reg}, are clearly possible.
   
   We conclude this section with a brief comment on RG, MOND, and electrostatics. RG was inspired by the behaviour of electric fields in matter, but the connection between RG and electrostatics does not go beyond the phenomenological formulation of the modified Poission equation~\eqref{eq:PoissonRG}. A completely different idea, still based on electrostatics, has instead been developed in a number of papers~\citep{DipolarMOND1,DipolarMOND2,DipolarMOND3,DipolarMOND4,DipolarMOND5}: the phenomenology described by MOND is interpreted by introducing, in addition to the standard CDM particles, a dark fluid subject to a polarisation in a gravitational field, similarly to the electrostatic polarisation of a dieletric medium. In this dipole dark matter model, the mechanism of gravitational polarisation is guaranteed by the presence of a vector field~\citep{DipolarMOND5}. This dipole dark matter model has no connection nor any similarity with RG; moreover, RG, at least at this stage of its development, clearly benefits from a much simpler framework in both its phenomenological and covariant formulations. A different issue is whether RG can indeed describe the observed properties of real systems, as we intend to investigate in the present work.

\section{Modelling the rotation curves alone of the DMS galaxies}
\label{sec:Only_RC}

To test whether RG can describe the dynamics of disk galaxies, we first consider the rotation curves of the 30 published galaxies of the DMS catalogue on their own \citep{DMSi,DMSii,DMSiii,DMSiv,DMSvi,DMSvii}.

For an axisymmetric mass density distribution $\rho(R,z)$, the Poisson equation~\eqref{eq:PoissonRG} returns 
the gravitational potential $\phi(R,z)$ that, in turn, yields the rotation curve,

\begin{equation}
\label{eq:RCespl}
        v(R,z=0)=\left[R\frac{\partial \phi(R,z)}{\partial R}\right]^{1/2} \; ,
\end{equation}
on the disk plane, $z=0$. 

We model each disk galaxy with a stellar disk, a stellar bulge, and an interstellar gas disk separated into an atomic and a molecular component. The stellar disk is described by a linear interpolation of the measured radial surface brightness and by an exponentially decreasing density profile along the vertical axis. The stellar bulge is described by a S\'ersic profile. The details of our model of the mass distribution are given in Appendix~\ref{sec:SBgas}.

To model the galaxy rotation curves, we estimate the RG potential by numerically solving the Poisson equation~\eqref{eq:PoissonRG} with a successive over relaxation algorithm described in Appendix~\ref{sec:SOR}. We then perform the numerical derivative of the potential to obtain the rotation curve from Eq.~\eqref{eq:RCespl}. In our model, the rotation curve depends on two parameters describing the galaxy, namely the disk mass-to-light 
ratio, $\Upsilon$, and the disk-scale height, $h_z$, and on the three parameters of the RG gravitational permittivity: $\epsilon_0$, $Q$, and $\rho_\mathrm{c}$. 

We adopt the same mass-to-light ratio for the bulge and the stellar disk because, in our sample, the galaxy luminosity is dominated by the disk (see Appendix~\ref{sec:SBgas}); assuming a different mass-to-light ratio for the bulge only introduces an additional free parameter without substantially improving the galaxy model.
We explore this five-dimensional parameter space with a Monte Carlo Markov Chain (MCMC) algorithm. 

We assume a Gaussian prior for the two galaxy parameters $\Upsilon$ and $h_z$ and a flat prior for the three RG parameters. Specifically:
\begin{enumerate}
        \item For the mass-to-light ratio $\Upsilon$ in the $K$-band, adopted for the measurement of the surface brightness of the DMS galaxies, we use a Gaussian prior; the first moment of the Gaussian is the value derived from the SPS models of~\citet{B&deJ01} applied to the DMS galaxies; for these galaxies the $B-K$ colour ranges from $2.7$ (\object{UGC 7244}) to $4.2$ (\object{UGC 4458})~\citep[see][Table 1]{DMSvi}. We set the second moment of the Gaussian to three times the maximum errors derived from the SPS models for each galaxy. This choice yields a second moment in the range of 0.21-0.36 dex.\footnote{
                Setting the second moment of the Gaussian to three times the error from the SPS models, rather than just the error, enables the
                MCMC analysis to explore a sufficiently extended  
                range of $\Upsilon$; in fact, the relative error on the SPS $\Upsilon$'s 
                is 16\%, on average, and, with the second moment of the Gaussian set to this value, the preferred value suggested by the MCMC analysis would often be forced to be close to the SPS value, independently of the theory of gravity we want to test. The same argument holds for the disk-scale height $h_{z,\text{SR}}$, whose relative error is 22\%, on average. }
                We set the Gaussian tail to zero where $\Upsilon<0$.
        \item For the disk-scale height $h_z$, we adopt a Gaussian prior with mean $h_{z,\text{SR}}$, where $h_{z,\text{SR}}$ is the disk-scale height derived from the relation, reported in Eq. \eqref{eq:hzhR} in Appendix \ref{sec:SBgas}, between the observed disk-scale heights and the disk-scale lengths, inferred from the observations of edge-on galaxies~\citep{DMSii}. We set the standard deviation of the Gaussian to three times the errors on $h_{z,\text{SR}}$,
                which basically coincide with the intrinsic scatter of the relation~\eqref{eq:hzhR}. On average, the error on $h_{z,\text{SR}}$ is $0.11$~kpc for the DMS galaxies. We set the Gaussian tail to zero where $h_z<0$.
        \item For the vacuum permittivity $\epsilon_0$, we adopt a flat prior in the range of $[0.10,1]$. In principle, the full allowed range is  $[0,1]$; however, we do not explore values smaller than $0.10$ because the boost of the gravitational field would yield unphysically large rotation velocities, of the order of $\sim$$600-1000$~km s$^{-1}$.
        \item For $Q$, we adopt a flat prior in the range  of $[0.01,2]$. This parameter regulates the steepness of the transition between the Newtonian and the RG regimes. Our range explores from very smooth ($Q=0.01$) to very steep transitions ($Q=2$).
        \item For $\log_{10} \rho_\mathrm{c}$, we adopt a flat prior in the range of $[-27,-23]$, with the critical density $\rho_\mathrm{c}$ in units of g~cm$^{-3}$. 
This range includes the two extreme values $-27$ and $-24$ considered by~\citet{M&D16}.
\end{enumerate}

We adopt the Metropolis-Hastings acceptance criterion in our MCMC algorithm. 
 The random variate ${\mathbf x}$ at step $i+1$ is drawn from the probability density $G({\mathbf x}\vert {\mathbf x}_i)$, which depends on the random variate ${\mathbf x}_i$ at the previous step. For the probability density $G({\mathbf x}\vert {\mathbf x}_i)$, we adopt a multi-variate Gaussian density distribution with mean value ${\mathbf x}_i$; its multiple standard deviations are $1/3$ the standard deviation of the Gaussian priors, for $\Upsilon$ and $h_z$, and 10\% of the prior ranges, for the three RG parameters. In our case, ${\mathbf x}$ is the five-dimensional vector ${\mathbf x}=(\Upsilon, h_z, \epsilon_0, Q, \log_{10} \rho_\mathrm{c})$.
We adopt the likelihood,
\begin{equation}
\label{eq:Px}
\mathcal{L}({\mathbf x})=\exp\left[-\frac{\chi^2_\mathrm{red,RC}({\mathbf x})}{2}\right]\; , 
\end{equation}
where,
\begin{equation}
\label{eq:chi2RC}
\chi^2_\mathrm{red,RC}({\mathbf x})={\frac{1}{n_\mathrm{dof,RC}}}\sum_{i=1}^{N_\mathrm{RC}}\frac{[v_\mathrm{mod}(R_i; {\mathbf x})-v_\mathrm{data}(R_i)]^2}{v_\mathrm{data,err}^2(R_i)}\; ,
\end{equation}
$N_\mathrm{RC}$ is the number of data points of the rotation curve, $v_\mathrm{data}$ are the velocity measures at the projected distance $R_i$ with their uncertainty $v_\mathrm{data,err}$, $v_\mathrm{mod}$ is estimated with Eq.~\eqref{eq:RCespl} and $n_\mathrm{dof,RC}=N_\mathrm{RC}-5$ is the number of degrees of freedom.
If $\mathit{p}({\mathbf x})$  is the product of the priors of the components of ${\mathbf x}$, the Metropolis-Hastings ratio is:
\begin{equation}
\label{eq:MHratio}
        A = \frac{\mathit{p}({\mathbf x})\times\mathcal{L}({\mathbf x})}{\mathit{p}({\mathbf x}_i)\times\mathcal{L}({\mathbf x}_i)}\frac{G({\mathbf x}\vert {\mathbf x}_i)}{G({\mathbf x}_i\vert {\mathbf x})} \; .
\end{equation}
If $A \ge 1$, we set ${\mathbf x}_{i+1}={\mathbf x}$, otherwise we set either ${\mathbf x}_{i+1}={\mathbf x}$, with probability $A$, or ${\mathbf x}_{i+1}={\mathbf x}_i$, with probability $1-A$.

For the chain starting points, we adopt the values found by the DMS collaboration for $\Upsilon$ and $h_z$ \citep[see][Table 1]{Angus15}; for the three RG parameters $\epsilon_0$, $Q$ and $\log_{10} \rho_\mathrm{c}$, we set $0.30$, $1.00$, and $-24.0$, respectively. We run the MCMC algorithm for 19000 steps, after a burn-in chain of 1000 steps. This number of steps guarantees the achievement of the chain convergence. We check the chain convergence using the Geweke diagnostic~\citep{Geweke92}: we compare the means of the first 10\% of the chain steps, after the burn-in, with the last 50\% of the chain steps; we compare the means with a Gaussian test, adopting the standard deviations of the two portions of the chains as the errors on the two means. In every case, the Gaussian test shows that the two means coincide at a significant level larger than $5\%$, suggesting that the chains converge.

\begin{figure*}
        \centering
        \includegraphics[width=17cm]{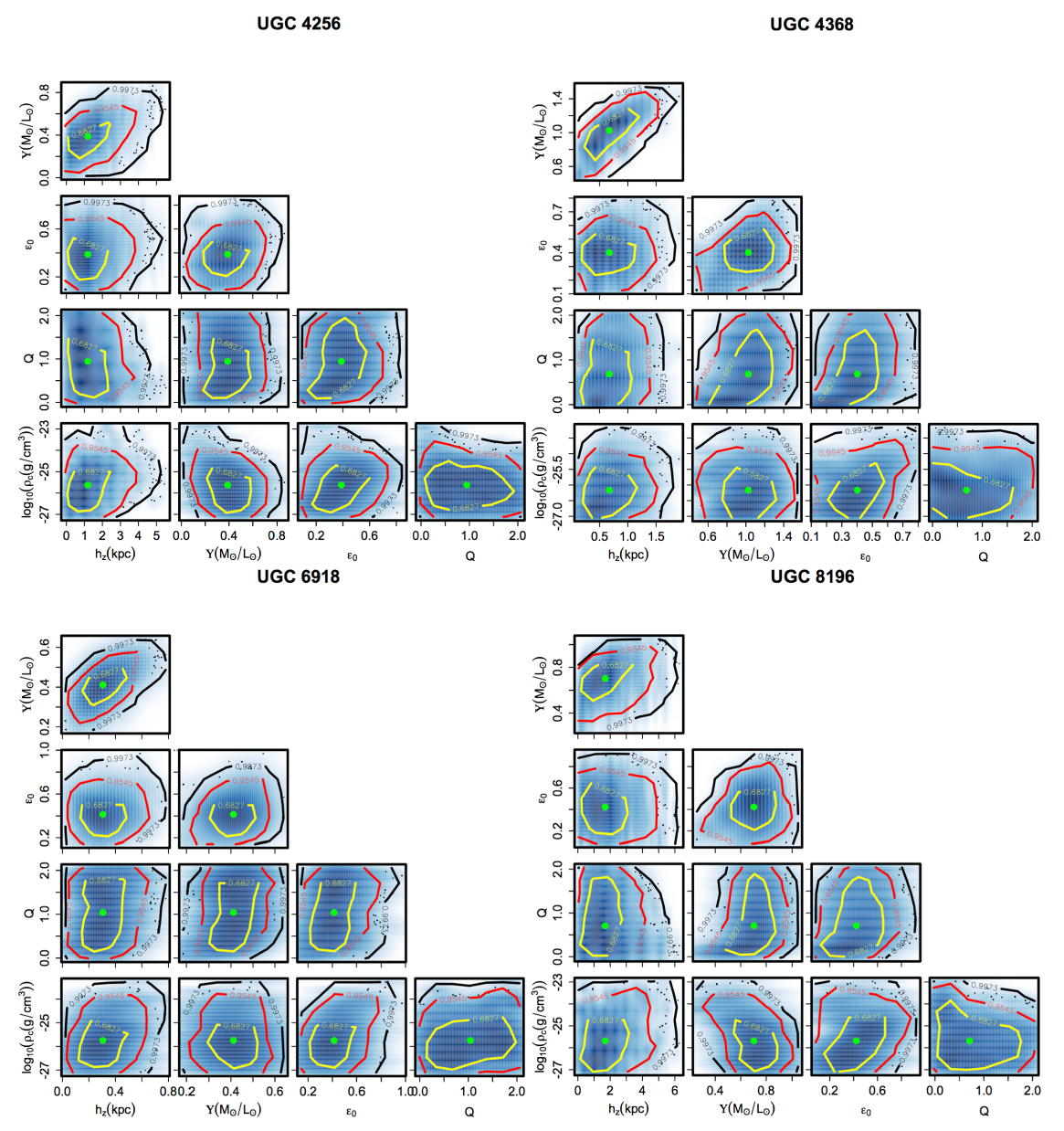}
        \caption{Examples of the posterior distributions for four galaxies. The quantities plotted are the two galaxy parameters and the three RG parameters estimated from the rotation curves alone. The green dots locate the median values and the yellow, red and black contours limit the $1\sigma$, $2\sigma$ and $3\sigma$ regions, respectively.}
        \label{fig:CP_v_4}
\end{figure*}

As an example, Fig. \ref{fig:CP_v_4} shows the posterior distributions for four galaxies. The posterior distributions for the remaining galaxies are 
qualitatively very similar. The posterior distributions show a single peak and we can thus adopt the medians of the posterior distributions as our parameter estimates; the range between the 15.9 and the 84.1 percentiles, thus including 68\% of the posterior cumulative distribution centred on the median, defines our $1\sigma$ uncertainty range on the parameter estimates. Table~\ref{tab:fit_RC_5_par} lists the medians of the parameters and their associated uncertainties.

We use these parameters to compute our rotation curve models. We collect all the figures showing our results in Appendix~\ref{sec:Figures}. We arrange the figures by galaxy, so that the outcomes of the various analyses we perform here can be compared more easily. 

The rotation curves estimated in this section are shown in sub-panels (d) of Figs. \ref{fig:Complete_analysis_1}-\ref{fig:Complete_analysis_7} as blue solid lines.
The red dots with error bars are the DMS data.
The vertical lines show the size of the bulge we adopt. In some galaxies, the presence of the bulge produces, at small radii, a relevant spike in the modelled rotation curve that is not present in the data. Other than these cases, the observed rotation curves are modelled relatively well. Because we model the surface brightness of the disk with a linear interpolation of the data  (Appendix~\ref{sec:SBgas}), the model rotation curves capture some of the features appearing in the measured rotation curves that have a correspondence in the surface brightness profile of the disk; the galaxies \object{UGC 1635}, \object{UGC 4555}, \object{UGC 6903}, or UGC 7244 are some examples of this correspondence. Nevertheless, the rotation curves of a few galaxies still have some features that are not described by the model, for example \object{UGC 4036}, \object{UGC 4622} or UGC \object{8196}.

The $\chi_{\rm red, RC}^2$ of Eq.~\eqref{eq:chi2RC}  are listed in Table~\ref{tab:fit_RC_5_par}; they quantify the quality of the RG model. In most cases, the large values of $\chi_{\rm red, RC}^2$ originate from a possible underestimation of the error bars of the data, as suggested by the visual inspection of the sub-panels (d) of Figs.~\ref{fig:Complete_analysis_1}-~\ref{fig:Complete_analysis_7}, rather than from an inappropriate modelling.

Table~\ref{tab:fit_RC_5_par} also lists the mass-to-light ratio $\Upsilon_{\rm SPS}$ that we estimate for each galaxy with the SPS models of \citet{B&deJ01} from the $B-K$ colours listed in Table 1 of~\citet{DMSvi}.\footnote{ We adopt the SPS models of~\citet{B&deJ01}, rather than more recent models of the relations between mass-to-light and colours~\citep[e.g.][]{McG&S14,Schombert19SPS} because~\citet{B&deJ01} specifically compute these relations for the $B-K$ colour, which is available in the DMS sample. } Specifically, we adopt the SPS model with a mass-dependent formation epoch with bursts and a scaled Salpeter initial mass function (IMF)~\citep[see][Table 1]{B&deJ01}. According to~\citet{B&deJ01}, this model better reproduces (1) the trends in colour-based stellar ages and metallicities~\citep{BeandBo00}, (2) the decrease in the colour-stellar mass-to-light ratio slope caused by modest bursts of star formation, and (3) it has an IMF consistent with maximum disk constraints. 

To somehow quantify the uncertainty on $\Upsilon_{\rm SPS}$, for the error bar we adopt the range covered by all the different models investigated  by~\citet{B&deJ01} based on a scaled Salpeter IMF~\citep[see][Table A3]{B&deJ01}. The resulting uncertainties are asymmetric and, except for three galaxies, the lower limit of the error bar is $0$ because, in these cases, our preferred SPS model yields the smallest mass-to-light ratio among all the other models with the same IMF. We neglect the models that adopt different IMFs, under the assumption that the scaled Salpeter IMF can reasonably be considered universal~\citep{B&deJ01}. We do not estimate the SPS mass-to-light ratio for the galaxy \object{UGC 3997}, since~\citet{DMSvi} does not provide its $B-K$ colour. 

\begin{figure*}
        \centering
        \includegraphics[width=17cm]{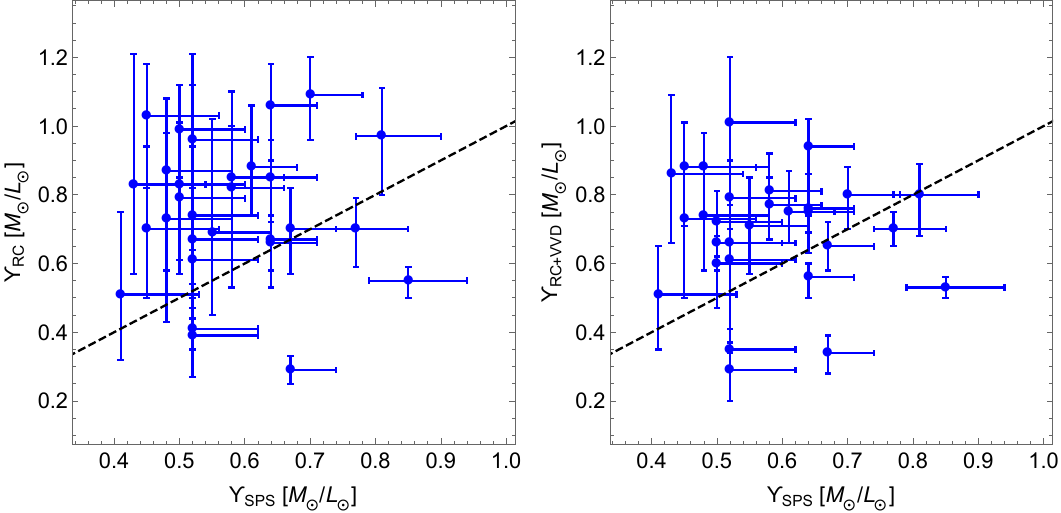}
        \caption{Mass-to-light ratios estimated with RG from the measured rotation curves alone ($\Upsilon_\text{RC}$ -- {\it left panel}) and with the rotation curves and the vertical velocity dispersion profiles  ($\Upsilon_\text{RC+VVD}$ -- {\it right panel}) compared with the mass-to-light ratios derived with the SPS model ($\Upsilon_\text{SPS}$), for each DMS galaxy. The black dashed line is the line of equality.}
        \label{fig:SP_YRC_YRCVVD_YSPS}
\end{figure*}

The two mass-to-light ratios for each galaxy, namely, the  $\Upsilon_{\rm SPS}$ and $\Upsilon$ derived from our RG model, are shown in the left panel of Fig.~\ref{fig:SP_YRC_YRCVVD_YSPS}. In Table~\ref{tab:fit_RC_5_par} we list the difference between these ratios in units of the uncertainties on the mass-to-light ratios. Our $\Upsilon$'s nearly cover the same range as the $\Upsilon_{\rm SPS}$'s and the two mass-to-light ratios for the same galaxy tend to agree with each other: for 23 out of 29 galaxies, they agree within $2\sigma$; for an additional five galaxies, they agree within $2.5$ to $3.5\sigma$; and for only one galaxy, they are discrepant at more than $6.5\sigma$.

\begin{figure*}
        \centering
        \includegraphics[width=17cm]{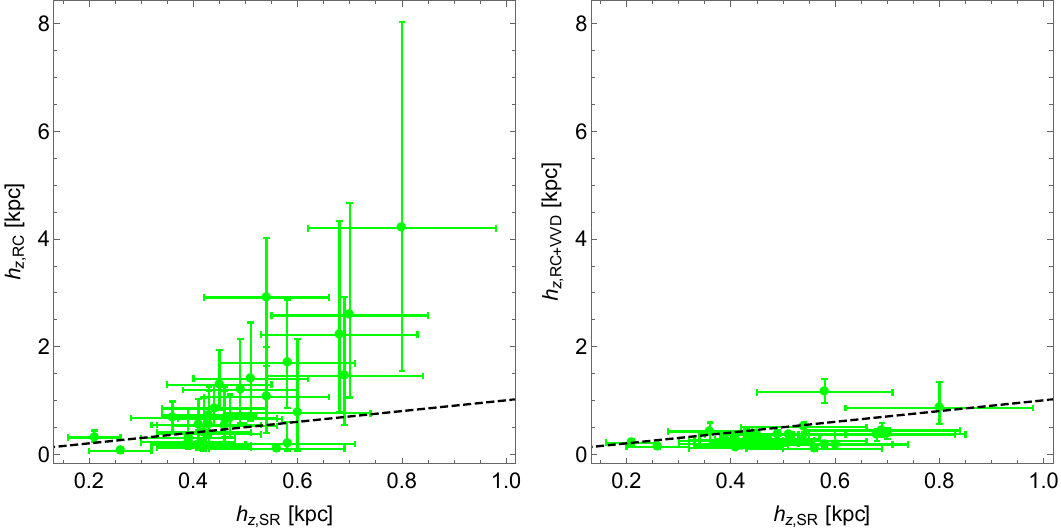}
        \caption{Disk-scale heights estimated with RG from the measured rotation curves alone ($h_{z,\text{RC}}$ -- {\it left panel}) and with the rotation curves and the vertical velocity dispersion profiles ($h_{z,\text{RC+VVD}}$ -- {\it right panel}) compared with the disk-scale heights inferred from the observations of edge-on galaxies ($h_{z,\text{SR}}$, see Eq.~\eqref{eq:hzhR}), for each DMS galaxy. The black dashed line is the line of equality.}
        \label{fig:SP_hzRC_hzRCVVD_hzSR}
\end{figure*}

The disk-scale height $h_z$ derived in RG tends to be larger than the scale height $h_{z,\text{SR}}$ calculated with Eq.~\eqref{eq:hzhR}, as shown in the left panel of Fig.~\ref{fig:SP_hzRC_hzRCVVD_hzSR}. Table~\ref{tab:fit_RC_5_par} also lists their ratios. Specifically, the RG $h_z$ is larger than $h_{z,\text{SR}}$ for 20 galaxies out of 30, and for nine galaxies $h_z/h_{z,\text{SR}}\ge 2$; among these nine galaxies, two have $h_z/h_{z,\text{SR}}\ge 5$. The ten galaxies
with $h_z/h_{z,\text{SR}}<1$ have thinner disks than expected because their gravitational potential wells are shallow: in fact, their central disk surface brightness $I_{\rm d0}$ and their disk-scale length $h_R$ are among the smallest in the sample, similarly to their rotation velocity and vertical velocity dispersion profiles.

\begin{table*}
        \caption{Parameters estimated from the rotation curves alone.}
        \label{tab:fit_RC_5_par}        
        \centering
        \begin{tabular}{lccccccccc}
                \hline\hline
                UGC & $\Upsilon$ $\Bigl[\frac{\text{M}_\odot}{\text{L}_\odot}\Bigr]$ & $\Upsilon_{\rm SPS}$ $\Bigl[\frac{\text{M}_\odot}{\text{L}_\odot}\Bigr]$ & $\frac{\Upsilon_{\rm fit}-\Upsilon_{\rm SPS}}{\sqrt{\sigma_{\Upsilon_{\rm fit}}^2+\sigma_{\Upsilon_{\rm SPS}}^2}      }$ &$h_z$ $[\text{kpc}]$ & $\frac{h_z}{h_{z,\text{SR}}}$ & $\epsilon_0$ & $Q$ & $\log_{10}\Bigl(\rho_\mathrm{c}$ $\Bigl[\frac{\text{g}}{\text{cm}^3}\Bigr] \Bigr)$&$\chi^2_\mathrm{red, RC}$\\
                (1) & (2) & (3) & (4) & (5) & (6) & (7) & (8) & (9) & (10) \\
                \hline
                448 & $1.06^{+0.12}_{-0.16}$ &$0.64^{+0.07}_{-0.00}$&$+2.88$& $0.52^{+0.65}_{-0.31}$ & $1.13$ & $0.62^{+0.24}_{-0.31}$ & $0.78^{+0.84}_{-0.65}$ & $-26.03^{+1.06}_{-0.58}$&$8.03$\\
                463 & $0.66^{+0.08}_{-0.08}$ &$0.64^{+0.07}_{-0.00}$&$+0.22$& $0.66^{+0.60}_{-0.34}$ & $1.43$ & $0.89^{+0.08}_{-0.14}$ & $0.92^{+0.73}_{-0.73}$ & $-25.37^{+1.82}_{-1.17}$&$10.93$\\
                1081 & $0.61^{+0.21}_{-0.17}$ &$0.52^{+0.10}_{-0.00}$&$+0.46$& $0.30^{+0.56}_{-0.18}$ & $0.77$ & $0.55^{+0.16}_{-0.16}$ & $1.34^{+0.40}_{-0.74}$ & $-23.72^{+0.55}_{-1.65}$&$16.34$\\
                1087 & $0.85^{+0.25}_{-0.32}$ &$0.58^{+0.08}_{-0.00}$&$+0.92$& $0.22^{+0.32}_{-0.13}$ & $0.56$ & $0.52^{+0.10}_{-0.11}$ & $0.78^{+0.56}_{-0.65}$ & $-25.78^{+1.81}_{-0.81}$&$10.67$\\
                1529 & $1.09^{+0.11}_{-0.13}$ &$0.70^{+0.08}_{-0.00}$&$+3.08$& $0.85^{+0.44}_{-0.31}$ & $1.93$ & $0.79^{+0.12}_{-0.16}$ & $0.86^{+0.83}_{-0.74}$ & $-25.86^{+1.09}_{-0.85}$&$42.03$\\
                1635 & $0.88^{+0.18}_{-0.14}$ &$0.61^{+0.07}_{-0.00}$&$+1.64$& $0.18^{+0.27}_{-0.10}$ & $0.44$ & $0.70^{+0.15}_{-0.16}$ & $0.60^{+0.79}_{-0.51}$ & $-24.48^{+1.19}_{-1.01}$&$8.58$\\
                1862 & $0.74^{+0.38}_{-0.33}$ &$0.52^{+0.10}_{-0.00}$ &$+0.61$& $0.06^{+0.06}_{-0.03}$ & $0.23$ & $0.64^{+0.19}_{-0.17}$ & $0.89^{+0.82}_{-0.75}$ & $-25.22^{+1.27}_{-1.34}$ &$23.16$\\
                1908 & $0.29^{+0.04}_{-0.04}$ &$0.67^{+0.07}_{-0.00}$&$-6.72$& $2.91^{+1.11}_{-1.27}$ & $5.37$ & $0.39^{+0.04}_{-0.04}$ & $1.86^{+0.13}_{-0.28}$ & $-23.86^{+0.10}_{-0.07}$&$32.02$\\
                3091 & $0.87^{+0.21}_{-0.29}$ &$0.48^{+0.10}_{-0.00}$&$+1.53$& $0.40^{+0.65}_{-0.23}$ & $0.95$ & $0.52^{+0.12}_{-0.22}$ & $1.06^{+0.75}_{-0.87}$ & $-23.58^{+0.48}_{-1.41}$&$11.62$\\
                3140 & $0.85^{+0.11}_{-0.13}$ &$0.64^{+0.07}_{-0.00}$&$+1.66$& $0.65^{+0.46}_{-0.34}$ & $1.55$ & $0.78^{+0.13}_{-0.17}$ & $0.77^{+0.82}_{-0.64}$ & $-25.63^{+1.70}_{-1.07}$&$8.75$\\
                3701 & $0.70^{+0.24}_{-0.20}$ &$0.45^{+0.11}_{-0.00}$&$+1.10$& $0.54^{+0.63}_{-0.37}$ & $1.26$ & $0.25^{+0.07}_{-0.06}$ & $0.87^{+0.67}_{-0.49}$ & $-23.79^{+0.32}_{-1.27}$&$2.13$\\
                3997 & $0.32^{+0.21}_{-0.10}$ &-&-& $0.19^{+0.41}_{-0.12}$ & $0.33$ & $0.23^{+0.14}_{-0.06}$ & $1.24^{+0.35}_{-0.20}$ & $-23.28^{+0.13}_{-0.21}$&$22.90$\\
                4036 & $0.79^{+0.20}_{-0.22}$ &$0.50^{+0.10}_{-0.00}$&$+1.34$& $0.54^{+0.48}_{-0.31}$ & $1.32$ & $0.55^{+0.19}_{-0.17}$ & $0.70^{+0.77}_{-0.55}$ & $-25.52^{+1.61}_{-1.07}$&$4.62$\\
                4107 & $0.82^{+0.18}_{-0.21}$ &$0.58^{+0.08}_{-0.00}$&$+1.18$& $0.17^{+0.25}_{-0.08}$ & $0.40$ & $0.81^{+0.11}_{-0.14}$ & $0.94^{+0.82}_{-0.62}$ & $-24.98^{+1.38}_{-1.45}$&$8.06$\\
                4256 & $0.39^{+0.11}_{-0.12}$ &$0.52^{+0.10}_{-0.00}$&$-1.00$& $1.19^{+0.95}_{-0.62}$ & $2.43$ & $0.39^{+0.12}_{-0.13}$ & $0.95^{+0.64}_{-0.57}$ & $-25.63^{+0.77}_{-0.79}$&$2.08$\\
                4368 & $1.03^{+0.15}_{-0.21}$ &$0.45^{+0.11}_{-0.00}$&$+3.06$& $0.67^{+0.31}_{-0.26}$ & $1.86$ & $0.40^{+0.09}_{-0.13}$ & $0.68^{+0.74}_{-0.53}$ & $-26.16^{+0.65}_{-0.58}$&$12.25$\\
                4380 & $0.96^{+0.25}_{-0.32}$ &$0.52^{+0.10}_{-0.00}$&$+1.50$& $1.06^{+0.93}_{-0.67}$ & $1.96$ & $0.47^{+0.18}_{-0.19}$ & $0.83^{+0.82}_{-0.74}$ & $-24.06^{+0.92}_{-2.06}$&$22.07$\\
                4458 & $0.55^{+0.04}_{-0.05}$ &$0.85^{+0.09}_{-0.06}$&$-3.18$& $4.20^{+3.84}_{-2.65}$ & $5.25$ & $0.77^{+0.15}_{-0.28}$ & $0.76^{+0.79}_{-0.56}$ & $-25.78^{+1.62}_{-0.95}$&$2.83$\\
                4555 & $0.73^{+0.25}_{-0.30}$ &$0.48^{+0.10}_{-0.00}$&$+0.88$& $0.12^{+0.15}_{-0.05}$ & $0.29$ & $0.91^{+0.05}_{-0.19}$ & $1.28^{+0.51}_{-0.86}$ & $-24.53^{+0.97}_{-1.10}$&$18.91$\\
                4622 & $0.69^{+0.33}_{-0.24}$ &$0.55^{+0.09}_{-0.00}$&$+0.48$& $2.58^{+2.09}_{-1.53}$ & $3.69$ & $0.45^{+0.18}_{-0.14}$ & $1.06^{+0.62}_{-0.95}$ & $-23.49^{+0.30}_{-1.24}$&$15.47$\\
                6903 & $0.67^{+0.27}_{-0.13}$ &$0.52^{+0.10}_{-0.00}$&$+0.73$& $0.10^{+0.07}_{-0.03}$ & $0.18$ & $0.37^{+0.26}_{-0.10}$ & $0.62^{+0.42}_{-0.39}$ & $-25.95^{+1.76}_{-0.59}$&$7.08$\\
                6918 & $0.41^{+0.06}_{-0.06}$ &$0.52^{+0.10}_{-0.00}$&$-1.41$& $0.31^{+0.13}_{-0.09}$ & $1.48$ & $0.41^{+0.12}_{-0.11}$ & $1.04^{+0.64}_{-0.60}$ & $-25.75^{+0.87}_{-0.72}$&$4.13$\\
                7244 & $0.51^{+0.24}_{-0.19}$ &$0.41^{+0.12}_{-0.00}$&$+0.44$& $0.72^{+0.53}_{-0.39}$ & $1.67$ & $0.23^{+0.06}_{-0.06}$ & $0.79^{+0.69}_{-0.42}$ & $-24.74^{+0.73}_{-1.20}$&$2.00$\\
                7917 & $0.97^{+0.14}_{-0.17}$ &$0.81^{+0.09}_{-0.04}$&$+0.92$& $1.45^{+1.47}_{-0.91}$ & $2.10$ & $0.85^{+0.11}_{-0.16}$ & $0.78^{+0.88}_{-0.64}$ & $-25.59^{+1.85}_{-1.06}$&$7.14$\\
                8196 & $0.70^{+0.09}_{-0.11}$ &$0.77^{+0.08}_{-0.03}$&$-0.60$& $1.69^{+1.18}_{-0.83}$ & $2.91$ & $0.42^{+0.16}_{-0.15}$ & $0.71^{+0.76}_{-0.48}$ & $-25.69^{+0.99}_{-0.86}$&$4.72$\\
                9177 & $0.83^{+0.38}_{-0.26}$ &$0.43^{+0.11}_{-0.00}$&$+1.23$& $2.22^{+2.11}_{-1.44}$ & $3.26$ & $0.46^{+0.19}_{-0.11}$ & $1.41^{+0.36}_{-0.63}$ & $-23.33^{+0.21}_{-0.69}$&$11.46$\\
                9837 & $0.67^{+0.18}_{-0.14}$ &$0.64^{+0.07}_{-0.00}$&$+0.18$& $0.76^{+1.38}_{-0.70}$ & $1.27$ & $0.29^{+0.18}_{-0.07}$ & $0.93^{+0.39}_{-0.34}$ & $-23.47^{+0.34}_{-0.20}$&$17.11$\\
                9965 & $0.83^{+0.18}_{-0.22}$ &$0.50^{+0.10}_{-0.00}$&$+1.60$& $0.37^{+0.38}_{-0.17}$ & $0.86$ & $0.78^{+0.15}_{-0.16}$ & $0.81^{+0.80}_{-0.63}$ & $-25.43^{+1.52}_{-1.05}$&$3.16$\\
                11318 & $0.70^{+0.12}_{-0.13}$ &$0.67^{+0.07}_{-0.00}$&$+0.22$& $1.40^{+1.05}_{-0.79}$ & $2.75$ & $0.66^{+0.23}_{-0.31}$ & $0.89^{+0.74}_{-0.63}$ & $-25.52^{+1.59}_{-1.00}$&$2.87$\\
                12391 & $0.99^{+0.13}_{-0.14}$ &$0.50^{+0.10}_{-0.00}$&$+3.30$& $1.29^{+0.64}_{-0.46}$ & $2.87$ & $0.37^{+0.26}_{-0.12}$ & $1.18^{+0.57}_{-0.66}$ & $-25.60^{+0.76}_{-0.80}$&$11.58$\\
                \hline
        \end{tabular}
    \tablefoot{Column 1: UGC number; Cols. 2, 5, 7, 8, 9:  medians of the posterior distributions of the model parameters estimated from the rotation curves alone; Col. 3: mass-to-light ratio derived with the SPS model; Col. 10: reduced chi-square, $\chi_{\rm red, RC}^2$, from Eq.~\eqref{eq:chi2RC}.}
\end{table*}

\section{Modelling the rotation curves and the vertical velocity dispersion profiles}
\label{sec:VVD_RC}

The DMS sample enables a more comprehensive investigation of the dynamics of disk galaxies because, in addition to the rotation curves, we have a measurement of their stellar vertical velocity dispersion profiles. In fact, the DMS galaxies 
are close to face-on: as illustrated in Appendix~\ref{sec:SBgas}, based on the Tully-Fisher relation (Eqs.~\eqref{eq:TF} and~\eqref{eq:iTF}), the estimated inclinations of the DMS galaxies are in the range of $5.8-45.3\deg$
\citep[see][Table 5]{DMSvi}.
Therefore, combining the two kinematic pieces of information can provide unique constraints on the dynamical properties of the galaxies and can represent a stringent test for theories of modified gravity. 

To model the stellar vertical velocity dispersion profile, we use the fact that our axisymmetric model of the galaxy, illustrated in Appendix~\ref{sec:SBgas}, is described by a two-integral distribution function $f(E,L_z)$ and thus the velocity dispersions, $\sigma(R,z)$, along the vertical and radial axes, $z$ and $R$, coincide~\citep{N&M76,Nipoti07}. The system also satisfies the Jeans equation, 
\begin{equation}
\label{eq:Jeans}
{\partial [\rho(R,z)\sigma^2(R,z)]\over \partial z} + \rho(R,z){\partial\phi(R,z)\over\partial z} = 0\; ,
\end{equation}
where $\rho(R,z)$ is the stellar density. The observed vertical velocity dispersion profile, weighted by the local stellar surface density $\Sigma_*(R)$, is
\begin{equation}
\label{eq:sigmaz2}
\sigma_z^2(R) = {1\over \Sigma_{*}(R)}\int_{-\infty}^{+\infty}\rho(R,z)\sigma^2(R,z) dz \; .
\end{equation}
By considering the contribution of the disk alone to $\Sigma_{*}(R)$ and $\rho(R,z)$, thus neglecting
the luminosity contribution of the bulge (see Appendix~\ref{sec:SBgas}), the stellar surface mass density profile is:
\begin{equation}
\label{eq:sigmastar}
\Sigma_{*}(R) = \int_{-\infty}^{+\infty} \rho_*(R,z)dz = \Upsilon_\mathrm{d} I_\mathrm{d}(R) \; ,
\end{equation}
where $\rho(R,z)=\rho_*(R,z)$ is our disk model, namely the surface brightness profile $I_\mathrm{d}(R)$ multiplied by $\exp(-|z|/h_z)/(2h_z)$ (Appendix~\ref{sec:SB}). Thus, combining Eqs.~\eqref{eq:sigmaz2},~\eqref{eq:sigmastar}, and~\eqref{eq:Jeans} yields:
\begin{equation}
\label{eq:VVD}
\sigma_z^2(R)=\frac{1}{h_z}\int_0^{+\infty}\left[\int_z^{+\infty}\exp\left(-\frac{|z'|}{h_z}\right)\frac{\partial\phi(R,z')}{\partial z'}dz'\right]dz\; .
\end{equation}

We model the two kinematic profiles, the rotation curve (Eq. \ref{eq:RCespl}), and the vertical velocity dispersion profile (Eq. \ref{eq:VVD}), with the same MCMC algorithm illustrated in Sect.~\ref{sec:Only_RC}. We adopt the same priors of Sect.~\ref{sec:Only_RC}.

The MCMC analysis considers the two kinematic profiles at the same time. We thus define the single figure of merit, 
\begin{equation}
\label{eq:chifinal}
\chi^2_\mathrm{red,tot}({\mathbf x})=\frac{\chi^2_{\rm RC}({\mathbf x})+\chi^2_{\rm VVD}({\mathbf x})}{n_\mathrm{dof,tot}} \; ,
\end{equation}
where $\chi^2_{\rm RC}$  is Eq.~\eqref{eq:chi2RC} multiplied by $n_\mathrm{dof,RC}$, and 
\begin{equation}
\label{eq:chi2VVD}
\chi^2_{\rm VVD}({\mathbf x})=\sum_{j=1}^{N_{\rm VVD}}\frac{[\sigma_{z,\text{mod}}(R_j, {\mathbf x})-\sigma_{z,\text{data}}(R_j)]^2}{\sigma_{z,\text{data,err}}^2(R_j)} \; 
\end{equation}
is derived from the vertical velocity dispersion profile, with $N_{\rm VVD}$ the number of data points of the vertical velocity dispersion profile,  $\sigma_{z,\text{data}}$ the velocity dispersion measured at the projected distance $R_j$ with their uncertainty $\sigma_{z,\text{data,err}}$, and $\sigma_{z,\text{mod}}$ the velocity dispersion model estimated with Eq.~\eqref{eq:VVD}; finally $n_\mathrm{dof,tot}=N_{\rm RC} + N_{\rm VVD}-5$ is the total number of degrees of freedom.

We estimate the vertical velocity dispersion error bars, $\sigma_{z,\text{data,err}}$, by summing their random ($\sim$$0.1-10$~km~s$^{-1}$) and systematic ($\sim$$1-10$~km~s$^{-1}$) uncertainties in quadrature \citep{DMSvii}.

\begin{figure*}
        \centering
        \includegraphics[width=17cm]{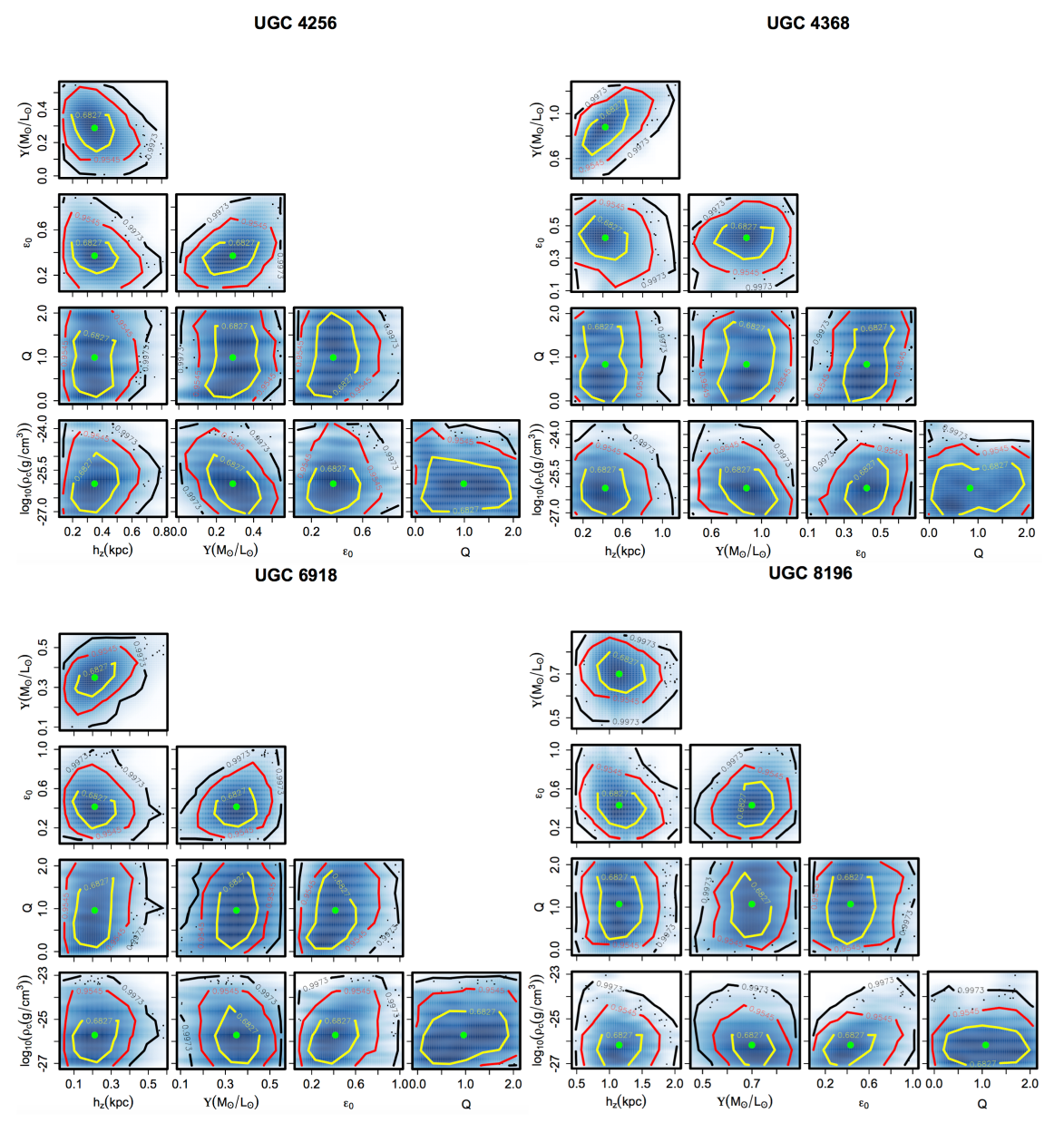}
        \caption{Four examples of posterior distributions of the two galaxy parameters and of the three RG parameters estimated from the rotation curves and the vertical velocity dispersion profiles at the same time. The green dots locate the median values and the yellow, red and black contours show the $1\sigma$, $2\sigma$ and $3\sigma$ levels, respectively.}
        \label{fig:CP_v_sigz_4}
\end{figure*}

 Figure~\ref{fig:CP_v_sigz_4} shows the posterior distributions for the same four galaxies shown in Fig.~\ref{fig:CP_v_4} for comparison. For the remaining galaxies, the posterior distributions are qualitatively similar. Because the 
 posterior distributions show a single peak, we adopt the medians of the posterior distributions as our parameter estimates as in Sect.~\ref{sec:Only_RC}.

Table~\ref{tab:fit_VVD_RC_orig_err_bars_v} lists the medians and the $15.9$ and the $84.1$ percentiles of the posterior distributions. We adopt these percentiles as the $1\sigma$ uncertainty range. 
The blue solid lines in the sub-panels (e) and (f) in Figs.~\ref{fig:Complete_analysis_1}-\ref{fig:Complete_analysis_7} in Appendix~\ref{sec:Figures}
show the model rotation curves and the vertical velocity dispersion profiles with the median parameters listed in Table~\ref{tab:fit_VVD_RC_orig_err_bars_v}. 
The red dots with error bars show the DMS data. As in Sect.~\ref{sec:Only_RC}, the features of the observed rotation curves are captured by the models in most cases although some discrepancies remain. Remarkably, the modelling of the rotation curves generally improves, like in \object{UGC 1087}, UGC 3997, and \object{UGC 9837}.

The right panel of Fig.~\ref{fig:SP_hzRC_hzRCVVD_hzSR} shows that the disk-scale heights $h_z$ required to model the rotation curves and the vertical velocity dispersion profiles are substantially smaller than the disk-scale heights $h_{z,\text{SR}}$ derived from Eq.~\eqref{eq:hzhR}, suggested by the observations of edge-on galaxies. In sub-panels (f) of Figs.~\ref{fig:Complete_analysis_1}-\ref{fig:Complete_analysis_7}, the cyan solid lines are the vertical velocity dispersion profiles when we adopt
the parameters of Table~\ref{tab:fit_VVD_RC_orig_err_bars_v} except for $h_z$ replaced by $h_{z,\text{SR}}$. 
Comparing the left and the right panels of Fig.~\ref{fig:SP_hzRC_hzRCVVD_hzSR} confirms that including the vertical velocity dispersion profiles in the modelling is responsible for requiring substantially thinner disks than $h_{z,\text{SR}}$; in fact, modelling the rotation curve alone may actually require disks thicker than expected from Eq.~\eqref{eq:hzhR}.

This comparison supports the conclusion that the discrepancies between $h_{z,\text{RC}+\text{VVD}}$ and $h_{z,\text{SR}}$ are derived
from the fact that the two scale heights are inferred from two different stellar populations:
a younger stellar population dominating the observed vertical velocity dispersion, hence $h_{z,\text{RC}+\text{VVD}}$, and
an older stellar population dominating the surface brightness in the edge-on galaxy sample, hence $h_{z,\text{SR}}$~\citep{Aniyan16}.

The general agreement between our results and those found by~\citet{Angus15} for MOND also confirms that, as anticipated in the introduction, if this disagreement between $h_{z,\text{RC}+\text{VVD}}$ and $h_{z,\text{SR}}$ is derived from an overlooked observational bias, it does not suggest a possible tension between the data and the theory of gravity, either MOND or RG.

This observational bias in the vertical velocity dispersion measure brings on consequences for the estimate of the
mass-to-light ratio. In fact, the velocity dispersion increases with the disk thickness and with the intensity of the gravitational field originated by the baryonic mass. By reducing the disk thickness, a larger mass must be attributed to the baryonic component to reproduce the observed velocity field. Therefore, our mass-to-light ratios are larger than the DMS values, as also occurs in MOND~\citep{Angus15}. In fact, our mass-to-light ratios agree with the MOND ratios within $2\sigma$ for 29 out of 30 galaxies and within $2.13\sigma$ for \object{UGC 11318}.

Our estimated mass-to-light ratios are corroborated by the comparison with the SPS values (Sect.~\ref{sec:Only_RC}). The right panel of Fig.~\ref{fig:SP_YRC_YRCVVD_YSPS} compares the mass-to-light ratios required to model the rotation curves and the vertical velocity dispersion profiles with the mass-to-light ratios derived from the SPS model. The RG mass-to-light ratios span the same range ($\sim$$0.3-1.2$~$M_\odot/L_\odot$) as the RG ratio derived by modelling the rotation curve alone (left panel of Fig.~\ref{fig:SP_YRC_YRCVVD_YSPS}), which nearly coincides with the range covered by the mass-to-light ratios derived from the SPS models. For each individual galaxy, our estimated mass-to-light ratio agrees with the mass-to-light ratio derived from the SPS model most of the time: for 20 out of 29 galaxies, the two values agree within $2\sigma$; for six galaxies, they agree within $2$ to $3\sigma$; and for the remaining three galaxies, they are consistent within $3$ to $5\sigma$. The largest discrepancy, $4.6\sigma$, occurs for \object{UGC 1908}; this discrepancy is smaller, however, than the $6.7\sigma$ discrepancy found in Sect.~\ref{sec:Only_RC} with the modelling of the rotation curve alone.

On the contrary, for $70\%$ of the galaxy sample, the mass-to-light ratios estimated by the DMS collaboration are at least a factor of 2 smaller than the SPS values and their disks turn out to be sub-maximal~\citep{Angus15,Aniyan16}. Despite these discrepancies, the large error bars associated with the DMS mass-to-light ratios make them agree within $3\sigma$ with the SPS values \citep[see][Table 1]{Angus15}.

\begin{table*}
        \caption{Parameters estimated from the rotation curves and the vertical velocity dispersion profiles.}
        \label{tab:fit_VVD_RC_orig_err_bars_v}  
        \centering
        \begin{tabular}{lccccccccc}
                \hline\hline
                UGC & $\Upsilon$ $\Bigl[\frac{\text{M}_\odot}{\text{L}_\odot}\Bigr]$ & $\Upsilon_{\rm SPS}$ $\Bigl[\frac{\text{M}_\odot}{\text{L}_\odot}\Bigr]$ & $\frac{\Upsilon_{\rm fit}-\Upsilon_{\rm SPS}}{\sqrt{\sigma_{\Upsilon_{\rm fit}}^2+\sigma_{\Upsilon_{\rm SPS}}^2}      }$ &$h_z$ $[\text{kpc}]$ & $\frac{h_z}{h_{z,\text{SR}}}$ & $\epsilon_0$ & $Q$ & $\log_{10}\Bigl(\rho_\mathrm{c}$ $\Bigl[\frac{\text{g}}{\text{cm}^3}\Bigr] \Bigr)$&$\chi^2_\mathrm{red,tot}$\\
                (1) & (2) & (3) & (4) & (5) & (6) & (7) & (8) & (9) & (10) \\
                \hline
                448 & $0.94^{+0.08}_{-0.20}$ &$0.64^{+0.07}_{-0.00}$ & $+2.06$ & $0.22^{+0.06}_{-0.05}$ & $0.48$ & $0.60^{+0.26}_{-0.37}$ & $0.57^{+0.72}_{-0.51}$ & $-25.82^{+1.07}_{-0.80}$&$5.32$\\
                463 & $0.56^{+0.04}_{-0.06}$ &$0.64^{+0.07}_{-0.00}$  & $-1.25$ & $0.25^{+0.07}_{-0.07}$ & $0.54$ & $0.90^{+0.07}_{-0.12}$ & $1.09^{+0.60}_{-0.70}$ & $-25.07^{+1.29}_{-1.42}$&$6.71$\\
                1081 & $0.66^{+0.13}_{-0.15}$ &$0.52^{+0.10}_{-0.00}$&$+0.94$& $0.24^{+0.06}_{-0.05}$ & $0.62$ & $0.53^{+0.19}_{-0.15}$ & $0.92^{+0.76}_{-0.59}$ & $-24.54^{+1.02}_{-1.57}$&$8.38$\\
                1087 & $0.81^{+0.11}_{-0.14}$ &$0.58^{+0.08}_{-0.00}$ &$+1.69$& $0.19^{+0.04}_{-0.03}$ & $0.49$ & $0.52^{+0.14}_{-0.13}$ & $0.85^{+0.63}_{-0.67}$ & $-25.53^{+1.20}_{-0.91}$&$3.53$\\
                1529 & $0.80^{+0.08}_{-0.10}$ &$0.70^{+0.08}_{-0.00}$&$+1.02$& $0.24^{+0.06}_{-0.05}$ & $0.55$ & $0.74^{+0.09}_{-0.07}$ & $1.11^{+0.55}_{-0.75}$ & $-25.02^{+0.96}_{-1.35}$ &$16.35$\\
                1635 & $0.75^{+0.12}_{-0.09}$ &$0.61^{+0.07}_{-0.00}$&$+1.20$& $0.11^{+0.02}_{-0.02}$ & $0.27$ & $0.66^{+0.18}_{-0.14}$ & $0.76^{+0.59}_{-0.51}$ & $-25.39^{+1.68}_{-1.29}$&$4.91$\\
                1862 & $1.01^{+0.19}_{-0.24}$ &$0.52^{+0.10}_{-0.00}$ &$+2.17$& $0.13^{+0.04}_{-0.03}$ & $0.50$ & $0.46^{+0.22}_{-0.17}$ & $0.84^{+0.74}_{-0.65}$ & $-24.28^{+0.94}_{-1.68}$ &$6.23$\\
                1908 & $0.34^{+0.05}_{-0.06}$ &$0.67^{+0.07}_{-0.00}$&$-4.57$& $0.50^{+0.09}_{-0.09}$ & $0.93$ & $0.37^{+0.09}_{-0.09}$ & $0.59^{+0.71}_{-0.32}$ & $-25.51^{+0.84}_{-0.98}$&$16.91$\\
                3091 & $0.74^{+0.15}_{-0.16}$ &$0.48^{+0.10}_{-0.00}$&$+1.55$& $0.15^{+0.03}_{-0.02}$ & $0.36$ & $0.53^{+0.18}_{-0.14}$ & $0.82^{+0.80}_{-0.66}$ & $-24.86^{+1.22}_{-1.58}$&$2.82$\\
                3140 & $0.75^{+0.06}_{-0.06}$&$0.64^{+0.07}_{-0.00}$& $+1.53$ & $0.28^{+0.04}_{-0.04}$ & $0.67$ & $0.78^{+0.13}_{-0.15}$ & $0.76^{+0.74}_{-0.56}$ & $-25.91^{+1.52}_{-0.83}$&$8.90$\\
                3701 & $0.73^{+0.28}_{-0.23}$ &$0.45^{+0.11}_{-0.00}$& $+1.05$ & $0.25^{+0.09}_{-0.07}$ & $0.58$ & $0.26^{+0.09}_{-0.07}$ & $0.78^{+0.79}_{-0.45}$ & $-24.15^{+0.64}_{-1.10}$&$2.09$\\
                3997 & $0.97^{+0.31}_{-0.23}$ &-& - &$0.18^{+0.06}_{-0.05}$&$0.31$&$0.47^{+0.17}_{-0.11}$&$1.27^{+0.51}_{-0.66}$&$-23.68^{+0.48}_{-1.39}$&$13.63$\\
                4036 & $0.72^{+0.09}_{-0.10}$ &$0.50^{+0.10}_{-0.00}$&$+1.97$& $0.26^{+0.05}_{-0.05}$ & $0.63$ & $0.58^{+0.19}_{-0.13}$ & $0.97^{+0.67}_{-0.68}$ & $-25.94^{+1.03}_{-0.78}$ &$5.20$\\
                4107 & $0.77^{+0.08}_{-0.10}$ &$0.58^{+0.08}_{-0.00}$&$+1.93$& $0.16^{+0.03}_{-0.03}$ & $0.38$ & $0.78^{+0.14}_{-0.14}$ & $0.62^{+0.82}_{-0.49}$ & $-25.45^{+1.82}_{-1.12}$ &$5.84$\\
                4256 & $0.29^{+0.08}_{-0.09}$ &$0.52^{+0.10}_{-0.00}$& $-2.23$ & $0.35^{+0.10}_{-0.07}$ & $0.71$ & $0.37^{+0.13}_{-0.11}$ & $0.99^{+0.65}_{-0.64}$ & $-25.99^{+0.74}_{-0.63}$&$2.10$\\
                4368 &$0.88^{+0.13}_{-0.17}$&$0.45^{+0.11}_{-0.00}$&$+2.66$&$0.42^{+0.17}_{-0.16}$& $1.17$ & $0.43^{+0.07}_{-0.07}$ &$0.84^{+0.78}_{-0.52}$&$-26.05^{+0.72}_{-0.64}$&$9.20$\\
                4380 &$0.79^{+0.11}_{-0.09}$&$0.52^{+0.10}_{-0.00}$ &$+2.41$&$0.24^{+0.04}_{-0.04}$&$0.44$&$0.54^{+0.18}_{-0.15}$&$0.28^{+0.92}_{-0.20}$&$-25.98^{+1.27}_{-0.73}$&$8.40$\\
                4458 &$0.53^{+0.03}_{-0.03}$&$0.85^{+0.09}_{-0.06}$& $-3.75$ & $0.86^{+0.48}_{-0.30}$ & $1.08$ & $0.77^{+0.15}_{-0.22}$ & $1.02^{+0.64}_{-0.72}$ & $-26.08^{+1.24}_{-0.73}$&$3.08$\\
                4555 &$0.88^{+0.10}_{-0.14}$&$0.48^{+0.10}_{-0.00}$&$+3.08$&$0.19^{+0.09}_{-0.06}$&$0.45$&$0.89^{+0.09}_{-0.10}$&$0.89^{+0.89}_{-0.60}$&$-25.33^{+1.69}_{-0.89}$&$11.55$\\
                4622 &$0.71^{+0.14}_{-0.14}$&$0.55^{+0.09}_{-0.00}$ &$+1.08$&$0.37^{+0.12}_{-0.09}$&$0.53$&$0.58^{+0.17}_{-0.12}$&$1.30^{+0.47}_{-0.53}$&$-25.14^{+1.05}_{-1.15}$&$6.06$\\
                6903 & $0.61^{+0.19}_{-0.27}$ &$0.52^{+0.10}_{-0.00}$&$+0.38$& $0.10^{+0.07}_{-0.03}$ & $0.18$ & $0.33^{+0.14}_{-0.10}$ & $1.11^{+0.84}_{-0.57}$ & $-25.04^{+1.38}_{-0.58}$&$8.47$\\
                6918 &$0.35^{+0.06}_{-0.06}$&$0.52^{+0.10}_{-0.00}$& $-2.18$ &$0.21^{+0.07}_{-0.06}$& $1.00$ & $0.42^{+0.15}_{-0.12}$&$0.96^{+0.66}_{-0.59}$&$-25.71^{+0.99}_{-0.79}$&$4.10$\\
                7244 & $0.51^{+0.14}_{-0.16}$ &$0.41^{+0.12}_{-0.00}$&$+0.62$& $0.37^{+0.10}_{-0.08}$ & $0.86$ & $0.22^{+0.07}_{-0.06}$ & $0.85^{+0.64}_{-0.49}$ & $-25.17^{+0.75}_{-1.16}$ &$3.41$\\
                7917 & $0.80^{+0.09}_{-0.12}$ &$0.81^{+0.09}_{-0.04}$&$-0.08$& $0.44^{+0.14}_{-0.11}$ & $0.64$ & $0.81^{+0.14}_{-0.19}$ & $1.01^{+0.61}_{-0.74}$ & $-25.86^{+1.02}_{-0.76}$ &$4.60$\\
                8196 &$0.70^{+0.05}_{-0.05}$&$0.77^{+0.08}_{-0.03}$& $-0.90$ & $1.15^{+0.25}_{-0.20}$& $1.98$&$0.43^{+0.17}_{-0.13}$&$1.07^{+0.60}_{-0.60}$&$-26.17^{+0.63}_{-0.56}$&$12.51$\\
                9177 & $0.86^{+0.23}_{-0.20}$ &$0.43^{+0.11}_{-0.00}$&$+1.89$& $0.36^{+0.11}_{-0.12}$ & $0.53$ & $0.55^{+0.20}_{-0.10}$ & $1.25^{+0.51}_{-0.62}$ & $-24.65^{+1.05}_{-1.78}$&$6.39$\\
                9837 & $0.76^{+0.13}_{-0.10}$ &$0.64^{+0.07}_{-0.00}$ & $+0.95$ & $0.18^{+0.05}_{-0.07}$ & $0.30$ & $0.35^{+0.06}_{-0.04}$ & $1.11^{+0.54}_{-0.41}$ & $-23.57^{+0.29}_{-0.16}$ & $3.24$\\
                9965 & $0.66^{+0.07}_{-0.08}$ &$0.50^{+0.10}_{-0.00}$&$+1.70$& $0.19^{+0.04}_{-0.03}$ & $0.44$ & $0.75^{+0.15}_{-0.17}$ & $0.88^{+0.56}_{-0.60}$ & $-25.67^{+1.32}_{-0.85}$&$2.21$\\
                11318 & $0.65^{+0.07}_{-0.07}$ &$0.67^{+0.07}_{-0.00}$&$-0.25$& $0.36^{+0.05}_{-0.04}$ & $0.71$ & $0.72^{+0.17}_{-0.29}$ & $1.37^{+0.44}_{-0.65}$ & $-25.68^{+0.81}_{-0.96}$ &$37.53$\\
                12391 & $0.60^{+0.08}_{-0.13}$ &$0.50^{+0.10}_{-0.00}$ &$+0.83$& $0.37^{+0.08}_{-0.07}$ & $0.82$ & $0.39^{+0.09}_{-0.12}$ & $0.61^{+0.71}_{-0.42}$ & $-25.67^{+0.99}_{-0.83}$&$8.02$\\
                \hline
        \end{tabular}
    \tablefoot{Column 1: UGC number; Cols. 2, 5, 7, 8, 9:  medians of the posterior distributions of the model parameters estimated from the rotation curves and the vertical velocity dispersion profiles; Col. 3: mass-to-light ratio derived with the SPS model; Col. 10: $\chi^2_\mathrm{red,tot}$ (Eq.~\eqref{eq:chifinal}).}
\end{table*}

\subsection{The observational bias in the vertical velocity dispersion profile}
\label{sec:VVD_RC_1pt55}

As mentioned above, the estimate of the vertical velocity dispersion profile is likely to be affected by 
an observational bias. Therefore, the disk thickness obtained by modelling this profile might give a severe underestimation of the real value. 
Here, we want to quantify how correcting the vertical velocity dispersions by an appropriate factor can give disk thicknesses that are consistent with the observations of edge-on galaxies and with mass-to-light ratios still consistent with the SPS models.

\citet{Aniyan16} model what we would observe in the disk of a typical external face-on galaxy by selecting a sample of giant stars simulated in the Milky Way from the on-line Besan\c{c}on model~\citep{Robin03}. This model describes the main structural components and stellar populations of the Milky Way and includes recipes for galactic reddening, star formation history, and dynamical evolution~\citep{Robin03,Aniyan16}. For their simulated star sample,~\citet{Aniyan16} choose giant stars of spectral type G8III - K5III, with luminosity in the range of $-3 \leq M_V \leq +3$, and colour in the range of $0.8 \leq B-V \leq 1.8$, within a cylindric volume, centred on the Sun and perpendicular to the Galactic disk, of radius 2 kpc and height 10 kpc. They assume a star formation and a dynamical history similar to the solar neighbourhood.

From the simulated sample, they extract the vertical velocity dispersion integrated vertically through the disk by considering either the entire stellar population ($\sigma_1$) or the dynamically hotter and older giant component alone ($\sigma_2$). They estimate that $\eta=\sigma_2/\sigma_1=1.55\pm0.02$. 
Now, observationally, the disk-scale heights are inferred from measures of the surface brightness that is dominated by the old hot giants, which have a larger vertical velocity dispersion, whereas the younger stellar population mostly contributes to the spectral signal used to derive the observed velocity dispersion. Therefore, according to the result of~\citet{Aniyan16}, assuming a single stellar population to interpret the vertical velocity dispersions and the disk-scale height at the same time introduces a bias that originates an underestimate of the disk thickness.

Here, we estimate this observational bias
with the ratio between the expected vertical velocity dispersions computed with the parameters inferred from the rotation curves alone (Table~\ref{tab:fit_RC_5_par})
and the measured vertical velocity dispersion profiles. 
We adopt the following procedure: (1) for each galaxy, we use Eq.~\eqref{eq:VVD} to compute the expected vertical velocity dispersion, according to the parameters of Table~\ref{tab:fit_RC_5_par}; (2) at each radial coordinate, we compute the ratio between the expected and the measured vertical velocity dispersions and we estimate the mean of these ratios along the radial profile of each of the 30 galaxies; (3) we then estimate the mean, $\eta'$, of these 30 means and their standard deviation. Thus, we obtain $\eta'=1.63\pm0.65$, which agrees within $0.12\sigma$ with the value $1.55\pm 0.02$ found by~\citet{Aniyan16}.
   

To quantify the effect of this bias on the estimate of the disk thickness, we now repeat the analysis presented in Sect.~\ref{sec:VVD_RC} for five galaxies where we artificially increase the vertical velocity dispersion profile by a factor of 1.63. We choose five galaxies with a $h_z$-to-$h_{z,\text{SR}}$ ratio smaller than $0.5$: UGC 1635, \object{UGC 3091}, \object{UGC 4107}, UGC 4555, and \object{UGC 9965}. We also model the kinematics of these galaxies in MOND where, as free parameters, we only have the galactic parameters $\Upsilon$ and $h_z$.
For the MCMC analysis, we use the same priors as above.

To model the galaxies in MOND, we derive the MOND potential by solving the Poisson equation in the QUMOND formulation of MOND~\citep{Milgrom10}:  
\begin{equation}
\label{eq:PoissonMOND}
        \nabla^2\phi=\nabla\cdot\left[\nu\left(\frac{|\nabla\phi_\mathrm{N}|}{a_0}\right)\nabla\phi_\mathrm{N}\right]\; ,
\end{equation}
where $\phi_\mathrm{N}$ is the Newtonian potential, $a_0=1.2\times 10^{-10}$~m~s$^{-2}$~$=$~$3600$~kpc$^{-1}$~$(\text{km }\text{s}^{-1})^2$ is the MOND critical acceleration, and $\nu$ is the interpolating function regulating the transition between the Newtonian and the MOND regimes. We use the `simple $\nu$-function',
\begin{equation}
\label{eq:simplenu}
\nu(y)=\frac{1}{2}\left(1+\sqrt{1+\frac{4}{y}}\right) \; .
\end{equation}

Tables \ref{tab:fit_VVDcorr_RC_RG} and \ref{tab:fit_VVDcorr_RC_MOND} list the medians and percentile ranges of the posterior distributions in RG and QUMOND, respectively. 
As expected, $h_z$ increases by a factor between $2.07$ (found for UGC 3091 in QUMOND) and $2.89$ (found for UGC 4555 in QUMOND) with respect to the values obtained in Sect.~\ref{sec:VVD_RC} (see Tables~\ref{tab:fit_VVD_RC_orig_err_bars_v},~\ref{tab:fit_VVDcorr_RC_RG}, and~\ref{tab:fit_VVDcorr_RC_MOND}). For UGC 1635 and UGC 3091, the disk-scale heights are still smaller than the values expected from the observations of edge-on galaxies, but their $h_z$-to-$h_{z,\text{SR}}$ ratios still increase to values greater than $0.5$.

The agreement between the RG mass-to-light ratios estimated in the current section and the SPS values worsens compared to Sect.~\ref{sec:VVD_RC}, but for four out of five galaxies, it is within $4\sigma$.
QUMOND mass-to-light ratios tend to be closer to the SPS values than the RG mass-to-light ratios: their agreement with their SPS expectations is within $3\sigma$ for four out of five galaxies.
However, since the QUMOND mass-to-light ratios uncertainties are smaller than in RG, the agreement between the QUMOND mass-to-light ratios and the SPS values is formally worse than in RG for two out of five galaxies (UGC 1635 and UGC 3091).

The sub-panels (i)-(l) of Figs.~\ref{fig:Complete_analysis_6}-\ref{fig:Complete_analysis_7} in Appendix~\ref{sec:Figures} show 
the RG and QUMOND models of the rotation curves and vertical velocity dispersion profiles (blue solid lines). The vertical velocity dispersion data are increased by the factor $1.63$  (red dots with error bars). RG describes the rotation curves slightly better than QUMOND whereas both theories describe the vertical velocity dispersion profiles equally well. Clearly, this better performance of RG follows from the version of RG we adopt at this stage that, for each galaxy, has three more free parameters than QUMOND.

Our QUMOND results of these five galaxies are comparable to the results of~\citet{Angus15}. Specifically, we also find that the QUMOND rotation curves of UGC 3091 and UGC 9965 properly describe the data, whereas the QUMOND rotation curves of UGC 1635, UGC 4107 and UGC 4555 tend to underestimate the inner profile and to overestimate the outer profile. Our vertical velocity dispersions are also comparable to \citet{Angus15}, although their models slightly uderestimate the most inner data point of UGC 1635 and UGC 3091.

\begin{figure*}
        \centering
        \includegraphics[width=17cm]{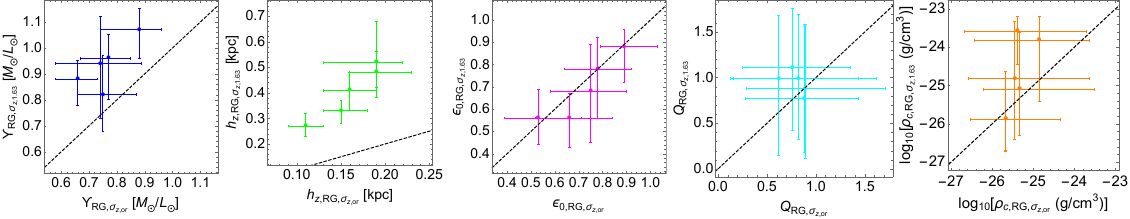}
        \caption{Comparison between the parameters estimated with RG from the two kinematic profiles of five DMS galaxies with the original values of $\sigma_z(R)$ (Sect.~\ref{sec:VVD_RC}) and the parameters estimated with RG from the two kinematic profiles with $\sigma_z(R)$ increased by the factor $1.63$ (Sect.~\ref{sec:VVD_RC_1pt55}). The black dashed lines show the lines of equality.}
        \label{fig:Scatter_plots_RG_sigzor_sigz1pt55}
\end{figure*}

\begin{figure*}
        \centering
        \includegraphics[width=17cm]{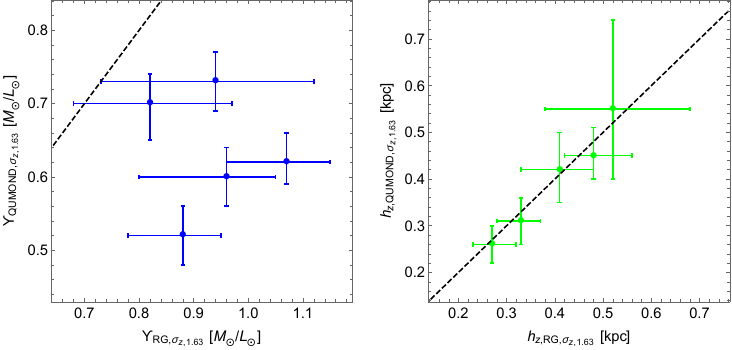}
        \caption{Comparison between the RG and QUMOND $\Upsilon$ and $h_z$ estimated from the two kinematic profiles of five DMS galaxies where $\sigma_z(R)$ is increased by the factor $1.63$ (Sect.~\ref{sec:VVD_RC_1pt55}). The black dashed lines show the lines of equality.}
        \label{fig:Scatter_plots_RG_MOND_sigz1pt55}
\end{figure*}

Figure~\ref{fig:Scatter_plots_RG_sigzor_sigz1pt55} compares the parameters estimated with RG with the original values of the vertical velocity dispersion profile, $\sigma_z(R)$ (Sect.~\ref{sec:VVD_RC}), with the parameters obtained with $\sigma_z(R)$ increased by the factor of $1.63$.
These figures show that the three RG parameters are not systematically affected by the increase of $\sigma_z(R)$, whereas both $\Upsilon$ and $h_z$  tend to be larger. 
Figure~\ref{fig:Scatter_plots_RG_MOND_sigz1pt55} compares $\Upsilon$ and $h_z$ obtained in RG and QUMOND with the $\sigma_z(R)$ values increased by the factor $1.63$. 
The disk-scale heights are comparable in the two theories of gravity whereas the mass-to-light ratios in RG are systematically larger than in QUMOND. This systematic difference will be relevant in describing the RAR we illustrate below.

\begin{table*}
        \caption{Parameters estimated in RG from the rotation curves and the vertical velocity dispersion profiles increased by the factor $1.63$.}
        \label{tab:fit_VVDcorr_RC_RG}   
        \centering
        \begin{tabular}{lccccccccc}
                \hline\hline
                UGC & $\Upsilon$ $\Bigl[\frac{\text{M}_\odot}{\text{L}_\odot}\Bigr]$ & $\Upsilon_{\rm SPS}$ $\Bigl[\frac{\text{M}_\odot}{\text{L}_\odot}\Bigr]$ & $\frac{\Upsilon_{\rm fit}-\Upsilon_{\rm SPS}}{\sqrt{\sigma_{\Upsilon_{\rm fit}}^2+\sigma_{\Upsilon_{\rm SPS}}^2}      }$ &$h_z$ $[\text{kpc}]$ & $\frac{h_z}{h_{z,\text{SR}}}$ & $\epsilon_0$ & $Q$ & $\log_{10}\Bigl(\rho_\mathrm{c}$ $\Bigl[\frac{\text{g}}{\text{cm}^3}\Bigr] \Bigr)$&$\chi^2_\mathrm{red,tot}$\\
                (1) & (2) & (3) & (4) & (5) & (6) & (7) & (8) & (9) & (10) \\
                \hline
                1635 & $0.82^{+0.15}_{-0.14}$ &$0.61^{+0.07}_{-0.00}$&$+1.35$& $0.27^{+0.05}_{-0.04}$ & $0.66$ & $0.56^{+0.11}_{-0.13}$ & $1.11^{+0.65}_{-0.69}$ & $-23.58^{+0.39}_{-0.86}$&$5.66$\\
                3091 & $0.94^{+0.18}_{-0.21}$ &$0.48^{+0.10}_{-0.00}$&$+2.23$& $0.33^{+0.04}_{-0.05}$ & $0.79$ & $0.56^{+0.13}_{-0.12}$ & $0.99^{+0.70}_{-0.67}$ & $-23.82^{+0.63}_{-1.60}$&$5.07$\\
                4107 & $0.96^{+0.09}_{-0.16}$ &$0.58^{+0.08}_{-0.00}$&$+2.79$& $0.41^{+0.07}_{-0.08}$ & $0.98$ & $0.78^{+0.14}_{-0.22}$ & $0.99^{+0.69}_{-0.84}$ & $-24.82^{+1.49}_{-1.59}$ &$7.82$\\
                4555 & $1.07^{+0.08}_{-0.11}$ &$0.48^{+0.10}_{-0.00}$&$+5.28$& $0.52^{+0.16}_{-0.14}$ & $1.24$ & $0.88^{+0.08}_{-0.16}$ & $0.88^{+0.70}_{-0.77}$ & $-25.10^{+1.47}_{-1.21}$&$13.77$\\
                9965 & $0.88^{+0.07}_{-0.10}$ &$0.50^{+0.10}_{-0.00}$&$+3.69$& $0.48^{+0.08}_{-0.06}$ & $1.12$ & $0.68^{+0.21}_{-0.23}$ & $0.77^{+0.81}_{-0.59}$ & $-25.88^{+1.24}_{-0.83}$&$4.42$\\
                \hline
        \end{tabular}
    \tablefoot{Column 1: UGC number; Cols. 2, 5, 7, 8, 9:  medians of the posterior distributions of the parameters estimated simultaneously, in RG, from the observed rotation curves and the vertical velocity dispersion profiles increased by the factor $1.63$; Col. 3: mass-to-light ratio derived with the SPS model; Col. 10: $\chi^2_\mathrm{red,tot}$ (Eq.~\eqref{eq:chifinal}).}
\end{table*}

\begin{table*}
        \caption{Parameters estimated in QUMOND from the rotation curves and the vertical velocity dispersion profiles increased by the factor $1.63$.}
        \label{tab:fit_VVDcorr_RC_MOND} 
        \centering
        \begin{tabular}{lcccccc}
                \hline\hline
                UGC & $\Upsilon$ $\Bigl[\frac{\text{M}_\odot}{\text{L}_\odot}\Bigr]$ & $\Upsilon_{\rm SPS}$ $\Bigl[\frac{\text{M}_\odot}{\text{L}_\odot}\Bigr]$ & $\frac{\Upsilon_{\rm fit}-\Upsilon_{\rm SPS}}{\sqrt{\sigma_{\Upsilon_{\rm fit}}^2+\sigma_{\Upsilon_{\rm SPS}}^2}      }$ &$h_z$ $[\text{kpc}]$ & $\frac{h_z}{h_{z,\text{SR}}}$&$\chi^2_\mathrm{red,tot}$\\
                (1) & (2) & (3) & (4) & (5) & (6) & (7) \\
                \hline
                1635 &$0.70^{+0.04}_{-0.05}$&$0.61^{+0.07}_{-0.00}$&$+1.41$&$0.26^{+0.04}_{-0.04}$&$0.63$&$7.58$\\
                3091 &$0.73^{+0.04}_{-0.04}$&$0.48^{+0.10}_{-0.00}$&$+3.90$&$0.31^{+0.05}_{-0.05}$&$0.74$&$6.09$\\
                4107 &$0.60^{+0.04}_{-0.04}$&$0.58^{+0.08}_{-0.00}$&$+0.35$&$0.42^{+0.08}_{-0.07}$&$1.00$&$12.91$\\
                4555 &$0.62^{+0.04}_{-0.03}$&$0.48^{+0.10}_{-0.00}$&$+2.19$&$0.55^{+0.19}_{-0.15}$&$1.31$&$27.83$\\
                9965 &$0.52^{+0.04}_{-0.04}$&$0.50^{+0.10}_{-0.00}$&$+0.31$&$0.45^{+0.06}_{-0.05}$&$1.05$&$5.74$\\
                \hline
        \end{tabular}
    \tablefoot{Column 1: UGC number; Cols. 2, 5: medians of the posterior distributions of the parameters estimated simultaneously, in QUMOND, from the observed rotation curves and the vertical velocity dispersion profiles increased by the factor $1.63$; Col. 3: mass-to-light ratio derived with the SPS model; Col. 7: $\chi^2_\mathrm{red,tot}$ (Eq.~\eqref{eq:chifinal}), with $n_{\rm dof,tot}=N_{\rm RC}+N_{\rm VVD}-2$ degrees of freedom.}
\end{table*}

\subsection{On the error bars of the rotation curves}
\label{sec:VVD_RC_10}

We conclude this section with a digression on the rotation curve error bars.
In their modelling of the DMS data  with MOND, rather than maintaining the original DMS error bars as we do here,~\citet{Angus15} arbitrarily increase the error bars from $\sim$$1-5$~km~s$^{-1}$ to $10$~km~s$^{-1}$. \citet{Angus15} were concerned by the possibility that disk warping could introduce systematic errors which can indeed affect the rotation curves;
by this approach,~\citet{Angus15} clearly intend to increase the statistical weight of the vertical velocity dispersion profiles which are more dependent on the disk thickness.

We verified that this modification is in fact unnecessary. By setting the rotation curves error bars 
to $10$~km~s$^{-1}$ and performing once again the MCMC analysis described above, we find that the new disk-scale
heights, $h_z$, substantially agree with our original results: the distributions of the ratios between the new and the original
$h_z$ is peaked around $\sim$$1$, with median $1.09^{+0.11}_{-0.20}$, where the uncertainties are the 15.9 and 84.1 percentiles. Similarly, 
the mass-to-light ratios remain unbiased, with their ratios between the new and the original values having median $0.95^{+0.09}_{-0.11}$.

\section{A universal combination of the RG parameters}
\label{sec:Fit_all}

In the formulation of RG, $\epsilon_0$, $Q$ and $\rho_\mathrm{c}$ are universal parameters. Here we show that, in principle,
a single set of these parameters could indeed be able to model the dynamics of our entire sample of disk galaxies. 

For this task, the ideal approach would be to use the MCMC algorithm to explore the 63-dimensional space of these three RG parameters and the $2\times 30$ parameters describing the galaxies, namely the mass-to-light
ratios $\Upsilon$ and the disk-scale heights $h_z$. However, this approach requires an extraordinary computational effort which, at this stage of our investigation of the RG viability, appears unreasonably large.  

\begin{figure*}
        \centering
        \includegraphics[width=17cm]{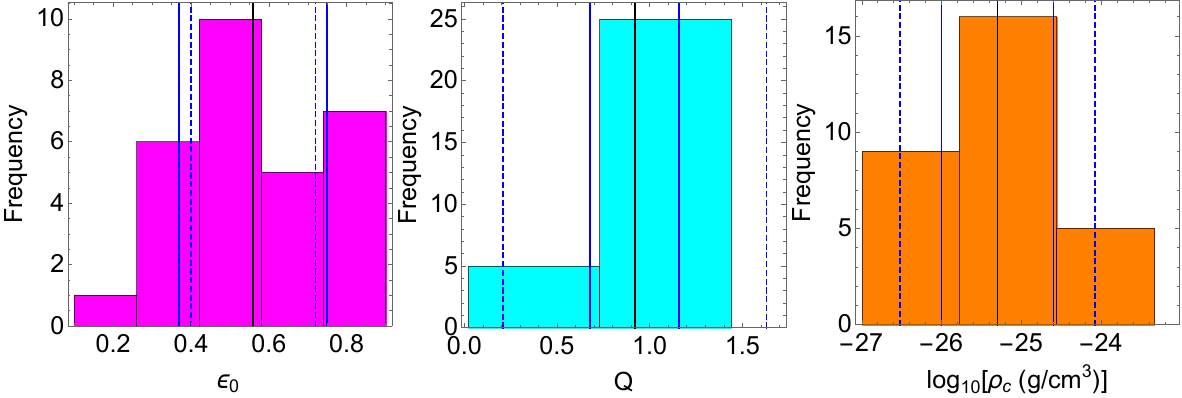}
        \caption{Distributions of the three RG parameters $\epsilon_0$, $Q$, and $\log_{10}\rho_\mathrm{c}$ listed in Table~\ref{tab:fit_VVD_RC_orig_err_bars_v}. The bin sizes are of the order of the mean uncertainties. The means of the distributions are shown as black solid lines; the two blue solid lines show the standard deviations of the distributions; the two blue dashed lines show the mean uncertainties of the parameters listed in Table~\ref{tab:fit_VVD_RC_orig_err_bars_v}.}
        \label{fig:histo_distrib_par_fit}
\end{figure*}

We prefer a simpler strategy. Our analyses above show that modelling each individual galaxy by keeping the three 
RG parameters free returns
values of these parameters that are roughly consistent from galaxy to galaxy. In fact, the $\epsilon_0$, $Q$ and $\log_{10}\rho_\mathrm{c}$ values listed in Table~\ref{tab:fit_VVD_RC_orig_err_bars_v}, whose distributions are shown in Fig.~\ref{fig:histo_distrib_par_fit}, have mean and
standard deviations $\epsilon_0=0.56\pm 0.19$, $Q=0.92\pm 0.24$, and $\log_{10}(\rho_\mathrm{c}/\mathrm{g\ cm}^{-3})=-25.30\pm 0.70$. These standard deviations 
are either smaller than
or comparable to
the mean uncertainties of the values listed in Table~\ref{tab:fit_VVD_RC_orig_err_bars_v}, which are $0.16$, $0.71$, and $1.22$, for $\epsilon_0$, $Q$ and $\log_{10}\rho_\mathrm{c}$, respectively.  
We can thus reasonably conclude that the different values that we find for different galaxies
can in principle be ascribed to statistical fluctuations.

Therefore, to check whether a single set of RG parameters could be able to describe the entire galaxy sample, rather than performing the computationally demanding exploration of the 63-dimensional parameter  space, we assume that the values of the mass-to-light
ratios $\Upsilon$ and the disk-scale heights $h_z$ for each galaxy, estimated in our previous analysis, are appropriate, and we only
explore the 3-dimensional space of the parameters $\epsilon_0$, $Q$ and $\log_{10}\rho_\mathrm{c}$ for the entire galaxy sample
at the same time. 

We adopt the priors for the three RG parameters as in Sects.~\ref{sec:Only_RC} and~\ref{sec:VVD_RC}. The figure of merit is now: 
\begin{equation}
\label{eq:chi2fitall}
\chi^2_\mathrm{red}({\mathbf x})=\sum_{i=1}^{N_\text{gal}}\chi^2_{\text{red,tot}, i}({\mathbf x}) \; ,
\end{equation}
where ${\mathbf x}=(\epsilon_0,Q,\log_{10} \rho_\mathrm{c})$, $N_\text{gal}$ is the number of DMS galaxies and $\chi^2_{\text{red,tot}, i}({\mathbf x})$ is Eq.~\eqref{eq:chifinal} where the number of free parameters is now equal to three instead of five.

As in Sects.~\ref{sec:Only_RC} and~\ref{sec:VVD_RC}, we run the MCMC for 19000 steps in addition to the 1000 burn-in steps.\footnote{To explore the parameter space, we consider all the galaxies at the same time and we thus need to run the Poisson solver $2\times 30$ times at each MCMC iteration. To reduce the computational effort, we parallelised our code with OpenMP, an application programming interface, which supports multiplatform shared memory multiprocessing programming in different languages. The C++ code we used is publicly available at \url{https://github.com/alpha-unito/astroMP}.}       To better assess the convergence of the chains we run three chains with three different starting points and we test their convergence with the variance ratio method of~\citet{Gel&Rub92}, described in Appendix~\ref{sec:convergence}. We find that 13000 steps are already sufficient to have the chains converging. We also test that each chain converges according to the Geweke diagnostic~\citep{Geweke92}.

\begin{figure*}
        \sidecaption
        \includegraphics[width=12cm]{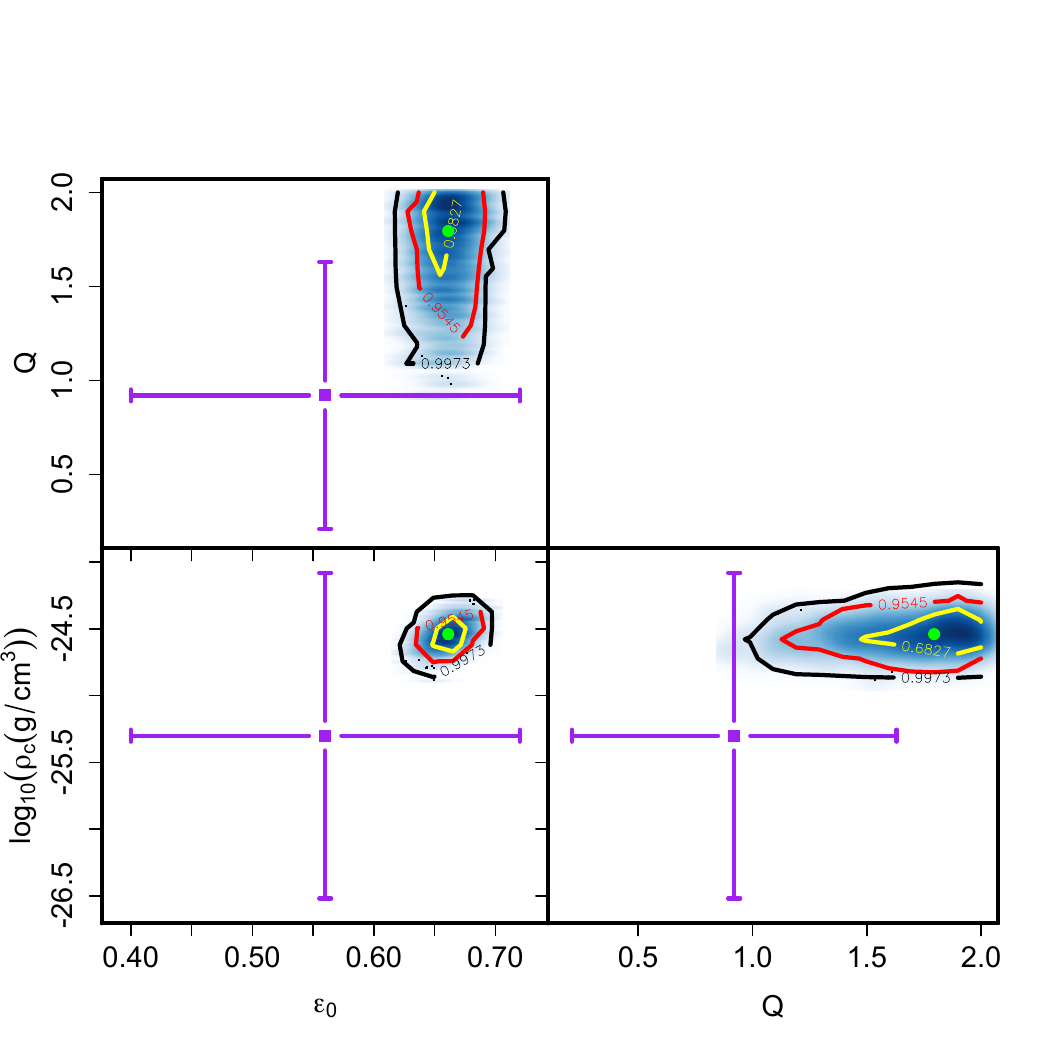}
        \caption{Posterior distributions of the three RG parameters. The green dots locate the median values and the yellow, red, and black contours show the $1\sigma$, $2\sigma$, and $3\sigma$ levels, respectively. The purple points and error bars show the means of the distributions of the RG parameters and their mean uncertainties
                found in Sect.~\ref{sec:VVD_RC} and reported in Fig.~\ref{fig:histo_distrib_par_fit}.}
        \label{fig:CP_Fit_all_gal}
\end{figure*}

Figure~\ref{fig:CP_Fit_all_gal} shows the posterior distributions of the three parameters. The green dots show the median values, and the yellow, red, and black curves show the $1$, $2$, and $3\sigma$ contour levels. The medians and 68\% confidence intervals are
$\epsilon_0=0.661^{+0.007}_{-0.007}$, $Q=1.79^{+0.14}_{-0.26}$ and $\log_{10}\rho_\mathrm{c}=-24.54^{+0.08}_{-0.07}$. The purple points show the means of the three distributions shown in Fig.~\ref{fig:histo_distrib_par_fit}; the error bars show the mean errors listed in Table~\ref{tab:fit_VVD_RC_orig_err_bars_v}.

As expected, the posterior distributions in Fig.~\ref{fig:CP_Fit_all_gal} show that considering all the DMS galaxies at the same time provides much tighter constraints on the RG parameters than in our previous analyses. Because of the large errors found in Sect.~\ref{sec:VVD_RC}, the universal parameters estimated here are consistent with our previous analyses: the means of the distributions of Fig.~\ref{fig:histo_distrib_par_fit} agree within $0.63$, $1.19$ and $0.62$ $\sigma$ with the medians, found here, of $\epsilon_0$, $Q$ and $\log_{10}\rho_\mathrm{c}$, respectively.

\begin{figure}
        \resizebox{\hsize}{!}{\includegraphics{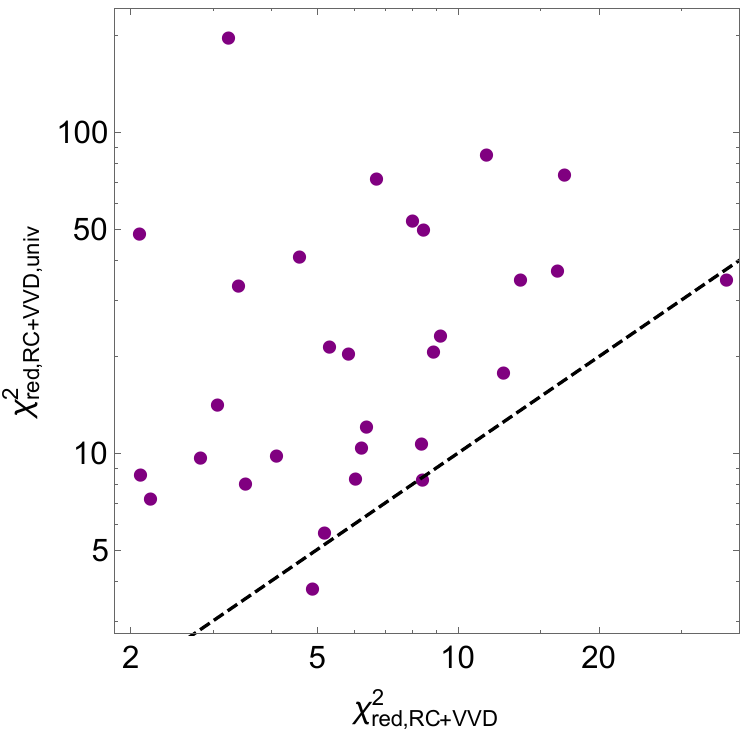}}
        \caption{Comparison between the reduced $\chi^2$ (Eq.~\eqref{eq:chifinal})
                for each galaxy by adopting the universal (Sect.~\ref{sec:Fit_all}) or the individual (Sect.~\ref{sec:VVD_RC}, Table~\ref{tab:fit_VVD_RC_orig_err_bars_v}) set of the RG parameters.
                The one-to-one line is shown as a black dashed line, for comparison.}
        \label{fig:SP_chi2red}
\end{figure}

However, the agreement between the data and the models of the rotation curves and the vertical velocity dispersion profiles derived with the mass-to-light ratios and disk-scale heights 
listed in Table~\ref{tab:fit_VVD_RC_orig_err_bars_v} and the universal combination of RG parameters found here worsens with respect to the agreement obtained with the individual  RG parameters found in Sect.~\ref{sec:VVD_RC} as shown in sub-panels (g) and (h) of Figs.~\ref{fig:Complete_analysis_1}-\ref{fig:Complete_analysis_7}. 
Figure~\ref{fig:SP_chi2red} compares the reduced $\chi^2$ found in Sect.~\ref{sec:VVD_RC} with the reduced $\chi^2$ found here.
Despite the presence of a number of galaxies whose $\chi^2$ substantially increases, the bulk of the sample maintains
its $\chi^2$ close to, albeit still larger than, the original $\chi^2$.

The increase of the $\chi^2$ compared to Sect.~\ref{sec:VVD_RC} is mainly due to the rotation curves. In fact, all the vertical velocity dispersion profiles are rather well interpolated with this universal combination of RG parameters: their $\chi^2$ are close to those of Sect.~\ref{sec:VVD_RC}, except for a few outliers, like UGC 3701 and UGC 3997. On the contrary, the rotation curves of only about half of the sample are still well described with this unique combination of RG parameters (e.g. \object{UGC 1081}, UGC 1635, UGC 4036, \object{UGC 4380} and UGC 9965), whereas the models worsen for the remaining sample.

        The good agreement of this universal combination of RG parameters with the parameters of Sect.~\ref{sec:VVD_RC} and the somewhat limited worsening of the $\chi^2$ shown in Fig.~\ref{fig:SP_chi2red} suggest that finding a unique set of RG 
        parameters that accurately describes the kinematics
of the DMS galaxies might be feasible.
This simple exercise in fact appears to indicate that properly exploring the full 63-dimensional parameter space 
might return substantially smaller $\chi^2$ 
with still reasonable, albeit different from the results
of Sect.~\ref{sec:VVD_RC},  mass-to-light ratios and disk-scale heights of the galaxies. 
We might also expect that the unique set of RG parameters will be statistically equivalent to the set we find here.

\section{The Radial Acceleration Relation}
\label{sec:RAR}
To test the viability of RG as a gravity theory describing the dynamics of disk galaxies, we need to consider an additional relevant observational piece of evidence that very clearly quantifies the mass discrepancy on galaxy scales: the RAR.~\citet{McGetal16} and~\citet{Lellietal17a} pointed out that the observed centripetal acceleration traced by the rotation curves, 
\begin{equation}
        \label{eq:gobs}
        g_\text{obs}(R)=v^2_\text{obs}(R)/R \; ,
\end{equation}
is tightly correlated with the Newtonian acceleration $g_\text{bar}(R)$ due to the baryonic matter distribution alone. 

\citet{McGetal16} found that the function: 
        \begin{equation}
                \label{eq:fittedRAR}
                g_\text{obs}(R)=\frac{g_\text{bar}(R)}{1-\exp\left(-\sqrt{\frac{g_\text{bar}(R)}{g_\dagger}}\right)}
        \end{equation}
provides a good fit for the entire SPARC sample, made of 153 galaxies~\citep{Lellietal16}. The fit has only one
free parameter $g_\dagger$ whose single value $g_\dagger=1.20\pm0.02$ (random) $\pm 0.24$ (systematic) $\times 10^{-10}$~m~s$^{-2}$ is appropriate for all the galaxies in the sample. This value also is consistent with the MOND acceleration scale $a_0$. The asymptotic limit of Eq.~\eqref{eq:fittedRAR} for small $g_\text{bar}$ returns the acceleration in the MOND regime $g_\text{obs}\sim \sqrt{g_\text{bar}(R) {a_0}}$. 

For the galaxies of the SPARC sample, this correlation has a relatively small root-mean-square scatter of $0.13$ dex, mostly due to possible variations of the stellar mass-to-light ratio from galaxy to galaxy and to observational uncertainties~\citep{McGetal16, Lellietal17a}. Indeed, for this result,~\citet{McGetal16} adopt a single $3.6$ $\mu$m mass-to-light ratio of $0.50$~M$_\odot/$L$_\odot$ for all the galaxy disks, under the assumption that the stellar mass-to-light ratio does not vary much in this band~\citep{McG&S14,McG&S15,Meidt14,S&McG14}. Similarly, for the galaxy bulges, which are present in 31 out of 153 galaxies,~\citet{McGetal16} adopt a mass-to-light ratio equal to $0.70$~M$_\odot/$L$_\odot$. 

To explore the intrinsic scatter of the RAR due to the mass-to-light ratio variations,~\citet{Lietal18} fit galaxy mass-to-light ratios to individual galaxies with Eq.~\eqref{eq:fittedRAR} and $g_\dagger$ fixed to $1.2\times 10^{-10}$~m~s$^{-2}$. They find a RAR tighter than~\citet{McGetal16}, with an intrinsic root-mean-square scatter of $0.057$ dex and mass-to-light ratios generally consistent with the SPS model predictions.

Here, we estimate the RAR for our DMS sample. The observed acceleration, $g_\text{obs}$, is directly derived from the measured rotation curve according to Eq.~\eqref{eq:gobs}. The Newtonian acceleration attributed to the baryonic matter alone, $g_\text{bar}=\vert -\partial\phi/\partial R \vert$, is derived from the numerical solution of the Newtonian Poisson equation~\eqref{eq:PoissonN} where the density, $\rho(R,z)$, is Eq.~\eqref{eq:rhotot}. For the galaxy disk-scale height, $h_z$, which appears in $\rho(R,z)$, we use the values derived from Eq.~\eqref{eq:hzhR} inferred from the observations of edge-on galaxies (see Appendix~\ref{sec:3D}). For the numerical solution of the Poisson equation, we adopt the successive over relaxation algorithm described in Appendix~\ref{sec:SOR}.

To estimate the mass density $\rho(R,z)$ from the galaxy surface brightness, we adopt the mass-to-light ratios derived with our MCMC analysis (Sect.~\ref{sec:VVD_RC}) and listed in Table~\ref{tab:fit_VVD_RC_orig_err_bars_v}. 
As discussed above, for 26 galaxies out of 29, these values agree within $3\sigma$ with the values of the SPS models \citep[see][Table 1]{B&deJ01} and for all the galaxies, the difference is within $5\sigma$.

\begin{figure*}
        \centering
        \includegraphics[width=17cm]{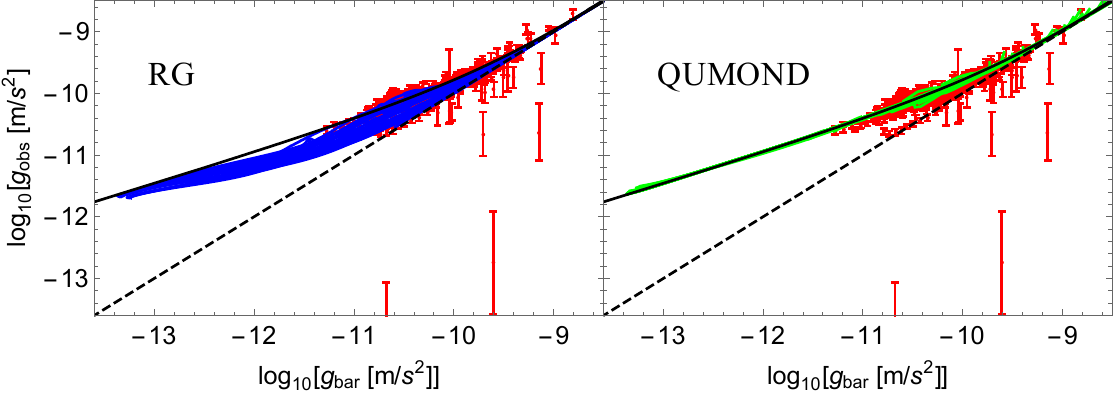}
        \caption{RG ({\it left panel}) and QUMOND ({\it right panel}) models of the RAR obtained for each galaxy with the parameters derived from the MCMC analysis of the rotation curves and vertical velocity dispersion profiles (Sect.~\ref{sec:VVD_RC}). Red points with error bars are the DMS measures. The black solid line is Eq.~\eqref{eq:fittedRAR}. The black dashed line is $g_\text{obs}=g_\text{bar}$.
    }
        \label{fig:RAR}
\end{figure*}

   Red symbols and error bars in both panels of Fig.~\ref{fig:RAR} show the RAR of all the DMS galaxies in our sample. The black curve is Eq.~\eqref{eq:fittedRAR}. The blue curves in the left panel of Fig.~\ref{fig:RAR} are the RAR of each DMS galaxy obtained from the RG parameters, the mass-to-light ratios and the disk-scale heights derived from our MCMC analysis of  the rotation curves and vertical velocity dispersion profiles (Sect.~\ref{sec:VVD_RC}). The blue curves are not fits to the observed RAR, but just the relations between the Newtonian $g_{\rm bar}$ and the expected RG centripetal acceleration $g_{\rm obs}$ based on the galaxy parameters estimated with our previous analysis. The dashed line shows the relation $g_\text{obs}=g_\text{bar}$ for comparison.

We also estimate the RAR expected in MOND, adopting the QUMOND formulation described in Sect.~\ref{sec:VVD_RC_1pt55}.
For QUMOND, we adopt the mass-to-light ratios $\Upsilon$ and the disk-scale heights $h_z$ that we derive for RG in Sect.~\ref{sec:VVD_RC}. These mass-to-light ratios agree within $2\sigma$ with the values estimated by~\citet{Angus15}, who model the rotation curves and vertical velocity dispersions in QUMOND with the simple interpolating function. Similarly, their values of $h_z$ agree with our estimates within $1\sigma$.
The QUMOND RAR curves are shown as green solid lines in the right panel of Fig.~\ref{fig:RAR}.

\begin{figure}
        \resizebox{\hsize}{!}{\includegraphics{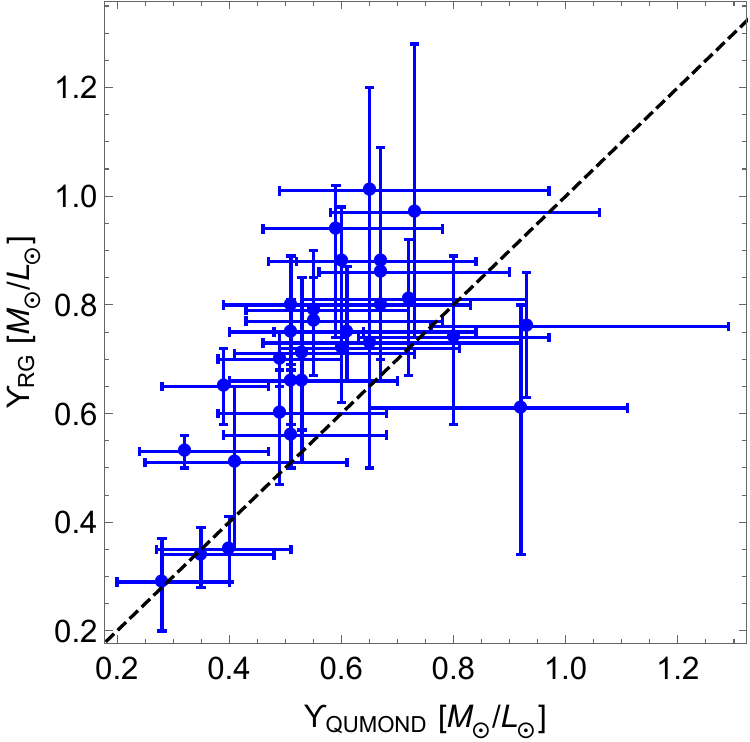}}
        \caption{Mass-to-light ratios estimated with RG, $\Upsilon_\mathrm{RG}$ (Table~\ref{tab:fit_VVD_RC_orig_err_bars_v}), and with QUMOND, $\Upsilon_\mathrm{QUMOND}$~\citep[][Table 1]{Angus15}. 
                The black dashed line is the line of equality.}
        \label{fig:SP_YRG_YMOND}
\end{figure}

RG correctly reproduces the asymptotic limits of the observed RAR and, on average, it interpolates the data, although it tends to underestimate the relation~\eqref{eq:fittedRAR} at low $g_\mathrm{bar}$; on the contrary, QUMOND properly reproduces the shape of the RAR relation~\eqref{eq:fittedRAR} along the entire range of $g_\mathrm{bar}$. The fact that RG slightly underestimates the observed RAR while at the same time providing a good fit to the kinematics of the individual galaxies, as shown in the previous sections, suggests that RG might attribute more mass to the luminous matter than QUMOND. In fact, Fig.~\ref{fig:SP_YRG_YMOND} shows that the RG mass-to-light ratios are systematically larger than in QUMOND, although the mass-to-light ratios in the two models agree with each other within $2\sigma$. This result is consistent with the left panel of Fig.~\ref{fig:Scatter_plots_RG_MOND_sigz1pt55} of Sect.~\ref{sec:VVD_RC_1pt55}.

\begin{figure}
        \resizebox{\hsize}{!}{\includegraphics{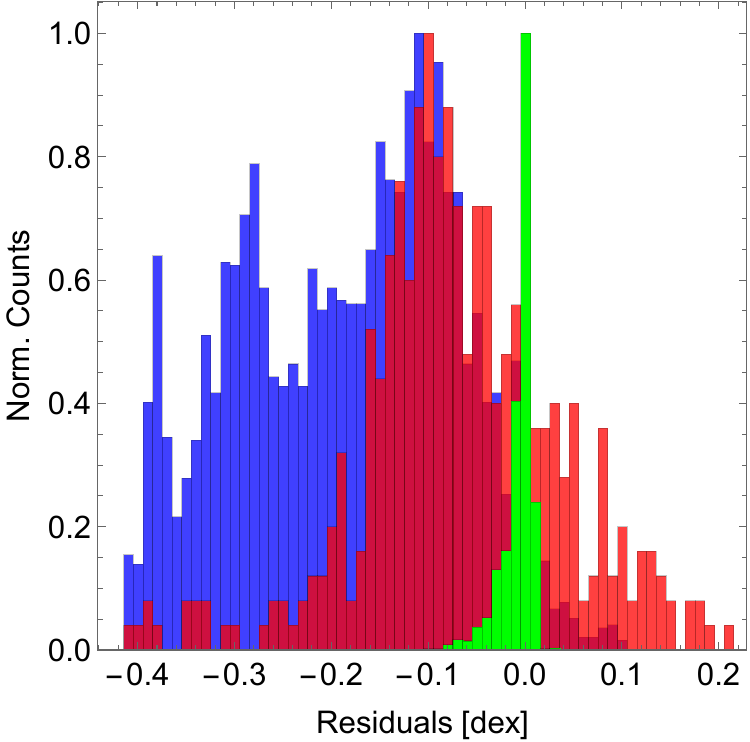}}
        \caption{Distributions of the residuals of the RG models (blue), the QUMOND models (green) and the data (red) of the RAR with respect to relation~\eqref{eq:fittedRAR}.}
        \label{fig:histo_distrib_res}
\end{figure}

A more serious issue for RG is the scatter of the curves along the $g_{\rm obs}$ axis. Figure~\ref{fig:histo_distrib_res} shows the distributions of the deviations of each curve from Eq.~\eqref{eq:fittedRAR}. We only consider the deviations of each curve from Eq.~\eqref{eq:fittedRAR} within the horizontal axis range of $\log_{10} [g_{\rm bar} (\mathrm{m}/\mathrm{s}^2)]=[-11.28,-8.81]$ covered by the data. This approach makes the comparison with the data more sensible. 
In passing, we note that we use the disk-scale heights $h_{z,\text{SR}}$ for the data and our estimated $h_z$ for the models.
These values can be different by a factor of two, as shown in the right panel of Fig.~\ref{fig:SP_hzRC_hzRCVVD_hzSR} and in Table~\ref{tab:fit_VVD_RC_orig_err_bars_v}. However, adopting, for the models,  $h_{z,\text{SR}}$ rather than $h_z$, leaves the distributions of the deviations basically unaffected.

The width of the residual distribution in Fig.~\ref{fig:histo_distrib_res} for the data, which quantifies the observed scatter of the RAR, is $0.12$ dex. We estimate this scatter by removing four outlying points, clearly visible in the right part of both panels of Fig.~\ref{fig:RAR}. These points belong to the galaxies UGC 1081, \object{UGC 1862}, UGC 3997, and UGC 6903, and they correspond to the innermost point of their rotation curves; these values are $17.00$, $0.05$, $1.19,$ and $17.69$~km~s$^{-1}$, which  are unusually small. If we include these four points, the root-mean-square scatter increases from $0.12$~dex to $0.32$~dex. The observed scatter of the DMS sample is thus comparable to the value $0.13$~dex found by~\citet{McGetal16} and~\citet{Lellietal17a}.

 The widths of the distributions of the residuals for the RG and QUMOND models shown in Fig.~\ref{fig:histo_distrib_res} are $0.11$ and $0.017$ dex, respectively. We can identify these widths with the intrinsic scatter of the RAR predicted by the two models. The small intrinsic scatter predicted by QUMOND, consistent with the expectations~\citep{Lellietal17a,BradaandMilgrom95}, is almost an order of magnitude smaller than the RG scatter. MOND actually appears with different formulations: in the version of modified inertia, which modifies the Newtonian second law of dynamics, the intrinsic scatter is predicted to be zero if the orbits are circular~\citep{Milgrom94}; similarly, in the version of modified gravity, which modifies the Poisson equation, like QUMOND does, the intrinsic scatter is predicted to be zero only for spherically symmetric systems~\citep{Bekenstein&Milgrom84}, whereas in flat systems, like disk galaxies, a small not-null intrinsic scatter should appear.

\begin{figure*}
        \centering
        \includegraphics[width=17cm]{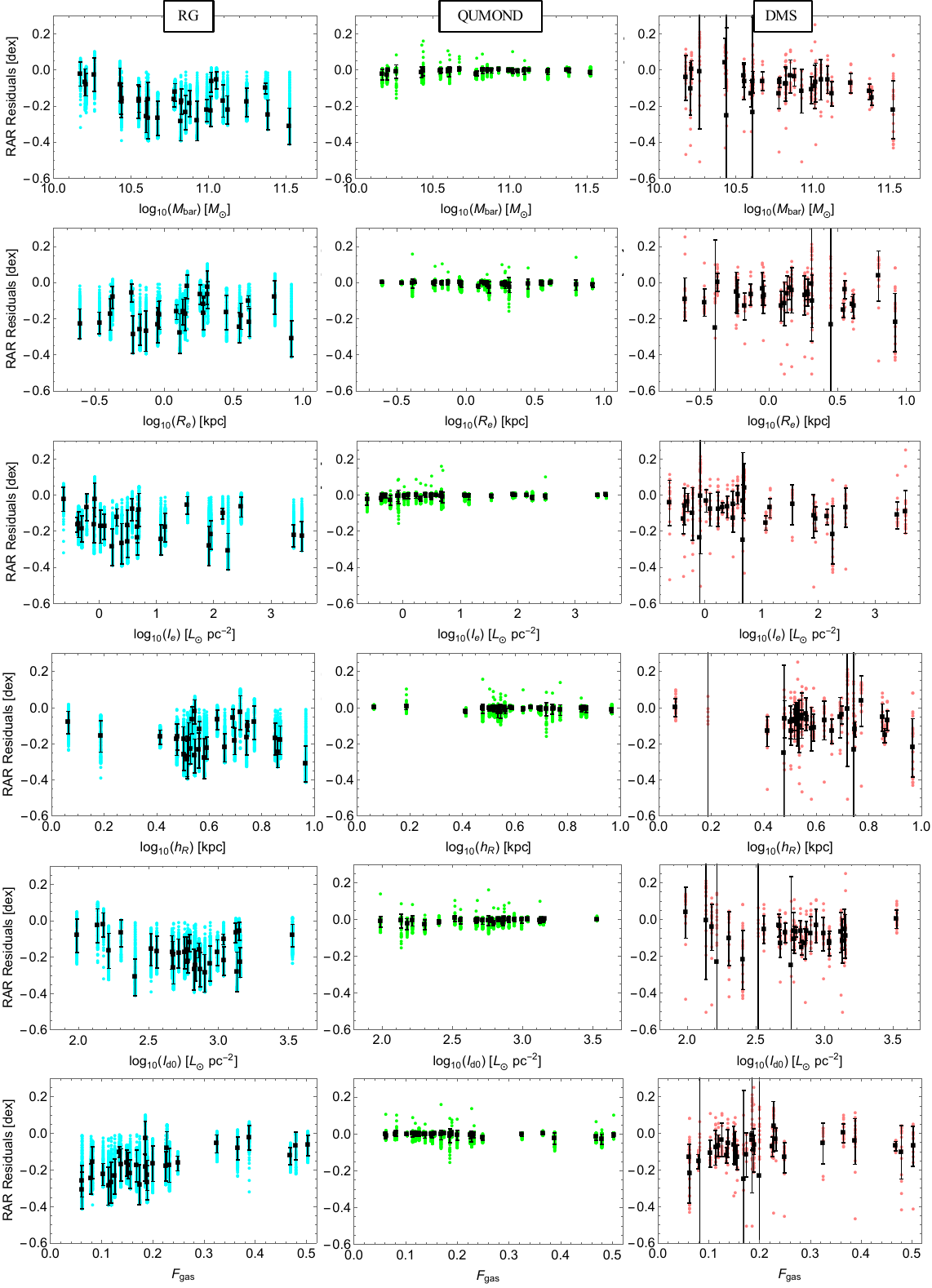}
        \caption{Residuals of the modelled or estimated RAR from the relation~\eqref{eq:fittedRAR} as a function of the global properties of the galaxies. {\it Left}, {\it middle}, and {\it right} columns: RG, QUMOND, and DMS residuals, respectively. From {\it top} to {\it bottom} the residuals are plotted against 
        the total baryonic mass, bulge effective radius, bulge effective surface brightness, disk-scale length, central disk surface brightness, and total gas fraction. 
        To guide the eye, solid squares with error bars show the means and standard deviations of binned residuals.}
        \label{fig:RAR_Residuals_Global_all}
\end{figure*}

\begin{figure*}
        \centering
        \includegraphics[width=17cm]{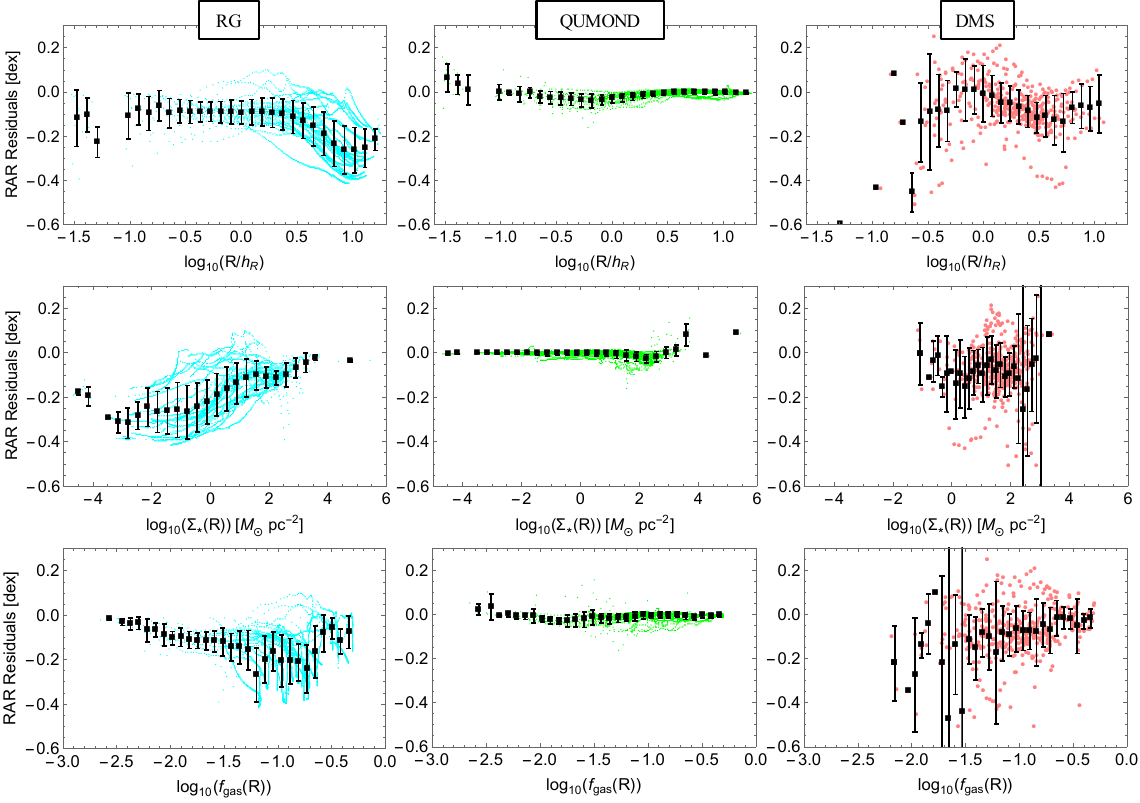}
        \caption{Same as Fig. \ref{fig:RAR_Residuals_Global_all} for three radially-dependent properties of the galaxies. From {\it top} to {\it bottom} the residuals are plotted against 
        the radius, the stellar surface density profile, and the gas fraction profile. }
        \label{fig:RAR_Residuals_Local_all}
\end{figure*}

To investigate the nature of the intrinsic scatter predicted by RG, we explore the possible correlation of the RAR residuals with the global and the radially-dependent properties of the galaxies. We plot these residuals in Figs.~\ref{fig:RAR_Residuals_Global_all} and~\ref{fig:RAR_Residuals_Local_all}. The first column shows, in cyan, the residuals of the RG models. The second and the third columns show the residuals for QUMOND, in green, and the DMS data, in pink.\footnote{ The residuals for the four outlying points of the DMS data that we mention above do not appear in the plots because they lie beyond the range of the vertical axis. They are however included in our statistical tests we describe below. These four outliers do not drive the correlations of the DMS data that we find: if we remove them when performing the statistical tests, the statistical significance of the correlations actually increases.}

Table~\ref{tab:Kendall_Spearman} lists the Kendall statistic $\tau$~\citep{Kendall1938} and the Spearman statistic $\rho$~\citep{Spearman1904} with their corresponding $p$-values, namely, the significance levels of the lack of correlation. For large size $N$ of the sample, the density distributions of the random variates $\tau$ and $\rho$ are excellently approximated
by the Gaussian distribution with zero mean and variance  $(4N+10)/(9N^2-9N)$ and $1/(N-1)$ for  $\tau$ and $\rho$, respectively \citep{Best1973,Kendall1975}. 
In our analysis of RG, $N=4560$ and, assuming Gaussian distributions, the corresponding standard deviations $\sigma$'s are $\langle \tau^2\rangle^{1/2}=0.0099$ and $\langle \rho^2\rangle^{1/2}=0.0148$. 

We can interpret the results of Table~\ref{tab:Kendall_Spearman} by arbitrarily choosing the threshold  $\log_{10}p=-3$: $p$-values smaller than the threshold of $p=10^{-3}$ indicate that the listed values of $\tau$ or $\rho$ have a probability smaller than $0.1\%$ of occurring by chance for an uncorrelated sample. For this probability, the values $\vert\tau\vert>0.049$ and $\vert\rho\vert>0.033$ are thus more than $3.3\sigma$ away from the expected means $\langle\tau\rangle=0$ and $\langle\rho\rangle=0$.
For RG, the only two parameters that have $p$-values larger than $10^{-3}$, namely, $-\log_{10}p<3$, and therefore their uncorrelation with the residuals is statistically significant, are the central surface brightness $I_{\mathrm{d}0}$ of the disk and its scale length $h_R$. The remaining parameters show significant correlations.
This result might appear at odds with the observed RAR because the observed residuals do not seem to correlate with the galaxy properties in the SPARC sample~\citep{Lellietal17a}.

This issue requires additional clarification. In fact, unlike the SPARC sample, the DMS sample also shows 
some correlations: the $p$-values listed in Table~\ref{tab:Kendall_Spearman} indicate that the RAR residuals are not
significantly correlated ($-\log_{10}p<3$) only with $h_R$, $I_{\rm d0}$, $R_{\rm e}$, and the stellar surface brightness profile, $\Sigma_*(R)$  (see also Figs.~\ref{fig:RAR_Residuals_Global_all} and~\ref{fig:RAR_Residuals_Local_all}).
Moreover, similarly to RG, correlations are found between the residuals of   
the QUMOND models and all the galaxy properties but the bulge central surface brightness
$I_{\rm e}$.\footnote{ In passing, we emphasise that the statistical significance of the
correlation is quantitatively supported by the $p$-values listed in Table~\ref{tab:Kendall_Spearman}: from a qualitative visual inspection of Figs.~\ref{fig:RAR_Residuals_Global_all} and~\ref{fig:RAR_Residuals_Local_all}, one might draw the incorrect conclusion that the residuals of QUMOND generally show a weaker correlation, if any, than the DMS data.}
For QUMOND, this result might not be surprising, however, because QUMOND is a modified-gravity version of MOND, where a non-null intrinsic scatter for the RAR, and thus correlated residuals, are expected for non-spherical systems, unlike modified-inertia versions of MOND, that predict a null intrinsic scatter, and thus uncorrelated residuals \citep{Bekenstein&Milgrom84}.

The correlations we find for the residuals of the DMS data might partly generate  the correlations we find for the RG residuals.
In addition, the correlations for the DMS data, at odds with the claimed uncorrelations for the SPARC sample, 
might suggest a difference between
the DMS and the SPARC samples. Unfortunately, we cannot quantify this difference here, if there even is any, 
because~\citet{Lellietal17a} only mention in their analysis that
the Kendall and Spearman coefficients
are in the range of $[-0.2,0.1]$ but they do not report their corresponding 
significance levels, namely their $p$-values. Therefore, we cannot assess the statistical significance of the
lack of correlation. In fact, many coefficients in Table~\ref{tab:Kendall_Spearman} from our analysis are in the range of $[-0.2,0.1]$,
but their $p$-values clearly indicate that they are at many $\sigma$'s away from the null expected means and, thus, they demonstrate the presence of a statistically significant correlation.

The significance levels listed in Table~\ref{tab:Kendall_Spearman} suggest that the correlations for the RG models are much stronger than 
for the DMS data for all the galaxy properties but $I_{\rm d0}$. In addition, RG shows a very strong correlation, at largely more than $5\sigma$, namely $-\log_{10} p>6.24$, with the radially-dependent properties of the galaxies, whereas the data show a significant correlation, between $4$ and $5\sigma$, namely $4.20<-\log_{10} p<6.24$, for $R$ and $f_{\rm gas}(R)$, but no correlation for $\Sigma_*(R)$. 

Therefore, a possible serious tension between RG and the data might indeed be present. However, given the possible 
tension between the DMS and the SPARC samples, which is yet to be confirmed, we conclude that this issue remains open at this stage of our testing of RG. 
Further investigations with multiple
data samples are necessary to clarify whether reproducing the observed properties of the RAR is indeed a challenge for
RG.

\begin{sidewaystable*}
                \caption{Correlations of the RAR residuals of RG models, QUMOND models, and DMS data with the galaxy properties.} 
                \label{tab:Kendall_Spearman}    
                \centering
                        \begin{tabular}{|l|cccc|cccc|cccc|}
                                \hline\hline
                                &&&&&&&&&&&& \\
                                {\bf Galaxy}& \multicolumn{4}{|c|}{\bf RG} & \multicolumn{4}{|c|}{\bf QUMOND} & \multicolumn{4}{|c|}{\bf DMS} \\
                                        {\bf property} & $\tau$ & $-\log_{10}p_\tau$ & $\rho$ & $-\log_{10}p_\rho$ & $\tau$ & $-\log_{10}p_\tau$ & $\rho$ & $-\log_{10}p_\rho$ & $\tau$ & $-\log_{10}p_\tau$ & $\rho$ & $-\log_{10}p_\rho$ \\
                                (1) & (2) & (3) & (4) & (5) & (6) & (7) & (8) & (9) & (10) & (11) & (12) & (13)\\
                                \hline
                                $h_R\text{ }[\mathrm{kpc}]$ &$-0.030$&$2.62$&$-0.038$&$1.96$&$-0.065$&$9.85$&$-0.094$&$9.72$ &$-0.042$& $0.68$ &$-0.060$&$0.66$\\
                                $I_{\mathrm{d}0}\text{ }\left[\frac{\mathrm{L}_\odot}{\mathrm{pc}^2}\right]$ &$+0.0083$&$0.39$&$+0.0052$&$0.14$&$+0.049$&$5.92$&$+0.073$&$6.10$&$-0.099$&$2.55$&$-0.15$&$2.64$\\
                                $M_\mathrm{bar}\text{ }[\mathrm{M}_\odot]$ &$-0.16$&$56.6$&$-0.23$&$54.4$ &$-0.053$&$6.77$&$-0.074$&$6.26$&$-0.27$&$15.7$&$-0.38$&$15.2$\\
                                $R_\mathrm{e}\text{ }[\mathrm{kpc}]$ &$+0.046$&$5.14$&$+0.061$&$4.18$ &$-0.094$&$19.0$&$-0.14$&$20.3$&$+0.0020$&$0.022$&$-0.012$&$0.086$\\
                                $I_\mathrm{e}\text{ }\left[\frac{\mathrm{L}_\odot}{\mathrm{pc}^2}\right]$ &$-0.10$&$22.6$&$-0.15$&$23.6$ &$-0.00023$&$0.0088$&$+0.011$&$0.31$&$-0.15$&$5.21$&$-0.23$&$5.41$\\
                                $F_\mathrm{gas}$ &$+0.33$&$237$&$+0.48$&$256$ &$-0.11$&$29.0$&$-0.17$&$31.6$&$+0.21$&$9.82$&$+0.31$&$10.1$\\
                                $R\text{ }[\mathrm{kpc}]$ &$-0.41$&$377$&$-0.59$&$419$ &$+0.15$&$53.9$&$+0.27$&$76.1$&$-0.16$&$6.10$&$-0.21$&$4.89$\\
                                $\Sigma_*(R)\text{ }\left[\frac{\mathrm{M}_\odot}{\mathrm{pc}^2}\right]$ &$+0.41$&$368$&$+0.57$&$395$ &$-0.17$&$62.6$&$-0.28$&$84.5$&$+0.033$&$0.49$&$+0.041$&$0.40$\\
                                $f_\mathrm{gas}(R)$ &$+0.073$&$12.9$&$+0.12$&$16.6$ &$+0.14$&$42.0$&$+0.20$&$43.2$&$+0.16$&$6.10$&$+0.24$&$6.12$\\
                                \hline
                        \end{tabular}
                    \tablefoot{Column 1: name of the galaxy property; Cols 2--5: correlations of the RAR residuals of RG models with the galaxy properties; Cols. 6--9: correlations of the RAR residuals of QUMOND models with the galaxy properties; Cols. 10--13: correlations of the RAR residuals of DMS data with the galaxy properties. $\tau$  and $\rho$ are the Kendall and Spearman correlation coefficients, respectively; $p_\tau$ and $p_\rho$ the  corresponding significance levels of the lack of correlation.}
\end{sidewaystable*}

\section{Conclusions}
\label{sec:conclusions}

We test the viability of RG, a theory of modified gravity that does not require the existence of dark matter to describe
the dynamics of cosmic structures. We test RG on the scale of galaxies with 
30 disk galaxies from the DMS~\citep{DMSi}. These galaxies appear almost face-on, having
an inclination between $5\deg$ and $46\deg$ with respect to the line of sight. 
We can thus model, using a MCMC approach, both the rotation curves and the vertical velocity dispersion profiles.

The Poisson equation in RG contains a universal monotonic function of the local mass density, the gravitational permittivity, 
for which we adopt a smooth step function containing three free parameters: the permittivity of the vacuum $\epsilon_0$, the critical density $\rho_\mathrm{c}$, which sets the transition between the RG and the Newtonian regimes, and the transition 
power index $Q$.  

By modelling each galaxy with two free parameters, the mass-to-light ratio, $\Upsilon$, and the disk-scale height, $h_z$,
and the additional three RG parameters, our MCMC analysis shows that RG is indeed able to describe the dynamics of these 
DMS galaxies with sensible values of $\Upsilon$ and $h_z$. Specifically, the mass-to-light ratios are consistent with those expected by SPS models~\citep{B&deJ01}; the agreement improves when we model  both the rotation curves and the vertical velocity dispersion profiles rather than the rotation curves alone. 

Similarly, the disk-scale heights, $h_z$, are consistent
with the disk thicknesses, $h_{z,\text{SR}}$, inferred from the observation of edge-on galaxies. When modelling both the rotation curves and the vertical velocity dispersion profiles, $h_z$ appears to prefer values that are a factor of $\sim$$2$ smaller than $h_{z,\text{SR}}$. However, this discrepancy might originate from an observational bias pointed out by~\citet{Milgrom15} and quantified by 
\citet{Aniyan16}: the estimate of $h_{z,\text{SR}}$ is based on near infrared photometry coming from old giant stars, with a larger vertical velocity dispersion and disk-scale height, whereas the vertical velocity dispersion profiles are estimated from integrated spectra, where the signal is dominated by young giants that have a smaller velocity dispersion. The observed velocity dispersion thus underestimates the velocity dispersion that would correspond to the observed $h_{z,\text{SR}}$.  We estimate a bias factor of $1.63$ for the DMS galaxies, a value that is consistent within $1\sigma$ with the value suggested by \citet{Aniyan16}.

In the formulation of RG, $\epsilon_0$, $Q$, and $\rho_\mathrm{c}$ should be universal parameters, rather than free parameters for
each individual galaxy as we assume above. To verify that a single set of these three parameters might indeed
model the entire galaxy sample, we adopt a simple strategy. We assume that the mass-to-light ratio, $\Upsilon$, and the disk-scale height, $h_z$, set by the previous analysis are appropriate and we only perform a MCMC exploration of the three-dimensional RG parameter space for the entire galaxy sample. The most likely set of the RG parameters is, within $1.2\sigma$, consistent with the distributions of the RG parameters  from the previous analysis. Although, as expected, the models of the rotation curves and of the vertical velocity dispersion profiles worsen, they still appear to be reasonable for most galaxies. This result suggests that a unique set of RG parameters describing the kinematics of the DMS galaxies with reasonable $\Upsilon$ and $h_z$ might indeed exist.

Finally, we show that RG can in principle describe the RAR, namely the relation between the observed acceleration and the Newtonian acceleration due to the baryonic matter alone~\citep{McGetal16}. The RAR expected in RG interpolates the data reasonably well and has the two asymptotic limits of the observed RAR.
However, the models slightly underestimate the observed accelerations at low Newtonian accelerations. 
Moreover, the intrinsic scatter
originating from the RG models appears not to be consistent with zero, as galaxy samples different from DMS seem to suggest. In fact, \citet{McGetal16} use the SPARC sample to show that the observed scatter of $0.13$ dex in the RAR, obtained by adopting the same mass-to-light ratio for all the galaxies, is comparable to the scatter of $0.12$ dex due to rotation curves, disk inclinations and galaxy distance uncertainties and to possible variations in mass-to-light ratios; this agreement leaves negligible room for intrinsic scatter.
        
        An additional tension might be the correlation between the galaxy properties and the residuals of the RG models from the RAR described
        by Eq.~\eqref{eq:fittedRAR}, which might appear at odds with the uncorrelations claimed by~\citet{Lellietal17a} for the SPARC sample. In particular, for the three radially-dependent properties of the galaxies, radius, surface brightness and gas fraction, the correlations are statistically significant at largely more than $5\sigma$. However, we also find significant correlations with radius and gas fraction, at more than $4\sigma$, for the residuals of the DMS data.
        Further investigations are thus required to settle the issue. 
        Modelling the SPARC data with RG, as we plan to do next, might shed light on whether the correlations we find here are, in fact, a feature
        of RG and, therefore represent a serious failure for RG or whether they are partly derived from the observed galaxy sample. 
        
\begin{acknowledgements}
        We sincerely thank the referee, whose detailed comments largely improved and clarified the presentation of our results, and especially our discussion of the RAR. We also thank Alistair Hodson, Michal B{\'i}lek, Luisa Ostorero, and Stefano Camera for inspiring discussions. We additionally thank Andrea Mignone for his crucial help with the Poisson solver, Emille Ishida for her help with the MCMC analysis with the software JAGS, and Marco Aldinucci for his help with the parallelisation of the code with OpenMP. We thank Compagnia di San Paolo (CSP) for funding the graduate-student fellowship of VC. We also acknowledge partial support from the INFN grant InDark and the Italian Ministry of Education, University and Research (MIUR) under the {\it Departments of Excellence} grant L.232/2016.
\end{acknowledgements}

        \bibliographystyle{aa}
        \bibliography{bib_paper1}
        
        \begin{appendix}
                
                \section{Modelling the mass distribution in the disk galaxies}
                \label{sec:SBgas}
                
                Here, we describe how we model (1) the surface brightness of the bulge and the stellar disk (Appendix \ref{sec:SB}); (2) the mass distribution of the gas (Appendix \ref{sec:gas}); and (3) the total three-dimensional mass distribution of the galaxy (Appendix \ref{sec:3D}).
                We adopt the Hubble constant $H_0=73$~km~s$^{-1}$~Mpc$^{-1}$, as per~\citet{DMSvi}.

                \subsection{Surface brightness of the bulge and the stellar disk}
                \label{sec:SB}
                
            The surface brightness of disk galaxies is the sum of the contribution of the bulge, which dominates the inner regions, and the contribution of the disk, which dominates the external regions. The surface brightness profile usually shows a distinct change of slope at the radius that separates the central region dominated by
                the bulge and the outer region dominated by the disk (see sub-panels (a) of Figs.~\ref{fig:Complete_analysis_1}-\ref{fig:Complete_analysis_7}). 
                
                To preserve the specific features of the distribution of the luminous matter, which in general correspond to features of the rotation curve, according to `Renzo's rule' \citep{Sancisi04}, we model the disk surface brightness with a linear interpolation of the surface brightness data points only beyond the central region. To estimate the disk contribution within the central region, 
                we fit the surface brightness data point beyond the central region with an exponential profile \citep{deV59,Freeman70}:
                \begin{equation}
                \label{eq:Id}
                I_\mathrm{d}(R)=I_\mathrm{d0}\exp\left(-\frac{R}{h_R}\right) \; ,
                \end{equation}
                where $h_R$ is the disk-scale length. 
                
                Therefore, to describe the disk surface brightness on the full radial range of the galaxy, we adopt the exponential model for the disk contribution within the central region,
                dominated by the bulge, and the linear interpolation of the data points in the outer region. We also adopt the exponential model 
                in the very outer region if the grid of the Poisson solver we use in Sects.~\ref{sec:Only_RC},~\ref{sec:VVD_RC},~\ref{sec:Fit_all}, and~\ref{sec:RAR} goes beyond the measured profile.
                
                We model the bulge surface brightness with a S\'ersic spherical profile: 
                \begin{equation}
                \label{eq:Ib}
                        I_\mathrm{b}(R)=I_\mathrm{e}\exp\left\{-7.67\left[\left(\frac{R}{R_\mathrm{e}}\right)^{\frac{1}{n_\mathrm{s}}}-1\right]\right\} \; ,
                \end{equation}
                with $R_\mathrm{e}$ the effective radius, $I_\mathrm{e}$ the surface brightness at radius $R_\mathrm{e}$, and $n_\mathrm{s}$ the S\'ersic index. The average bulge-to-total luminosity ratio in the $K$-band in the DMS galaxies is $0.09$ (see Table 1 in~\citealt{DMSvi}); therefore, ignoring the triaxiality of the bulge should introduce negligible systematic errors on the modelling of the galaxy luminosity. 
                
                For the bulge, we adopt the S\'ersic model rather than interpolating the surface brightness data points as we do for the disk, for two reasons: (1) the seeing affects the central region more than the outer region, as we detail below, and (2) unlike a two-dimensional disk, the bulge cannot be trivially deprojected without an analytical approximation. The deprojection is a required step to model the galaxy mass distribution we illustrate in Appendix~\ref{sec:3D}.
        
                To model the surface brightness profile, we also consider the seeing that affects the measurements made with the $3.5$~m diameter ground telescope at the Calar Alto Observatory. We model the seeing with a Gaussian point spread function,  
                \begin{equation}
                \label{eq:PSF}
                        p_{\rm PSF}(R)=\frac{1}{2\pi\sigma^2}\exp\left[-\frac{(R-R_{\text{peak}})^2}{2\sigma^2}\right] \; ,
                \end{equation} 
                where $\sigma$ is the average effective seeing, as listed in Tables 3 and 4 of~\citet{DMSvi},
                and $R_{\text{peak}}$ is the location of the surface brightness peak. 
                
                According to~\citet{DMSvi}, we only convolve the bulge profile (Eq.~\ref{eq:Ib}) with Eq.~\eqref{eq:PSF}:
                \begin{equation}
                \label{eq:conv}
                I_{\rm obs}(R)=\int_{-\infty}^{+\infty}I_{\rm true}(R^\prime)p_{\rm PSF}(R^\prime-R)\text{ }dR^\prime \; .
                \end{equation}
                Ignoring the effect of the seeing on the disk should introduce negligible systematic errors because the disk is a factor of ten more luminous than the bulge, on average, and spatially more extended. 

                The galaxy disks in the DMS sample appear almost face-on, with small inclination angles $i$. If we neglect the dust extinction within the galaxy, the observed surface brightness of the disk is brighter than the intrinsic surface brightness because
                the observed flux appears to come from an area $\cos i$ times smaller than the actual disk area. Therefore, to derive the light distribution of the disk from the measured surface brightness, $\mu_K$, we need to correct for the inclination of the disk with respect to the line of sight. 
                
                By assuming the Tully-Fisher relation~\citep{Verheijen01},
                \begin{equation}
                        \label{eq:TF}
                        \log_{10} v_{\rm TF}=-0.30103+(5.12-M_K)/11.3 \; ,
                \end{equation}
                \citet{DMSvi} estimate the inclination angle, 
                \begin{equation}
                \label{eq:iTF}
                \sin i_{\rm TF}=\frac{v_{\rm obs}}{v_{\rm TF}} \; ,
                \end{equation}
                from the observed luminosity, $M_K$, and the observed rotation velocity, $v_{\rm obs}$. The face-on surface brightness of the disk is thus:                
                \begin{equation}
                        \label{eq:inclination}
                        \mu^i_K=\mu_K-2.5C_k\log(\cos i_{\rm TF}) \; ,
                \end{equation}
                where $C_k=1$, for a transparent disk, as assumed by~\citet{DMSvi}. 
                
                For the conversion of the surface brightness from the astronomical units mag~arcsec$^{-2}$ to units L$_\odot$~pc$^{-2}$, we use the equation~\citep{B&M98}:
                \begin{equation}
                \label{eq:SBconversion}
                \mu\left(\frac{\text{mag}}{\text{arcsec}^2}\right)=M_{\odot,K}+21.572-2.5\log_{10}I\left(\frac{\text{L}_\odot}{\text{pc}^2}\right) \; ,
                \end{equation}
                where $M_{\odot,K}=3.28$ is the absolute magnitude of the Sun in the $K$-band.
                This equation neglects the expansion of the Universe because all the DMS galaxies are nearby, with the farthest galaxy (UGC 4622) at $z=0.043$, namely at the distance $178.2$~Mpc~\citep[see][Table 1]{DMSvi}.
        
            We now illustrate the steps we adopt to model the entire surface brightness profile. We apply this procedure to the surface brightness profiles that were already corrected for inclination with Eq.~\eqref{eq:inclination} by~\citet{DMSvi}.
        
            As anticipated, most observed surface brightness profiles in the DMS sample show a distinct change of slope that suggests how extended the bulge is. We remove the central data points at radii smaller than the location of the slope change. The removed data points for each galaxy are shown in sub-panels (a) of Figs.~\ref{fig:Complete_analysis_1}-\ref{fig:Complete_analysis_7} as green dots. 
            We assume that the remaining profile, given by the red dots in sub-panels (a) of Figs.~\ref{fig:Complete_analysis_1}-\ref{fig:Complete_analysis_7}, is only due to the surface brightness of the disk. We linearly interpolate this remaining profile. The disk contribution in the most inner region, where the bulge contribution is dominant, is estimated with 
            the exponential model of Eq.~\eqref{eq:Id}, with the parameters of the model set by the least-square best fit to the data points of the outer region alone. 
        
             We now subtract the extrapolated surface brightness profile of the disk from the observed total surface brightness profile of the
                central region, namely the green data points in sub-panels (a) of Figs.~\ref{fig:Complete_analysis_1}-\ref{fig:Complete_analysis_7}.
                                The remaining profile is the surface brightness profile of the bulge that we now model with Eq.~\eqref{eq:conv}. 
                                Because of the presence of the convolution integral, to estimate the three parameters  $I_\mathrm{e}$, $R_\mathrm{e}$ and $n_\mathrm{s}$ of the bulge, it is more convenient to apply a MCMC approach. We run the MCMC algorithm with JAGS\footnote{\url{https://sourceforge.net/projects/mcmc-jags/}}, a free software that adopts the Gibbs sampling algorithm to generate the Markov chains. 
                                
                                We adopt Gaussian priors on the three free parameters; we set the Gaussian tails to zero for the unphysical negative values of the parameters. The Gaussian dispersions are set to $1/\sqrt{200}$ L$_\odot$~pc$^{-2}$, $1/\sqrt{200}$~arcsec, and $1/\sqrt{200}$ for  $I_\mathrm{e}$, $R_\mathrm{e}$ and $n_\mathrm{s}$, respectively.
                                Assuming a Gaussian prior avoids the choice of an upper limit required in a uniform prior. To determine the means of the Gaussian priors, we compare the surface brightness profiles with the model profiles with a number of different choices of the parameters set by hand. We pick up the set of parameters that qualitatively best reproduce the data. This simple approach is sufficient to set the means of the Gaussian priors to reasonable values. The burn-in chain has $10,000$ steps and we then run the chain for $100,000$ steps.
                                
                        For the best parameter estimates and their uncertainties, we adopt the medians and the standard deviations of the posterior distributions obtained from the MCMC runs. This choice is justified by the fact that the posterior distributions show a single peak and are basically symmetric.
                        
                                By adopting this procedure, we find that two galaxies, UGC 1862 and UGC 9965, have no surface brightness in excess to the disk in the central region and we thus consider them bulgeless, a finding that is in agreement with~\citet{DMSvi}. \citet{DMSvi} find two additional bulgeless galaxies:  UGC 3091 and UGC 7244; however,
                                for these galaxies, we do find a non-negligible light excess in the central region.
        
        Table~\ref{tab:dbfits} lists the parameters estimated from the MCMC analysis for the bulge profiles and the best-fit parameters of the exponential profile used to estimate the disk surface brightness in the central region. The uncertainties on the parameters $I_{\rm d0}$ and $h_R$ are derived from the covariance matrix obtained from the least-square fit. Sub-panels (a) of Figs.~\ref{fig:Complete_analysis_1}-\ref{fig:Complete_analysis_7} show the 30 measured profiles, corrected for inclination, in green the region where the bulge contribution is dominant, and in red the region where the disk contribution is dominant; the models are in blue and are the sum of Eq.~\eqref{eq:conv} and the surface brightness of the disk.
                
                To convert the fit parameters from angular (arcsec) to physical (kpc) radial units we use the relation:
                \begin{equation}
                \label{eq:arcsectokpc}
                R(\text{kpc})=4.84814\times 10^{-6}D(\text{kpc})R(\text{arcsec}) \; ,
                \end{equation}
                where $4.84814\times 10^{-6}$ is the conversion factor from arcsec to radians and $D(\text{kpc})$ is the galaxy distance reported in Table 1 of~\citet{DMSvi}. This relation is strictly valid in a non-expanding Euclidean geometry, but it can be applied to our DMS galaxies because they all are at redshift smaller than $0.043$.

                \begin{table*}
                        \caption{Parameters of the model of the surface brightness.}
                        \label{tab:dbfits}
                        \centering   
                        \begin{tabular}{lccccccc}
                                \hline\hline
                                UGC & $I_{\rm d0}$ $\left[\frac{{\rm L}_\odot}{{\rm pc}^2}\right]$ & $h_R$ [arcsec] & $h_R$ [kpc] & $I_\mathrm{e}$ $\left[\frac{{\rm L}_\odot}{{\rm pc}^2}\right]$ & $R_\mathrm{e}$ [arcsec] & $R_\mathrm{e}$ [kpc] & $n_\mathrm{s}$\\
                                (1) & (2) & (3) & (4) & (5) & (6) & (7) & (8) \\
                                \hline
                                448 &$603\pm44$&$12.29\pm0.47$& $3.89\pm0.19$ &$2500.00\pm0.07$&$1.08\pm0.05$& $0.34\pm0.02$ &$6.47\pm0.07$\\
                                463 &$1357\pm23$&$13.17\pm0.16$& $3.81\pm0.14$ &$83.00\pm0.07$&$4.50\pm0.07$& $1.30\pm0.05$ &$2.03\pm0.01$\\
                                1081 &$573\pm12$&$14.79\pm0.20$& $3.00\pm0.16$ &$4.69\pm0.07$&$2.02\pm0.07$& $0.41\pm0.02$ &$0.47\pm0.02$\\
                                1087 &$598\pm25$&$10.41\pm0.33$& $3.01\pm0.14$ &$0.44\pm0.02$&$4.72\pm0.07$& $1.36\pm0.05$ &$0.46\pm0.03$\\
                                1529 &$872\pm29$&$12.19\pm0.27$& $3.64\pm0.15$ &$4.70\pm0.07$&$3.01\pm0.07$& $0.90\pm0.04$ &$0.61\pm0.02$\\
                                1635 &$477\pm17$&$14.11\pm0.37$& $3.19\pm0.17$ &$3.16\pm0.07$&$2.97\pm0.07$& $0.67\pm0.03$ &$0.46\pm0.02$\\
                                1862 &$329.7\pm8.5$&$17.29\pm0.39$& $1.54\pm0.18$ &-&-&-&-\\
                                1908 &$1406\pm42$&$9.21\pm0.19$& $4.91\pm0.14$ &$35.00\pm0.07$&$1.09\pm0.06$& $0.58\pm0.03$ &$0.48\pm0.03$\\
                                3091 &$467\pm36$&$9.30\pm0.53$& $3.33\pm0.21$ &$1.06\pm0.06$&$3.97\pm0.07$& $1.42\pm0.05$ &$0.60\pm0.03$\\
                                3140 &$1426\pm68$&$11.23\pm0.31$& $3.38\pm0.15$ &$3749.96\pm0.07$&$0.82\pm0.04$& $0.25\pm0.01$ &$4.64\pm0.07$\\
                                3701 &$151\pm16$&$16.7\pm1.3$& $3.51\pm0.32$ &$0.24\pm0.01$&$7.02\pm0.07$& $1.47\pm0.07$ &$ 0.37\pm0.02$\\
                                3997 &$165\pm25$&$13.7\pm1.2$& $5.52\pm0.50$ &$0.82\pm0.05$&$7.00\pm0.07$& $2.82\pm0.08$ &$0.98\pm0.04$\\
                                4036 &$985\pm12$&$12.87\pm0.08$& $3.18\pm0.13$ &$1.24\pm0.06$&$3.70\pm0.07$& $0.91\pm0.04$ &$0.44\pm0.03$\\
                                4107 &$669\pm16$&$13.26\pm0.23$& $3.29\pm0.15$ &$2.50\pm0.06$&$2.99\pm0.07$& $0.74\pm0.04$ &$0.49\pm0.02$\\
                                4256 &$1327\pm29$&$11.78\pm0.14$& $4.27\pm0.13$ &$305.00\pm0.07$&$5.03\pm0.06$& $1.82\pm0.06$ &$3.55\pm0.02$\\
                                4368 &$726\pm18$&$9.49\pm0.07$& $2.59\pm0.10$ &$0.42\pm0.03$&$4.50\pm0.07$& $1.23\pm0.05$ &$0.34\pm0.04$\\
                                4380 &$473\pm36$&$9.78\pm0.45$& $4.98\pm0.25$ &$0.50\pm0.04$&$7.03\pm0.07$& $3.58\pm0.08$ &$0.66\pm0.04$\\
                                4458 &$254\pm22$&$28.0\pm1.3$& $9.27\pm0.52$ &$180.00\pm0.07$&$24.99\pm0.07$& $8.29\pm0.26$ &$3.26\pm0.01$\\
                                4555 &$799\pm13$&$10.99\pm0.10$& $3.29\pm0.12$ &$1.71\pm0.07$&$1.99\pm0.07$& $0.60\pm0.03$ &$0.37\pm0.03$\\
                                4622 &$520\pm27$&$8.59\pm0.26$& $7.42\pm0.24$ &$14.20\pm0.07$&$1.07\pm0.06$& $0.92\pm0.05$ &$0.55\pm0.03$\\
                                6903 &$137\pm11$&$34.5\pm2.4$& $5.23\pm0.52$ &$0.84\pm0.05$&$13.50\pm0.07$& $2.05\pm0.14$ &$0.79\pm0.04$\\
                                6918 &$3372\pm40$&$10.96\pm0.08$& $1.16\pm0.12$ &$3.79\pm0.07$&$4.02\pm0.07$& $0.42\pm0.05$ &$0.43\pm0.02$\\
                                7244 &$202\pm18$&$10.81\pm0.32$& $3.43\pm0.15$ &$0.61\pm0.04$&$6.51\pm0.07$& $2.06\pm0.07$ &$0.70\pm0.04$\\
                                7917 &$686.0\pm9.8$&$14.53\pm0.10$& $7.25\pm0.17$ &$12.00\pm0.07$&$7.01\pm0.07$& $3.50\pm0.09$ &$1.13\pm0.01$\\
                                8196 &$1085\pm54$&$9.63\pm0.25$& $5.59\pm0.18$ &$144.00\pm0.07$&$7.00\pm0.07$& $4.06\pm0.09$ &$2.57\pm0.02$\\
                                9177 &$362\pm23$&$11.06\pm0.44$& $7.10\pm0.31$ &$3.12\pm0.07$&$2.99\pm0.07$& $1.92\pm0.06$ &$0.60\pm0.02$\\
                                9837&$97\pm11$&$28.2\pm2.3$& $5.91\pm0.58$ &$4.98\pm0.07$&$30.00\pm0.07$& $6.28\pm0.33$ &$2.69\pm0.02$\\
                                9965 &$739\pm15$&$10.25\pm0.19$& $3.51\pm0.14$ &-&-&-&-\\
                                11318 &$1090\pm27$&$11.04\pm0.19$& $4.56\pm0.14$ &$90.00\pm0.07$&$9.99\pm0.07$& $4.13\pm0.11$ &$3.46\pm0.02$\\
                                12391 &$622\pm22$&$11.31\pm0.30$& $3.66\pm0.15$ &$2.04\pm0.06$&$5.95\pm0.07$& $1.93\pm0.06$ &$0.97\pm0.02$\\
                                \hline
                        \end{tabular}
                    \tablefoot{Column 1: UGC number; Cols. 2, 3: best-fit parameters of the exponential profile~\eqref{eq:Id}, which are used to model the disk surface brightness in the central regions of the galaxies; Cols. 5, 6, 8: MCMC parameters for the bulge surface brightness profile; Cols. 4, 7: disk-scale length, $h_R$, and bulge effective radius, $R_\mathrm{e}$, in physical units (kpc).}
                \end{table*}

                \subsection{Gas surface mass density}
                \label{sec:gas}
                
                The gas component of each galaxy is distributed in a disk-like structure that is thinner and more extended than the stellar disk. In addition, both the atomic and the molecular gas contribute to the gas component.
        
   The observed atomic profile is set to $\Sigma_{\rm atom}=1.4 \Sigma_{\rm HI}$~\citep{DMSvii}, where $\Sigma_{\rm HI}$ is the measured HI gas surface mass density estimated from 21-cm radio synthesis observations~\citep[see][Sect. 2.5]{Martinsson11}.     
   Similarly to the surface brightness of the disk, we linearly interpolate the data points of the surface mass density of the atomic gas.
   
                \citet{Martinsson11} measured $\Sigma_{\rm HI}$ only for 24 galaxies out of the original sample of 30 galaxies. For the remaining six (UGC 1081, \object{UGC 1529}, UGC 1862, UGC 1908, UGC 3091, \object{UGC 12391}),~\citet{Martinsson11} modelled the $\Sigma_{\rm HI}$ profiles with a Gaussian with mean and dispersion taken from the average of the other galaxies with measured $\Sigma_{\rm HI}$~\citep[see Sect. 3.1 of][for details]{DMSvii}, obtaining synthetic data. For these six galaxies, we clearly adopt their synthetic Gaussian profiles.
                
                The molecular gas surface mass density, $\Sigma_{\text{mol}}$, is indirectly derived from 24-$\mu$m \textit{Spitzer} observations, based on the CO line detection \citep[see][Sect. 3.2]{DMSvii}. Again, we linearly interpolate the measures of the surface mass density of the molecular gas.
                
In sub-panels (b) and (c) of Figs.~\ref{fig:Complete_analysis_1}-\ref{fig:Complete_analysis_7}, the red dots with error bars are the measures of the surface mass density of the atomic and molecular gas; the blue lines show the linearly interpolated profiles.

        \subsection{Three-dimensional mass density model}
        \label{sec:3D}
        
        We model the total baryonic mass density as the sum of the disk mass density, $\Upsilon_\mathrm{d} j_\mathrm{d}(R,z)$, the bulge 
        mass density, $\Upsilon_\mathrm{b} j_\mathrm{b}(r)$, with $r=(R^2+z^2)^{1/2}$, and the atomic and molecular gas mass densities, $\rho_{\text{atom}}(R,z)$ and $\rho_{\text{mol}}(R,z)$:
        \begin{equation}
        \label{eq:rhotot}
                \rho(R,z)=\Upsilon_\mathrm{d} j_\mathrm{d}(R,z)+\Upsilon_\mathrm{b} j_\mathrm{b}(r)+\rho_{\text{atom}}(R,z)+\rho_{\text{mol}}(R,z) \; ,
        \end{equation}
        where $\Upsilon_\mathrm{d}$ and $\Upsilon_\mathrm{b}$ are the mass-to-light ratios of the stellar populations of the disk and the bulge, respectively. We assume that the mass-to-light ratios are independent of $R$ and $z$. In addition, we set $\Upsilon_\mathrm{b}=\Upsilon_\mathrm{d}=\Upsilon$ because the bulge is, on average, an order of magnitude less luminous than the disk, as mentioned in Appendix \ref{sec:SB}.
        Equation~\eqref{eq:rhotot} is the total mass distribution of the galaxy in modified gravity models where the dark matter component is assumed to be absent.

        We model the three-dimensional luminosity density of the disk by multiplying the disk surface brightness profile, $I_\mathrm{d}(R)$, which 
        is the sum of the exponential model and the interpolated profile, by an exponentially decreasing density profile along the vertical axis $z$, 
    \begin{equation}
    \label{eq:jd}
    j_\mathrm{d}(R,z)=\frac{I_\mathrm{d}(R)}{2h_z}\exp\left(-\frac{|z|}{h_z}\right) \; ,
    \end{equation}   
    where the disk-scale height $h_z$ is a free parameter.
    The factor $1/(2h_z)$ provides the correct normalisation.
        In the main body of the paper, we compare our estimate of $h_z$ with the scale height $h_{z,\text{SR}}$ obtained from the relation derived in~\citet{DMSii} from a combined sample of 60 edge-on late-type galaxies, 
        \begin{equation}
        \label{eq:hzhR}
                \log_{10}\left(\frac{h_R}{h_{z,\text{SR}}}\right)=0.367\log_{10}\left(\frac{h_R}{\text{kpc}}\right)+0.708 \pm 0.095 \; ,
        \end{equation}
    where the term $\pm 0.095$ quantifies the $\sim$$25\%$ intrinsic scatter.
        We estimate the uncertainty on $h_{z,\text{SR}}$ as the sum in quadrature of the intrisic scatter of the relation~\eqref{eq:hzhR} and the uncertainty on $h_R$. The latter contribution is negligible with respect to the former, so the error on $h_{z,\text{SR}}$ nearly coincides with its intrinsic scatter.
        
        We model the three-dimensional luminosity density of the bulge with the Abel integral: 
        \begin{equation}
        \label{eq:jb}
        j_\mathrm{b}(r)=-\frac{1}{\pi}\int_{r}^{\infty}\frac{dI_\mathrm{b}}{dR}\frac{dR}{\sqrt{R^2-r^2}} \; ,
        \end{equation}
        where $I_\mathrm{b}(R)$ is the surface brightness profile modelled as described in Appendix~\ref{sec:SB}, $R$ is the radius projected on the sky and $r$ is the three-dimensional radius. The equation above assumes a spherically symmetric bulge. As mentioned in Appendix~\ref{sec:SB}, neglecting the triaxial structure of the bulge should introduce negligible systematic errors in the mass estimates because the galaxy luminosity is dominated by the disk~\citep{Angus15}.
        As anticipated, adopting an analytical model for the two-dimensional surface brightness of the bulge, rather than interpolating the data points as we do for the disk, facilitates its deprojection into three dimensions.
        
        We consider both the atomic and molecular gas distributions as razor-thin disks~\citep{DMSvii},
        \begin{equation}
        \label{eq:rhogas}
                \rho_{\rm atom, mol}(R,z)=\Sigma_{\rm atom, mol}(R)\delta(z) \; ,
        \end{equation}
        where $\Sigma(R)$ is the linearly interpolated mass surface density of the gas disk and $\delta(z)$ is the Dirac $\delta$ function.

                \section{Successive Over Relaxation Poisson solver}
                \label{sec:SOR}
                
                \subsection{Numerical solution of the Poisson equation}
                \label{sec:SOR_intro}
                
                For the theories of gravity we consider here, deriving the gravitational potential $\phi$ that originates from the mass
                density distribution $\rho$ requires solving the Poisson equation,
        \begin{equation}
                \label{eq:genericalPoisson}
                \nabla^2\phi(R,z)+ S(\rho;R,z)=0 \; ,
                \end{equation} 
                where the source term $S$ is a generic function of the density $\rho$. For axisymmetric disk galaxies, we limit the equation to cylindrical coordinates $R$ and $z$. We only consider static models and Eq.~\eqref{eq:genericalPoisson} is thus an elliptic partial differential equation independent of time.
                
                We solve the Poisson equation with a successive over relaxation (SOR) algorithm, an iterative procedure based on 
                the Jacobi and the Gauss-Seidel algorithms~\citep{Young54}. We find the solution on a rectangular grid of size $L_R\times L_z$, with $(N_R+1)\times (N_z+1)$ grid points and steps $\Delta_R=L_R/N_R$ and $\Delta_z=L_z/N_z$ in the two dimensions (Table~\ref{tab:num_details}).
        
                Given the solution of the gravitational potential $\phi^n_{i,k}$ at the $n$-th iteration on the grid point of indexes $(i,k)$, the solution at the following $(n+1)$-th iteration is:
                \begin{equation}
                \label{eq:SOR2Dcyl}
                \begin{aligned}
                        \phi^{n+1}_{i,k}={} &\phi^n_{i,k}(1-\omega_\text{SOR})\\
                        &+\frac{\omega_\text{SOR}}{2R_i(\Delta_R^2+\Delta_z^2)}\left[\phi^{n+1}_{i+1,k}\left(R_i+\frac{1}{2}\Delta_R\right)\Delta_z^2 \right.\\
                &+\phi^{n+1}_{i-1,k}\left(R_i-\frac{1}{2}\Delta_R\right)\Delta_z^2+(\phi^{n+1}_{i,k+1}+\phi^{n+1}_{i,k-1})R_i\Delta_R^2\\
                        &\left.+\text{ }R_iS_{i,k} \Delta_R^2\Delta_z^2\right] \; ,
                \end{aligned}
                \end{equation}
                where $S_{i,k}$ is the source term on the grid and $\omega_\text{SOR}$ is a parameter in the range of $(0,2)$ to guarantee the convergence. The value of $\omega_\text{SOR}$ which guarantees the fastest convergence is $\omega_\text{SOR}=2/(1+\pi/N)$ for a square grid and $\omega_\text{SOR}=2/(1+\pi/N_\text{min})$ for a rectangular grid, where $N_\text{min}$ is the smallest number between $N_R$ and $N_z$. In general, it is computationally convenient to choose the number of grid points $N$ such that $\omega_\text{SOR}$ is in the range of $(1,2)$; with this choice, the number of iterations necessary to reach convergence is linearly proportional to $N$, whereas for $\omega_\text{SOR}$ in the range of $(0,1)$ the number of iterations is proportional to $N^2$~\citep{Young54}.  
                
                We adopt $L_R=2\times12h_R$ for almost all galaxies and $L_z=2\times100h_{z,\text{SR}}$ for all galaxies, where $h_{z,\text{SR}}$ is derived from Eq.~\eqref{eq:hzhR}. With this choice, the grid domain is substantially larger than the galaxy size, and we can adopt the asymptotic behaviour of the gravitational potential to set the proper boundary conditions, as we describe in Appendix~\ref{sec:SOR_BC} below. For \object{UGC 4368} and \object{UGC 6918}, we adopt $L_R=2\times20h_R$ and $L_R=2\times18h_R$, respectively, because $L_R=2\times12h_R$ is smaller than the extension of the measured rotation curve. UGC 4458 has $h_R=9.27$~kpc and, with this large scale length, we adopt $L_R=2\times10h_R$, which is already sufficient to reach the asymptotic behaviour of the gravitational potential. 
                
                The radial resolution of the rotation curves and of the vertical velocity dispersion profiles measured for each galaxy in the DMS sample differs from galaxy to galaxy. For each galaxy, we thus choose $N_R$ and $N_z$ that yield both $\omega_\text{SOR}$ in the range of $(1,2)$ and the numerical resolution $\Delta_R=L_R/N_R$ and $\Delta_z=L_z/N_z$ comparable to the observed resolution.
                
                For most galaxies, the grid where we compute the mass distribution and the galactic potential is more extended than the measured surface brightness of the disk and the measured gas surface mass density. To estimate the mass distribution in these regions, we extrapolate the disk surface brightness with Eq.~\eqref{eq:Id} with the parameters listed in Table~\ref{tab:dbfits}; we instead set to zero the gas mass density because its contribution to the galaxy mass is negligible in these outer regions.
                
                \begin{table*}
                        \caption{Parameters of the computational grid for the Poisson solver.}
                        \label{tab:num_details} 
                        \centering
                        \begin{tabular}{lccccccc}
                                \hline\hline
                                UGC & $L_R$ [kpc]& $L_z$ [kpc] & $N_R$ & $N_z$ & $\Delta_R$ [kpc] & $\Delta_z$ [kpc] & $h_{z,\text{SR}}$ [kpc]\\
                                (1) & (2) & (3) & (4) & (5) & (6) & (7) & (8) \\
                                \hline
                                448 & $93.36$ & $92.00$ & $331$ & $326$ & $0.28$ & $0.28$ &$0.46\pm0.10$\\
                            463 & $91.44$ & $92.00$ & $355$ & $358$ & $0.26$ & $0.26$ &$0.46\pm0.10$\\
                            1081 & $72.00$ & $78.00$ & $399$ & $390$ & $0.18$ & $0.20$ &$0.39\pm0.09$\\
                            1087 & $72.24$ & $78.00$ & $281$ & $304$ & $0.26$ & $0.26$ &$0.39\pm0.09$\\
                            1529 & $87.36$ & $88.00$ & $329$ & $330$ & $0.27$ & $0.27$ &$0.44\pm0.10$\\
                            1635 & $76.56$ & $82.00$ & $367$ & $394$ & $0.21$ & $0.21$ &$0.41\pm0.09$\\
                            1862 & $36.96$ & $52.00$ & $341$ & $306$ & $0.11$ & $0.17$ &$0.26\pm0.06$\\
                            1908 & $117.84$ & $108.00$ & $301$ & $276$ & $0.39$ & $0.39$ &$0.54\pm0.12$\\
                            3091 & $79.92$ & $84.00$ & $287$ & $302$ & $0.28$ & $0.28$ &$0.42\pm0.09$\\
                            3140 & $94.08$ & $94.00$ & $303$ & $314$ & $0.27$ & $0.27$ &$0.42\pm0.10$\\
                            3701 & $84.24$ & $86.00$ & $387$ & $394$ & $0.22$ & $0.22$ &$0.43\pm0.10$\\
                            3997 & $132.48$ & $116.00$ & $369$ & $324$ & $0.36$ & $0.36$ &$0.58\pm0.13$\\
                            4036 & $76.32$ & $82.00$ & $347$ & $374$ & $0.22$ & $0.22$ &$0.41\pm0.09$\\
                            4107 & $78.96$ & $84.00$ & $357$ & $380$ & $0.22$ & $0.22$ &$0.42\pm0.09$\\
                            4256 & $102.48$ & $98.00$ & $317$ & $304$ & $0.32$ & $0.32$ &$0.49\pm0.11$\\
                            4368 & $103.60$ & $72.00$ & $395$ & $274$ & $0.26$ & $0.26$ &$0.36\pm0.08$\\
                            4380 & $119.52$ & $108.00$ & $321$ & $290$ & $0.37$ & $0.37$ &$0.54\pm0.12$\\
                            4458 & $185.40$ & $160.00$ & $475$ & $360$ & $0.39$ & $0.44$ &$0.80\pm0.18$\\
                            4555 & $78.96$ & $84.00$ & $295$ & $314$ & $0.27$ & $0.27$ &$0.42\pm0.09$\\
                            4622 & $178.08$ & $140.00$ & $401$ & $316$ & $0.44$ & $0.44$ &$0.70\pm0.15$\\
                            6903 & $125.52$ & $112.00$ & $371$ & $330$ & $0.30$ & $0.30$ &$0.56\pm0.13$\\
                            6918 &  $41.76$ & $42.00$ & $397$ & $398$ & $0.11$ & $0.11$ &$0.21\pm0.05$\\
                            7244 & $82.32$ & $86.00$ & $293$ & $306$ & $0.28$ & $0.28$ &$0.43\pm0.09$\\
                            7917 & $174.00$ & $138.00$ & $395$ & $314$ & $0.44$ & $0.44$ &$0.69\pm0.15$\\
                            8196 & $134.16$ & $116.00$ & $377$ & $326$ & $0.36$ & $0.36$ &$0.58\pm0.13$\\
                            9177 & $170.40$ & $136.00$ & $383$ & $306$ & $0.44$ & $0.44$ &$0.68\pm0.15$\\
                            9837 & $141.84$ & $120.00$ & $365$ & $310$ & $0.33$ & $0.33$ &$0.60\pm0.14$\\
                            9965 & $84.24$ & $86.00$ & $237$ & $242$ & $0.36$ & $0.36$ &$0.43\pm0.10$\\
                            11318 & $109.44$ & $102.00$& $297$ & $278$ & $0.37$ & $0.37$ &$0.51\pm0.11$\\
                            12391 & $87.84$ & $90.00$ & $349$ & $356$ & $0.25$ & $0.25$ &$0.45\pm0.10$\\
                                \hline
                        \end{tabular}
                    \tablefoot{Column 1: UGC number; Col. 2: size of the computational grid in the $R$-direction; Col. 3: size of the computational grid in the $z$-direction; Col. 4: grid points in the $R$-direction; Col. 5: grid points in the $z$-direction; Col. 6: grid step in the $R$-direction; Col. 7: grid step in the $z$-direction; Col. 8: disk-scale height derived with Eq.~\eqref{eq:hzhR}.}
                \end{table*}
                
            We centre the computational domain on the origin $R=z=0$. The coordinate $R$ appears at the denominator in Eq.~\eqref{eq:SOR2Dcyl}. Therefore, to avoid divergences, we choose the grid so that the $R=0$ axis is not a grid strand, unlike the $z=0$ axis. We can thus compute both the rotation curve in the plane $z=0$ and the vertical velocity dispersion at $z=0$ for any $R\neq 0$.
                
                We set the initial values of the potential to $\phi^0_{i,k}=0$ over the entire domain except the boundaries (see Appendix~\ref{sec:SOR_BC})  and stop the iteration when: 
                \begin{equation}
                \label{eq:finalerr}
                        \varepsilon^{n+1}=\frac{1}{(N_R-1)(N_z-1)}\sum_{i,k}\vert\phi^{n+1}_{i,k}-\phi^n_{i,k}\vert <10^{-9} \; .
                \end{equation}

We test our algorithm with mass density distributions where the Poisson equation in Newtonian gravity can be solved analytically: the Miyamoto-Nagai disk (see Eqs.~\eqref{eq:rhoMN} and~\eqref{eq:phiMN}), Satoh disk, logarithmic potential, Plummer sphere, isochrone potential, Hernquist sphere, and Navarro-Frenk-White potential~\citep{B&T08}. We compare the numerical and the analytical potentials as a function of $R$ and $z$.

Within the half-scale length from the centre, the numerical solution is within 1\% of the analytic solution for the Miyamoto-Nagai and the Satoh disks, $0.05$\% for the logarithmic potential, $0.25$\% for the Plummer sphere, $0.13$\% for the isochrone potential, 4\% for the Hernquist sphere, and 2\% for the Navarro-Frenk-White potential. The agreement improves outwards: beyond two scale lengths, it is smaller than $0.5$\% for the Miyamoto-Nagai and Satoh disks, Hernquist and Navarro-Frenk-White spheres, $0.1$\% for the  Plummer sphere, $0.05$\% for the isochrone potential, and $0.015$\% for the logarithmic potential.

                \subsection{The source term $S(R,z)$}
                \label{sec:Source}
                In this work, we consider two gravity theories: MOND and RG.
                
                In the QUMOND formulation of MOND~\citep{Milgrom10}, we have:
                \begin{equation}
                \label{eq:SMOND}
                S_\text{MOND}(R,z)=-\nabla\cdot\left[\nu\left(\frac{\vert\nabla\phi_\mathrm{N}\vert}{a_0}\right)\nabla\phi_\mathrm{N}\right] \; ,
                \end{equation}
        where $\phi_\mathrm{N}$ is the Newtonian gravitational potential due to the mass density distribution of the baryonic matter, and $\nu(y)$  is the MOND interpolating function, with $y=\vert \nabla\phi_\mathrm{N}\vert /a_0$. Here, we adopt the simple $\nu$-function, given by Eq.~\eqref{eq:simplenu}~\citep[see][Eq. (50) with $n=1$]{Famaey12}.
  
                In cylindrical coordinates, Eq.~\eqref{eq:SMOND} becomes:
                \begin{equation}
                \label{eq:SMONDtransformed2}
                \begin{aligned}
                        S_\text{MOND}(R,z) = {} & -\left(\frac{\nu}{R}\frac{\partial\phi_\mathrm{N}}{\partial R}+\frac{\partial\nu}{\partial R}\frac{\partial\phi_\mathrm{N}}{\partial R}+\nu\frac{\partial^2\phi_\mathrm{N}}{\partial R^2} \right.\\
                &\left. +\frac{\partial\nu}{\partial z}\frac{\partial\phi_\mathrm{N}}{\partial z}+\nu\frac{\partial^2\phi_\mathrm{N}}{\partial z^2}\right)\; .
                \end{aligned}
                \end{equation}

                Deriving the MOND gravitational potential clearly requires to solve the Poisson equation twice: first, to estimate the Newtonian potential $\phi_\mathrm{N}$, where, in Eq.~\eqref{eq:SOR2Dcyl}, we use the standard source term:     
                        \begin{equation}
                \label{eq:SN}
                S_\mathrm{N}(R,z)=-4\pi G\rho(R,z) \; ,
                \end{equation}
                and, subsequently, to compute $\phi_{\rm MOND}$ with the source term $S_\text{MOND}(R,z)$ of Eq.~\eqref{eq:SMONDtransformed2}.
                
                To derive the source term in the RG case, we need to recast the RG Poisson equation~\eqref{eq:PoissonRG},
                \begin{equation}
                        \nabla\cdot[\epsilon(\rho)\nabla\phi_\text{RG}]=4\pi G \rho \;,
        \end{equation}
        as: 
        \begin{equation}
                \label{eq:PoissonRG1stmember}
                \nabla\epsilon(\rho)\cdot\nabla\phi_\text{RG}+\epsilon(\rho)\nabla^2\phi_\text{RG} = 4\pi G\rho \; ,
                \end{equation}
                so that the source term is: 
                \begin{equation}
                \label{eq:PoissonRGtranformed}
                        S_{\rm RG}(R,z)=-\frac{4\pi G \rho(R,z)-\nabla\epsilon(\rho)\cdot\nabla\phi_\text{RG}(R,z)}{\epsilon(\rho)} \; .
                \end{equation}
                
                Here, the source term contains the unknown $\phi_\text{RG}$. At each iteration step, in the source term, we insert the potential $\phi_\text{RG}$ estimated at the previous step.
                
            The form of Eq.~\eqref{eq:PoissonRGtranformed} requires some additional care in the numerical algorithm: the source term increases when the vacuum permittivity $\epsilon_0$  decreases, and the Poisson solver does not necessarily converge with the optimal value $\omega_\text{SOR}=2/(1+\pi/N)\sim1.97-1.98$, for our typical $N_R$ and $N_z$. To make the Poisson solver converge for $\epsilon_0$ in the flat prior range of $[0.10-1]$,
            we need to set $\omega_\text{SOR}$ to values smaller than $\sim$$1.97-1.98$. For example $\omega_\text{SOR}=2/(1+\pi/25)$ guarantees convergence for every $\epsilon_0$ in our flat prior range. Yet, setting $\omega_\text{SOR}$ to this unique value would increase the total computational time by at least a factor of 4. We thus vary $\omega_\text{SOR}$ from $2/(1+\pi/25)$, for $\epsilon_0$ close to $0.10$, to $2/(1+\pi/200)$, for $\epsilon_0$ close to $1$, according to the explored value of $\epsilon_0$ within the flat prior range.
                
                In MOND, this issue is not present and we can use the same value of $\omega_\text{SOR}$ for all the parameter combinations, namely $2/(1+\pi/N)$ for a square grid or $2/(1+\pi/N_\text{min})$ for a rectangular grid.
                
                Numerically, we compute all the derivatives present in all the above equations with the leap-frog method. The first and second derivatives of a function $f$ at a point $x$ are 
                \begin{equation}
                \label{eq:LeapFrogfirstder}
                \frac{\partial{f}}{\partial{x}}=\frac{f(x+\Delta x)-f(x-\Delta x)}{2\Delta x} 
                \end{equation}
                and 
                \begin{equation}
                \label{eq:LeapFrogsecder}
                \frac{\partial^2f}{\partial{x^2}}=\frac{f(x+\Delta x)+f(x-\Delta x)-2f(x)}{\Delta x^2} \; ,
                \end{equation}
                where $\Delta x$ is the grid step. With the grid steps we adopt, the leap-frog method guarantees relative errors smaller than $5$\% on the estimated derivatives.

                \subsection{Boundary conditions}
                \label{sec:SOR_BC}
                Elliptic equations, like our Poisson equations, require boundary conditions to be set. To facilitate the choice of the boundary conditions we put the disk galaxy at the centre of the grid domain and choose the size of the domain substantially larger than the galaxy size. We could use the axial symmetry of the problem and only consider a fourth of that domain, with two boundary conditions along the $R$ and $z$ axes. This choice would clearly reduce the computation time by a factor of four, but it would pose the non-trivial problem of setting the boundary conditions on the two axes cutting through the galaxy centre. We thus prefer to compute the potential in all the four quadrants and to set the boundary conditions far from the galaxy centre.
                
                Since the bulge and gas components are not dynamically dominant in the DMS galaxies~\citep{DMSi}, we use only the disk component to set the boundary conditions. In the external regions of the domain we model this component with a double exponential disk (Eq.~\ref{eq:jd}), since we use the exponential profile~\eqref{eq:Id} to extrapolate the disk surface brightness in these regions. In Newtonian gravity, this mass distribution generates a gravitational potential that does not have a close analytic expression, but it is well approximated by the sum of the gravitational potentials of three Miyamoto-Nagai disks \citep{Smith15}. The Miyamoto-Nagai disk has mass density distribution~\citep{M&N75}:
                \begin{equation}
                \label{eq:rhoMN}
                        \rho_{\rm MN}(R,z)= \left(\frac{b^2M_{\text{MN}}}{4\pi}\right)\frac{aR^2+(a+3\sqrt{z^2+b^2})(a+\sqrt{z^2+b^2})^2}{[R^2+(a+\sqrt{z^2+b^2})^2]^{5/2}(z^2+b^2)^{3/2}} \; ,
                \end{equation}
            and generates the Newtonian gravitational potential~\citep[e.g.][]{B&T08},
                \begin{equation}
                \label{eq:phiMN}
                \phi_{\text{MN}}(R,z)=-\frac{GM_{\text{MN}}}{\sqrt{R^2+(a+\sqrt{z^2+b^2})^2}} \; ,
                \end{equation}
                where $M_{\text{MN}}$ is the disk mass and $a$ and $b$ are its scale length and scale height.
                
                The gravitational potential of a double exponential disk can thus be approximated by the equation: 
                \begin{equation}
                \label{eq:phiexp}
                \begin{split}
                \phi_{\text{exp}}(R,z)=&-\frac{GM_1}{\sqrt{R^2+(a_1+\sqrt{z^2+b^2})^2}}\\
                &-\frac{GM_2}{\sqrt{R^2+(a_2+\sqrt{z^2+b^2})^2}}\\
                &-\frac{GM_3}{\sqrt{R^2+(a_3+\sqrt{z^2+b^2})^2}} \; ,\\
                \end{split}
                \end{equation}
                where $M_1$, $M_2$ and $M_3$ are the three Miyamoto-Nagai disk masses, $a_1$, $a_2$, $a_3$ are their scale lengths, and $b$ is the scale height which is equal in all the three disks. The three disks are only mathematical artefacts to generate the physical gravitational potential, but they do not correspond to three physical disks. Indeed, the mass $M_2$ of the second disk is always negative.
                
                The parameters of the Miyamoto-Nagai disks are related to the parameters of the double exponential disk (Eq.~\ref{eq:jd}) 
                and its total mass $M_\mathrm{d}$ according to the equations~\citep{Smith15}
                \begin{equation}
                \label{eq:bhRratio}
                \frac{b}{h_R} = -0.269\left(\frac{h_z}{h_R}\right)^3+1.080\left(\frac{h_z}{h_R}\right)^2+1.092\frac{h_z}{h_R} \; ,
                \end{equation}
                \begin{equation}
                \label{eq:M1Mdratio}
                \begin{split}
                \frac{M_1}{M_\mathrm{d}}=&-0.0090\left(\frac{b}{h_R}\right)^4+0.0640\left(\frac{b}{h_R}\right)^3\\
                &-0.1653\left(\frac{b}{h_R}\right)^2+0.1164\frac{b}{h_R}+1.9487 \; ,\\
                \end{split}
                \end{equation}
                \begin{equation}
                \label{eq:M2Mdratio}
                \begin{split}
                \frac{M_2}{M_\mathrm{d}}=&0.0173\left(\frac{b}{h_R}\right)^4-0.0903\left(\frac{b}{h_R}\right)^3\\
                &+0.0877\left(\frac{b}{h_R}\right)^2+0.2029\frac{b}{h_R}-1.3077 \; ,\\
                \end{split}
                \end{equation}
                \begin{equation}
                \label{eq:M3Mdratio}
                \begin{split}
                \frac{M_3}{M_\mathrm{d}}=&-0.0051\left(\frac{b}{h_R}\right)^4+0.0287\left(\frac{b}{h_R}\right)^3\\
                &-0.0361\left(\frac{b}{h_R}\right)^2-0.0544\frac{b}{h_R}+0.2242 \; ,\\
                \end{split}
                \end{equation}
                \begin{equation}
                \label{eq:a1Mdratio}
                \begin{split}
                \frac{a_1}{h_R}=&-0.0358\left(\frac{b}{h_R}\right)^4+0.2610\left(\frac{b}{h_R}\right)^3\\
                &-0.6987\left(\frac{b}{h_R}\right)^2-0.1193\frac{b}{h_R}+2.0074 \; ,\\
                \end{split}
                \end{equation}
                \begin{equation}
                \label{eq:a2Mdratio}
                \begin{split}
                \frac{a_2}{h_R}=&-0.0830\left(\frac{b}{h_R}\right)^4+0.4992\left(\frac{b}{h_R}\right)^3\\
                &-0.7967\left(\frac{b}{h_R}\right)^2-1.2966\frac{b}{h_R}+4.4441 \; ,\\
                \end{split}
                \end{equation}
                \begin{equation}
                \label{eq:a3Mdratio}
                \begin{split}
                \frac{a_3}{h_R}=&-0.0247\left(\frac{b}{h_R}\right)^4+0.1718\left(\frac{b}{h_R}\right)^3\\
                &-0.4124\left(\frac{b}{h_R}\right)^2-0.5944\frac{b}{h_R}+0.7333 \; .\\
                \end{split}
                \end{equation}
                
        In Newtonian gravity, we impose that the gravitational potential equals Eq.~\eqref{eq:phiexp} on the borders of the rectangular domain. 
        In MOND, when the acceleration drops below the critical acceleration, $a_0=3600$~kpc$^{-1}$~(km~s$^{-1}$)$^2$, the gravitational field has the asymptotic behaviour~\citep{Milgrom83}: 
                \begin{equation}
                \label{eq:gasinptoticMOND}
                g=\sqrt{a_0|g_\mathrm{N}|} \; ,
                \end{equation}
        which, in cylindrical components, becomes:       
                \begin{equation}
                \label{eq:gRasinptoticMOND}
                \frac{\partial\phi}{\partial{R}}=\sqrt{a_0\left\vert\frac{\partial\phi_\mathrm{N}}{\partial{R}}\right\vert} \; ,
                \end{equation}
                        and
                \begin{equation}
                \label{eq:gzasinptoticMOND}
                \frac{\partial\phi}{\partial{z}}=\sqrt{a_0\left\vert\frac{\partial\phi_\mathrm{N}}{\partial{z}}\right\vert} \; .
                \end{equation}
                To solve the MOND Poisson equation, we thus impose the Neumann boundary conditions~\eqref{eq:gRasinptoticMOND} and~\eqref{eq:gzasinptoticMOND}, with $\partial\phi_\mathrm{N}/\partial R$ and $\partial\phi_\mathrm{N}/\partial z$ derived analytically from Eq.~\eqref{eq:phiexp}.
                
                As illustrated in~\citet{M&D16}, at sufficiently large distances from the galaxy centre, RG recovers the MOND asymptotic behaviour of the radial gravitational field. We thus impose the Neumann condition~\eqref{eq:gRasinptoticMOND} on the four borders of the domain for the radial component of the gravitational field. For the vertical component, we resort to the fact that the field lines are refracted and, at large distances from the source, they are almost parallel to the disk plane. We thus set:
                \begin{equation}
                        \label{eq:RefractionLines}
                        \frac{\partial\phi}{\partial{z}}=0
                \end{equation}
                on the domain borders.
                
                In the analysis of the DMS galaxies, both the disk-scale height $b$ and the disk mass $M_\mathrm{d}$, which appear in Eqs.~\eqref{eq:phiexp}-\eqref{eq:a3Mdratio}, are initially unknown. These parameters depend on $h_z$ and $\Upsilon$, which are two free parameters of the fit. Therefore, we update the boundary conditions derived from Eqs.~\eqref{eq:gRasinptoticMOND} and~\eqref{eq:gzasinptoticMOND} at each step of the MCMC chain.

                \section{Convergence tests of the MCMC chains}
                \label{sec:convergence}
                
                \subsection{The variance ratio method}
                \label{sec:vrm}

                The variance ratio method of~\citet{Gel&Rub92} is a convergence diagnostic test for monitoring the convergence of MCMC chains: in other words, it estimates how close to convergence a chain is and if the convergence can be improved with additional steps of the chain~\citep{Brooks&Roberts98}.
                
                The test compares the output of $m\geq2$ independent chains that have different starting points. Each chain has $n_0+n$ elements, where the first $n_0$ are the discarded burn-in chain. Each chain $i$, at each step $t$, returns an estimate of a given parameter of interest $\theta_i({\mathbf x}^t)\equiv \theta_i^t$, where ${\mathbf x}$ are
                the variables of the problem updated at each chain step $t$. For each chain, we compute the mean of the parameter estimates,
                \begin{equation}
                \label{eq:mean_chain_i}
                \bar{\theta}_i=\frac{1}{n}\sum_{t=n_0+1}^{n_0+n}\theta^t_i \; ,
                \end{equation}
                the mean of the means of the $m$ chains, 
                \begin{equation}
                \label{eq:mean_means_chains_i}
                \bar{\theta}=\frac{1}{m}\sum_{i=1}^{m}\bar{\theta}_i \; ,
                \end{equation}
                and their variance,
                \begin{equation}
                \label{eq:var_means}
                \frac{B}{n}=\frac{1}{m-1}\sum_{i=1}^{m}(\bar{\theta}_i-\bar{\theta})^2 \; .
                \end{equation}
                
                In addition, we compute the variance of $\theta_i^t$ within each chain,
                \begin{equation}
                \label{eq:within_seq_var}
                s^2_i=\frac{1}{n-1}\sum_{t=n_0+1}^{n_0+n}(\theta^t_i-\bar{\theta}_i)^2 \; ,
                \end{equation}
                and their mean,
                \begin{equation}
                \label{eq:mean_vars}
                W=\frac{1}{m}\sum_{i=1}^{m}s^2_i \; .
                \end{equation}
                
                \citet{Gel&Rub92} assume that the mean of the posterior distribution of the
                parameter $\theta$ is $\hat{\mu}=\bar{\theta}$ and that its variance is:
                \begin{equation}
                \label{eq:sigmahat2}
                \hat{\sigma}^2=\frac{n-1}{n}W+\frac{B}{n} \; .
                \end{equation}
                \citet{Gel&Rub92} model the variability of $\hat{\mu}$ and $\hat{\sigma}^2$ due to sampling with an approximate Student's $t$ distribution 
                with mean $\hat{\mu}=\bar{\theta}$, variance
                \begin{equation}
                        \hat{V}=\hat{\sigma}^2+{1\over m}{B\over n}\; ,
               \end{equation}
               and number of degrees of freedom $\nu =2\hat{V}^2/\hat{\text{var}}(\hat{V})$, where
                \begin{equation}
                \label{eq:varhatVhat}
                \begin{aligned}
                \hat{\text{var}}(\hat{V})={} & \left(\frac{n-1}{n}\right)^2\frac{1}{m}\hat{\text{var}}(s^2_i)+\left(\frac{m+1}{mn}\right)^2\frac{2}{m-1}B^2\\
                &+2\frac{(m+1)(n-1)}{mn^2}\frac{n}{m}\left[\hat{\text{cov}}(s^2_i,\bar{\theta}^2_i)-2\bar{\theta}\hat{\text{cov}}(s^2_i,\bar{\theta}_i)\right] \; ,
                \end{aligned}
                \end{equation}
                and where the variance $\hat{\text{var}}(s^2_i)$ and the covariances $\hat{\text{cov}}(s^2_i,\bar{\theta}^2_i)$ and $\hat{\text{cov}}(s^2_i,\bar{\theta}_i)$ are estimated with the $m$ values $\bar{\theta}_i$ and $s_i^2$.
                
        To monitor the convergence of the chains,~\citet{Gel&Rub92} compute the \textit{potential scale reduction factor} 
                \begin{equation}
                \label{eq:PSRF}
                        \hat{R}=\frac{\hat{V}}{W}\frac{\nu}{\nu-2}\; .
                \end{equation}
                If $\hat{R}\to 1$ as $n\to \infty$, the estimated variance $\hat{V}$ of the expected posterior distribution of $\theta$ is increasingly closer to
                the variance $W$ estimated from the chains for an increasingly large number of steps. In other words, values of $\hat{R}\gg 1$ suggest that 
                the estimate of the target distribution can be improved with additional steps, whereas values of $\hat{R}\sim1$ suggest that the chains are 
                close to the target distribution.

                \subsection{The MCMC convergence for the universal combination of the RG parameters}
                \label{sec:convergence_RG}
        
        In our MCMC analysis we adopt a burn-in chain with $n_0=1000$ and a chain with $n=19000$ iterations. To assess the convergence of the chains, in Sect.~\ref{sec:Fit_all} we adopt the variance ratio method.
        
        We run the test for the first $n=13000$ iterations. If the test is positive for this $n$, we are confident that running the chains for $n=19000$ iterations will provide a posterior distribution close to the target distribution. We consider $m=3$ chains.
        
        The values we obtain for the potential scale reduction factor, $\hat{R}$, for each RG parameter, $\epsilon_0$, $Q$ and $\rho_\mathrm{c}$, are $1.01$, $1.04$ and $1.004$, respectively. According to the interpretation of \citet{Gel&Rub92}, these values suggest that our distributions estimated with $n=13000$ iterations already are reasonably close to the target distributions. The results we show in the text are for $n=19000$ iterations
        and we are thus confident that our values for $\epsilon_0$, $Q$ and $\rho_\mathrm{c}$ are 
        robust estimates of the target values.

\section{Figures of the individual DMS galaxies}
\label{sec:Figures}

\begin{figure*}
        \centering
        \includegraphics[width=14cm]{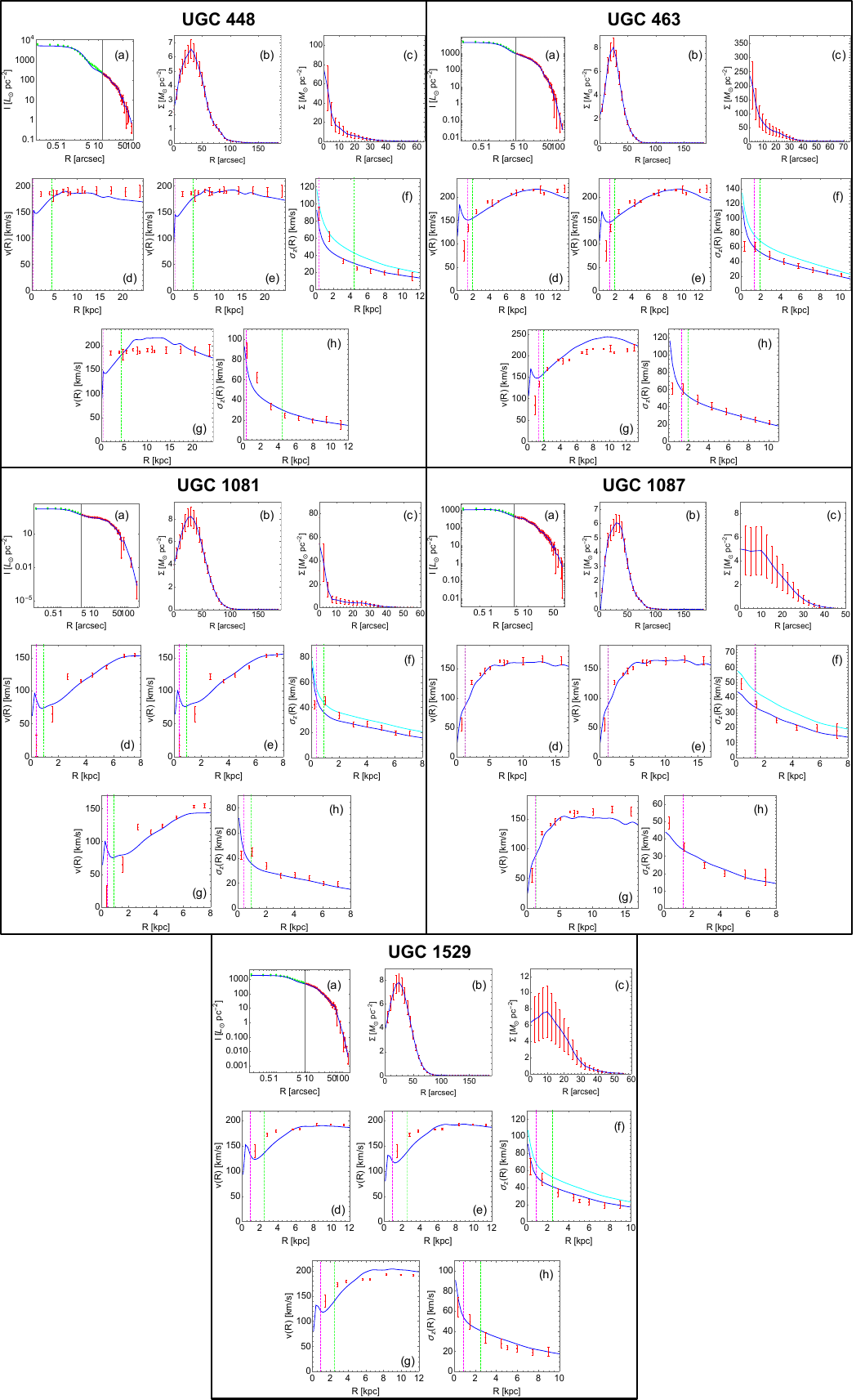}
        \caption{Surface brightness, surface mass densities of the atomic and molecular gas profiles and kinematic profiles of the DMS galaxies and their modelling according to our different analyses (see text of Appendix~\ref{sec:Figures}).}
        \label{fig:Complete_analysis_1}
\end{figure*}

\begin{figure*}
        \centering
        \includegraphics[width=14cm]{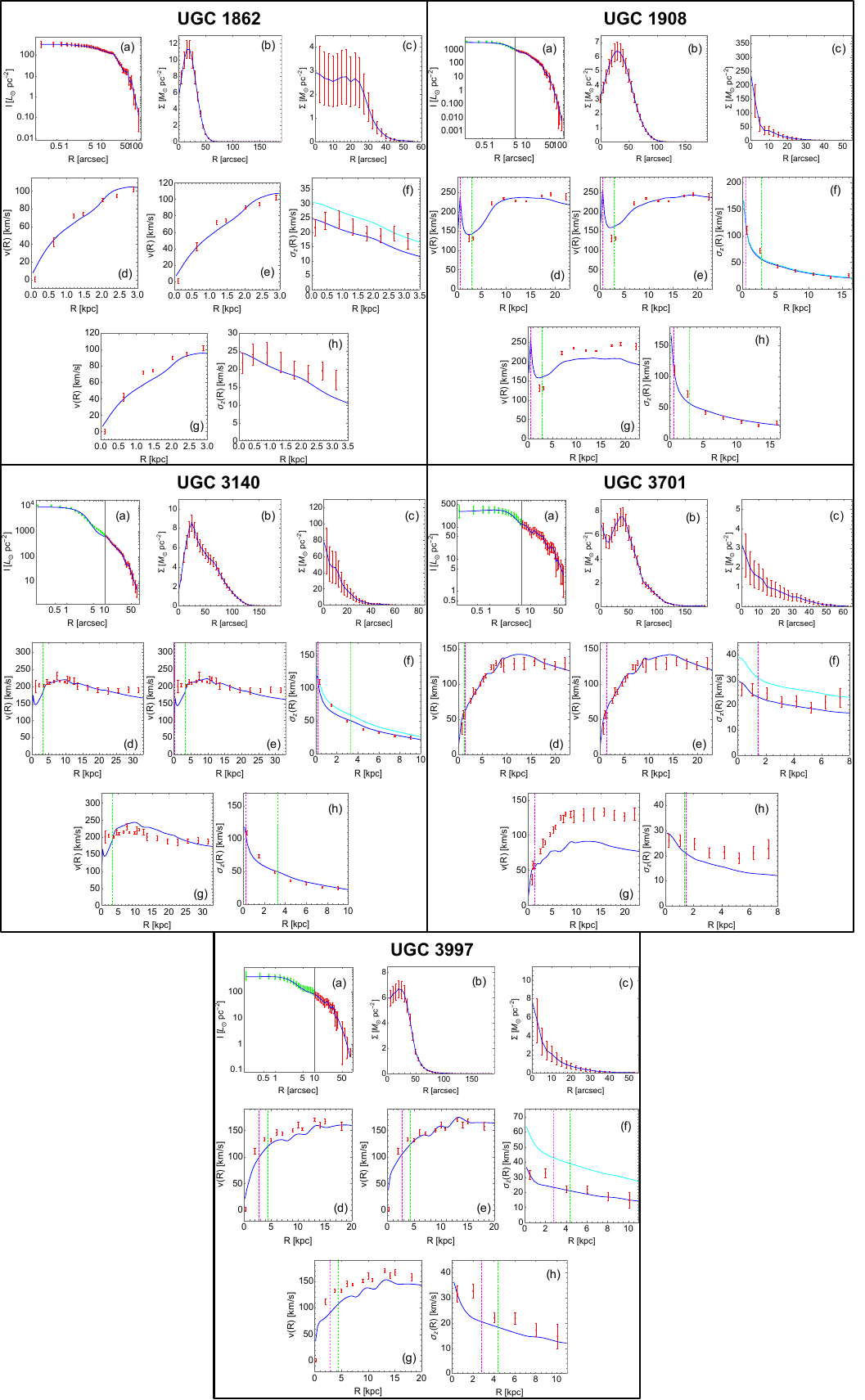}
        \caption{Same as in Fig.~\ref{fig:Complete_analysis_1}.}
        \label{fig:Complete_analysis_2}
\end{figure*}

\begin{figure*}
        \centering
        \includegraphics[width=14cm]{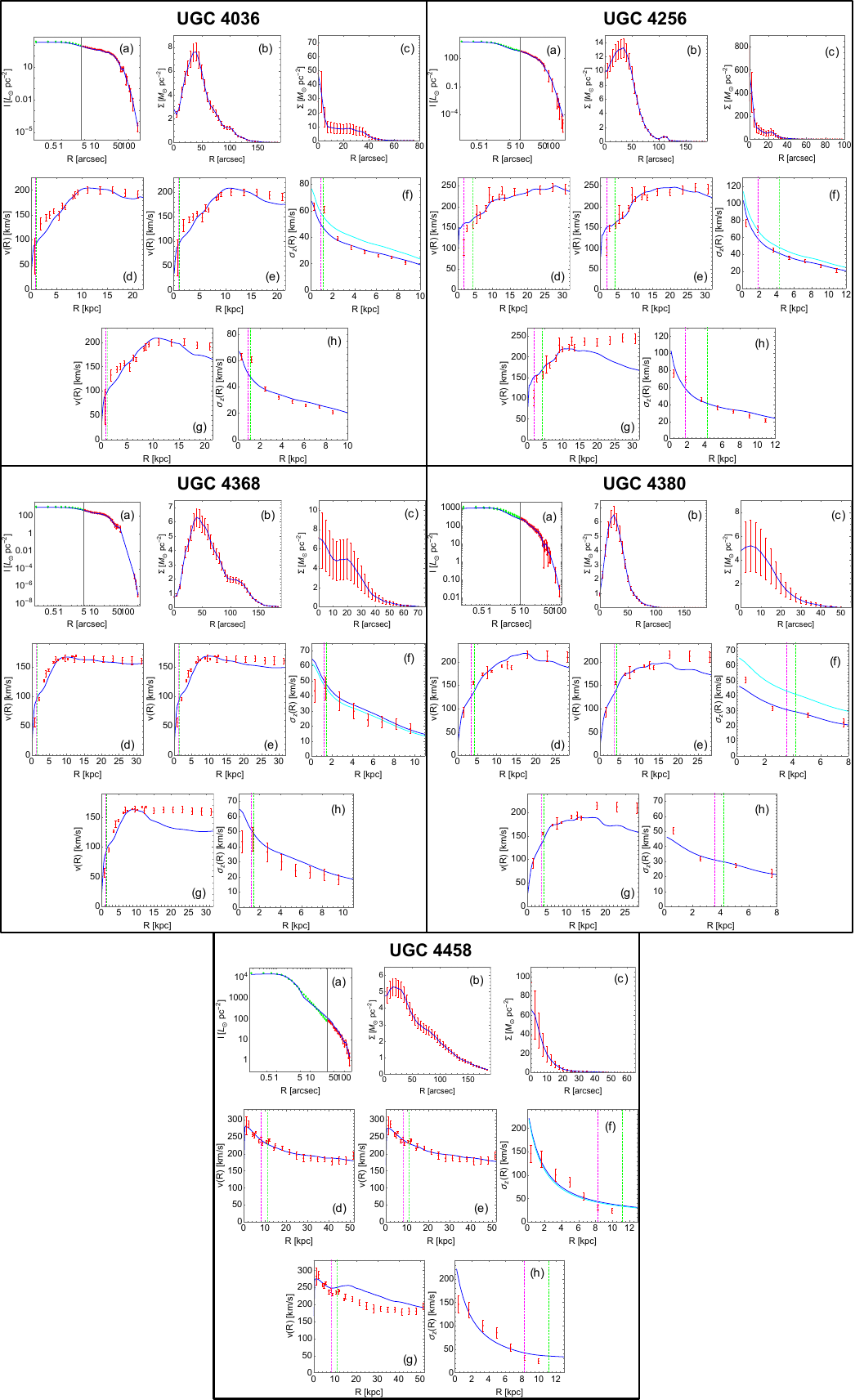}
        \caption{Same as in Fig.~\ref{fig:Complete_analysis_1}.}
        \label{fig:Complete_analysis_3}
\end{figure*}

\begin{figure*}
        \centering
        \includegraphics[width=14cm]{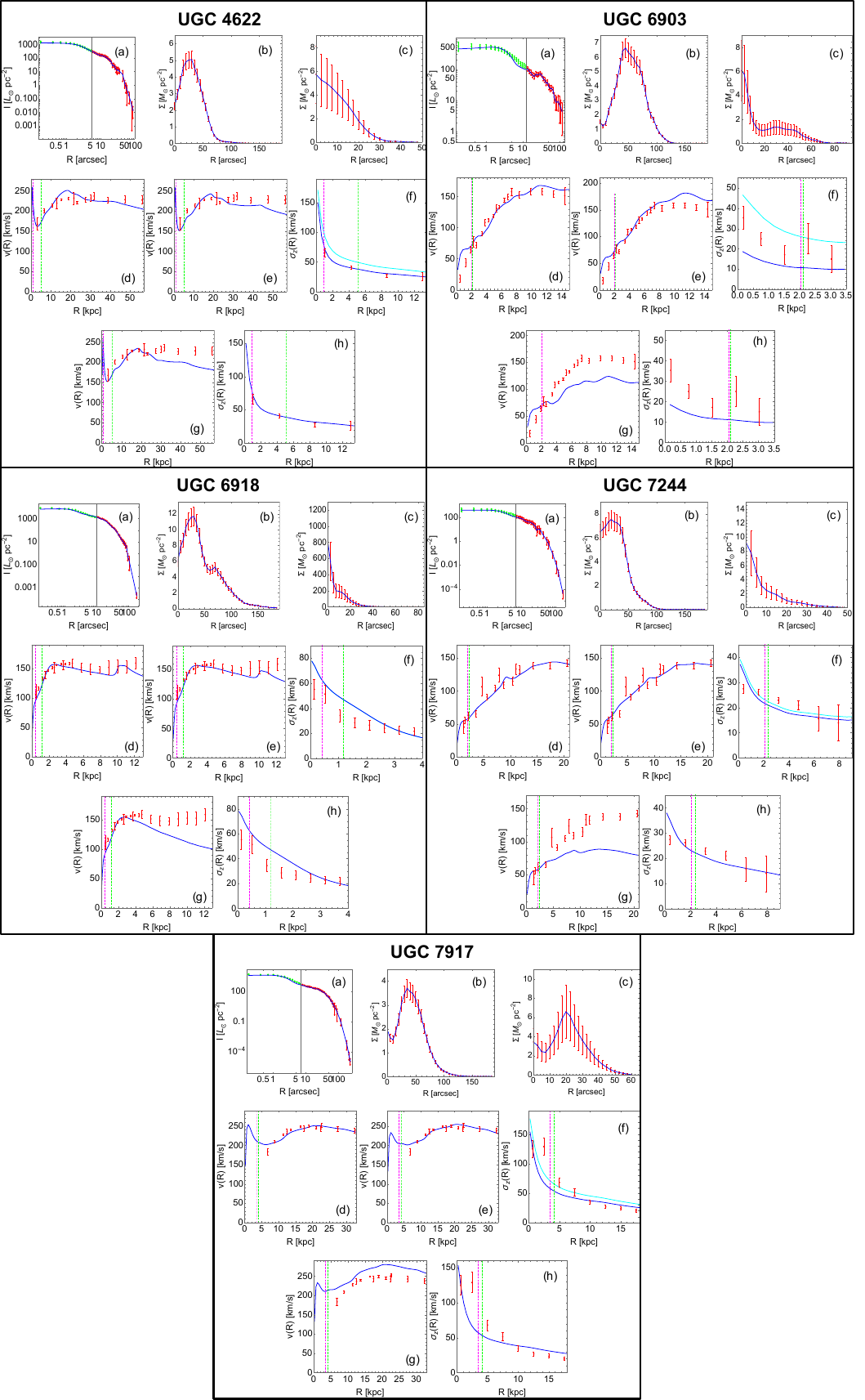}
        \caption{Same as in Fig.~\ref{fig:Complete_analysis_1}.}
        \label{fig:Complete_analysis_4}
\end{figure*}

\begin{figure*}
        \centering
        \includegraphics[width=14cm]{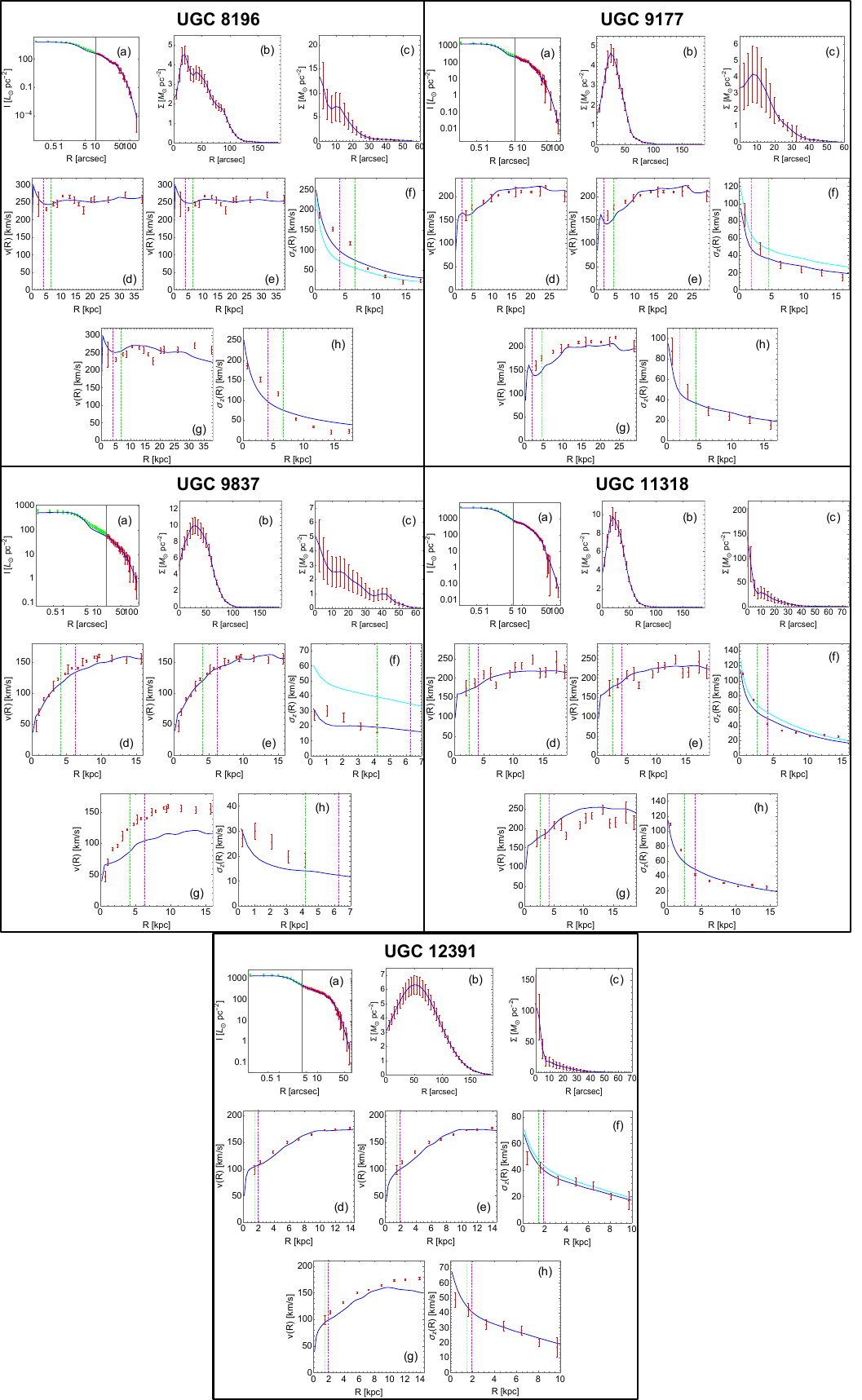}
        \caption{Same as in Fig.~\ref{fig:Complete_analysis_1}.}
        \label{fig:Complete_analysis_5}
\end{figure*}

\begin{figure*}
        \centering
        \includegraphics[width=14cm]{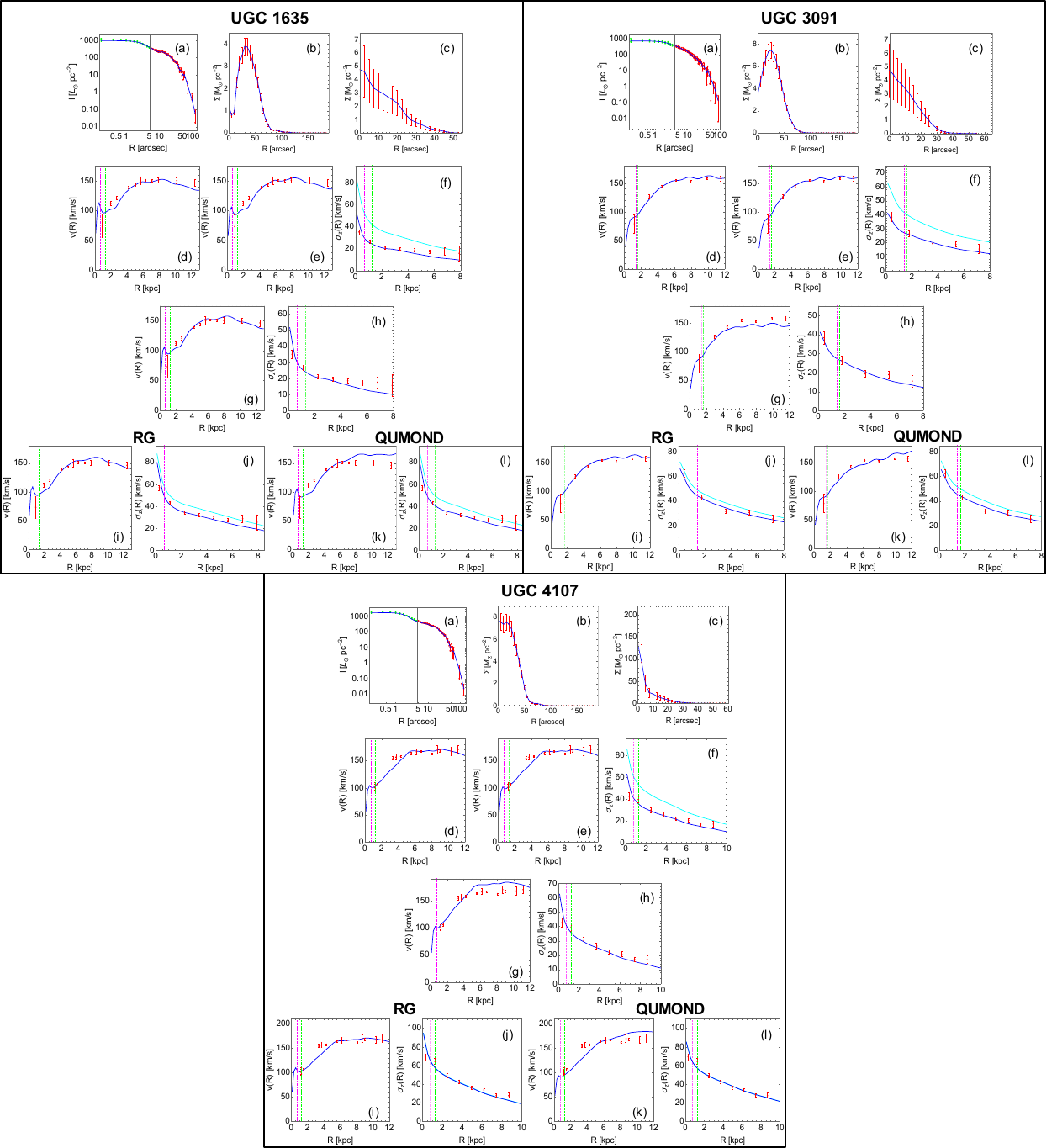}
        \caption{Same as in Fig.~\ref{fig:Complete_analysis_1} with the additional analyses with the vertical velocity dispersion profiles artificially
        increased by the factor $1.63$ in RG and QUMOND.}
        \label{fig:Complete_analysis_6}
\end{figure*}

\begin{figure*}
        \centering
        \includegraphics[width=14cm]{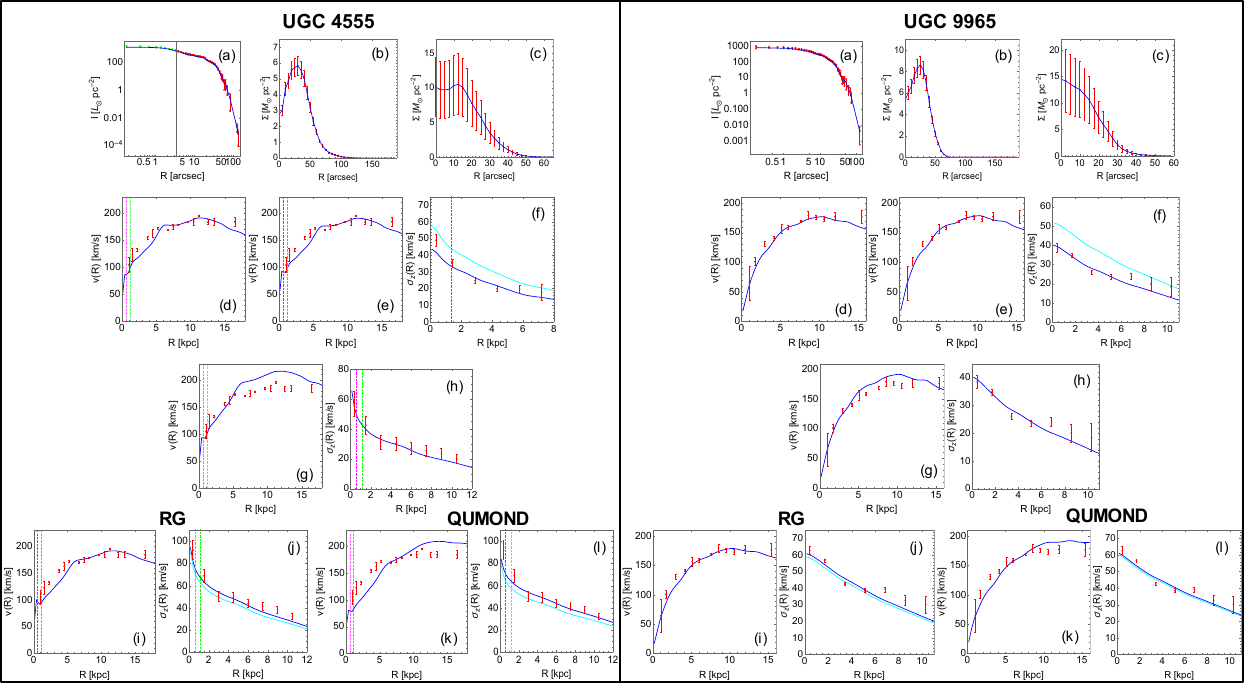}
        \caption{Same as in Fig.~\ref{fig:Complete_analysis_6}.}
        \label{fig:Complete_analysis_7}
\end{figure*}

In this Appendix, we collect all the figures showing the relevant quantities of each DMS galaxy.
Figures \ref{fig:Complete_analysis_1}-~\ref{fig:Complete_analysis_7} show the profiles of the surface brightness, the surface mass density of the atomic and molecular gas and the kinematic profiles. Each galaxy appears in an individual panel. Figures~\ref{fig:Complete_analysis_6} and~\ref{fig:Complete_analysis_7} show the five galaxies that we analysed both in RG and in QUMOND. Each panel contains 8 sub-panels in Figs.~\ref{fig:Complete_analysis_1}-~\ref{fig:Complete_analysis_5} and 12 sub-panels in Figs. \ref{fig:Complete_analysis_6}-\ref{fig:Complete_analysis_7}.

The sub-panels (a) show the surface brightness. The filled circles with error bars are the data in the $K$-band, corrected for inclination~\citep{DMSvi} and the blue solid curves are our models (Appendix~\ref{sec:SB}). The green data points are removed before performing the fit with the exponential profile~\eqref{eq:Id}. We only use this exponential model to estimate the disk surface brightness both in the central region, to separate the disk and the bulge contributions, and in the outer regions not covered by the data but still within the numerical domain
of our Poisson solver. The grey vertical lines show the radius for our bulge-disk decomposition. 

The sub-panels (b) show the surface mass density profile of the atomic gas. The filled circles with error bars are the estimates according to~\citet{DMSvii} and the blue solid curves are our linear interpolations (Appendix~\ref{sec:gas}). The sub-panels (c) show the surface mass density profile of the molecular gas (Appendix~\ref{sec:gas}). Symbols and lines are as in the sub-panels (b).

The sub-panels (d) show the rotation curves. The filled circles with error bars are the data; the blue solid lines show the RG models  
whose parameters are the medians of their posterior distributions estimated from the rotation curve alone (Sect.~\ref{sec:Only_RC}).
The dashed magenta vertical lines show the bulge effective radius, whereas the dashed green vertical lines show the bulge radius we adopt in our disk-bulge decomposition for the surface brightness fit (Appendix~\ref{sec:SB}). For the galaxy UGC 1087, these two radii coincide. The dashed green vertical lines coincide with the grey lines of sub-panels (a). These two vertical lines are repeated in all sub-panels (d)-(h).

The sub-panels (e) again show the rotation curves and the sub-panels (f) show the vertical velocity dispersion profiles. In the sub-panels (e), the data
are the same as in the sub-panels (d). In the sub-panels (f), the data are the filled circles with error bars. In the sub-panels (e) and (f), the blue solid lines show the RG models
whose parameters are the medians of their posterior distributions estimated from the rotation curve and the vertical velocity dispersion profile at the same time (Sect.~\ref{sec:VVD_RC}). In the sub-panels (f), the  cyan solid lines show the vertical velocity dispersion profile when we adopt the same parameters as for the blue lines except for the disk-scale height $h_z$; for the cyan solid lines, $h_z$ is the value derived from Eq.~\eqref{eq:hzhR}. For the galaxy UGC 6918 (Fig.~\ref{fig:Complete_analysis_4}), the cyan line overlaps the blue line because the estimated $h_z$ coincides with the value derived from Eq.~\eqref{eq:hzhR}.

The sub-panels (g) and (h) again show the rotation curves and the vertical velocity dispersion profiles. In these sub-panels, the blue 
solid lines show the RG models where the mass-to-light ratios and the disk-scale heights are set to the values derived in Sect.~\ref{sec:VVD_RC} and listed in Table~\ref{tab:fit_VVD_RC_orig_err_bars_v}, whereas the values of the three RG parameters $\epsilon_0$, $Q$ , and $\rho_{\rm c}$ 
are set to those of the unique combination found in Sect.~\ref{sec:Fit_all}.

Figures~\ref{fig:Complete_analysis_6} and~\ref{fig:Complete_analysis_7} show four additional sub-panels. The sub-panels (i) and (k) show the rotation curves and the sub-panels (j) and (l) show the velocity dispersion profiles. The measured velocity dispersion profiles in these sub-panels are artificially increased by the factor $1.63$, which is our estimate of the observational bias suggested by~\citet{Aniyan16} (Sect.~\ref{sec:VVD_RC_1pt55}). The blue curves are the models whose parameters are the medians of their posterior distributions estimated from the rotation curve and the vertical velocity dispersion profile at the same time (Sect. \ref{sec:VVD_RC_1pt55}). The sub-panels (i) and (j) show the RG models and the sub-panels (k) and (l) show the QUMOND models. The cyan lines in the sub-panels (j) and (l) show the models where the scale height $h_z$ is set by Eq.~\eqref{eq:hzhR}. For the galaxy UGC 4107 (Fig.~\ref{fig:Complete_analysis_6}), the cyan lines in sub-panels (j) and (l) overlap the blue lines because the estimated $h_z$ in both RG and QUMOND are almost identical to the values obtained with Eq.~\eqref{eq:hzhR}.
 The vertical magenta and green lines are the same as in the sub-panels (d)-(h).

\end{appendix}
  
\end{document}